\documentclass[11pt]{article}

\pdfoutput=1
\usepackage{graphics}
\usepackage{subfigure}
\usepackage{epsfig}
\usepackage{latexsym}
\usepackage{graphicx}
\usepackage{color}
\usepackage{amsmath,amssymb}
\usepackage{cite}
\usepackage[colorlinks,urlcolor=black,linkcolor = black,citecolor = black,]{hyperref}
\usepackage{braket}
\usepackage{booktabs}
\usepackage{caption}
\usepackage{amsfonts}
\usepackage{dsfont}

\makeatletter
\def\section{\@startsection {section}{1}{\z@}{-3.5ex plus -1ex minus -.2ex}{2.3ex plus .2ex}{\large\bf}}
\def\subsection{\@startsection{subsection}{2}{\z@}{-3.25ex plus -1ex
minus -.2ex}{1.5ex plus .2ex}{\normalsize\bf}}
\makeatother
\newcommand{\captionfonts}{\small}
\makeatletter  % Allow the use of @ in command names
\long\def\@makecaption#1#2{%
  \vskip\abovecaptionskip
  \sbox\@tempboxa{{\captionfonts #1: #2}}%
  \ifdim \wd\@tempboxa >\hsize
    {\captionfonts #1: #2\par}
  \else
    \hbox to\hsize{\hfil\box\@tempboxa\hfil}%
  \fi
  \vskip\belowcaptionskip}
\makeatother   % Cancel the effect of \makeatletter
\catcode`@=11
\def\marginnote#1{}
\newcount\hour
\newcount\minute
\newtoks\amorpm
\hour=\time\divide\hour
by60
\minute=\time{\multiply\hour by60 \global\advance\minute
by-\hour}
\edef\standardtime{{\ifnum\hour<12 \global\amorpm={am}
\else\global\amorpm={pm}\advance\hour by-12 \fi
 \ifnum\hour=0
\hour=12 \fi
 \number\hour:\ifnum\minute<10
0\fi\number\minute\the\amorpm}}
\edef\militarytime{\number\hour:\ifnum\minute<10
0\fi\number\minute}
\def\draftlabel#1{{\@bsphack\if@filesw
{\let\thepage\relax
 \xdef\@gtempa{\write\@auxout{\string
\newlabel{#1}{{\@currentlabel}{\thepage}}}}}\@gtempa
 \if@nobreak
\ifvmode\nobreak\fi\fi\fi\@esphack}
\gdef\@eqnlabel{#1}}
\def\@eqnlabel{}
\def\@vacuum{}
\def\draftmarginnote#1{\marginpar{\raggedright\scriptsize\tt#1}}
\def\draft{\oddsidemargin
0.0truein
 \def\@oddfoot{\sl preliminary draft \hfil
\rm\thepage\hfil\sl\today\quad\militarytime}
 \let\@evenfoot\@oddfoot
\overfullrule 3pt
 \let\label=\draftlabel
\let\marginnote=\draftmarginnote
\def\@eqnnum{(\theequation)\rlap{\kern\marginparsep\tt\@eqnlabel}
\global\let\@eqnlabel\@vacuum}
}
\catcode`@=12
\newcommand{\beq}{\begin{eqnarray}}
\newcommand{\eeq}{\end{eqnarray}}

\newcommand{\s}{\newline \vspace*{-3.5mm}}
\allowdisplaybreaks

\begin{document}

\thispagestyle{empty}

\begin{center}
\hfill DESY 17-174\\
\hfill KA-TP-39-2017 \\

\begin{center}

\vspace{0.2cm}

{\LARGE\bf The CP-Violating 2HDM in Light of a \\
Strong First Order Electroweak Phase Transition and \\[0.3cm]
Implications for Higgs Pair Production}
\end{center}

\vspace{0.4cm}

\renewcommand{\thefootnote}{\fnsymbol{footnote}}
{\bf P.~Basler$^{\,1\,}$}\footnote{E-mail:
  \texttt{philipp.basler@kit.edu}}, {\bf
  M.~M\"uhlleitner$^{\,1\,}$}\footnote{E-mail:
  \texttt{milada.muehlleitner@kit.edu}} and {\bf J.~Wittbrodt$^{\,2\,}$}\footnote{E-mail: \texttt{jonas.wittbrodt@desy.de}}\\

\vspace{0cm}

${}^1\!\!$
{\em Institute for Theoretical Physics, Karlsruhe Institute of
    Technology,} \\
{\em Wolfgang-Gaede-Str.~1, 76131 Karlsruhe, Germany}\\[3mm]
${}^2\!\!$
{\em Deutsches Elektronen-Synchrotron DESY, Notkestra{\ss}e 85, D-22607
Hamburg, Germany}
\\

\end{center}

\vspace{0.2cm}
\centerline{\bf Abstract}
\begin{quote}
We investigate the strength of the electroweak phase transition (EWPT)
within the CP-violating 2-Higgs-Doublet Model (C2HDM). The 2HDM is a
simple and well-studied model, which can feature CP violation at
tree level in its extended scalar sector. This makes it, in
contrast to the Standard Model (SM), a promising candidate for explaining the baryon
asymmetry of the universe through electroweak baryogenesis. We apply a
renormalisation scheme which allows efficient scans of the C2HDM
parameter space by using the loop-corrected masses and mixing matrix
as input parameters. This procedure enables us to investigate the
possibility of a strong first order EWPT required for baryogenesis and
study its phenomenological implications for the LHC. Like in the
CP-conserving (real) 2HDM (R2HDM) we find that a strong EWPT favours
mass gaps between the non-SM-like Higgs bosons. These lead to
prominent final states comprised of gauge+Higgs bosons or pairs of
Higgs bosons. In contrast to the R2HDM, the CP-mixing of the C2HDM
also favours approximately mass degenerate spectra with dominant
decays into SM particles. The requirement of a strong EWPT further
allows us to distinguish the C2HDM from the R2HDM using the signal
strengths of the SM-like Higgs boson. We additionally find that a
strong EWPT requires an enhancement of the SM-like trilinear Higgs
coupling at next-to-leading order (NLO) by up to a factor of 2.4 compared to
the NLO SM coupling, establishing another link
between cosmology and collider phenomenology. We provide several C2HDM
benchmark scenarios compatible with a strong EWPT and all experimental
and theoretical constraints. We include the dominant branching ratios
of the non-SM-like Higgs bosons as well as the Higgs pair production
cross section of the SM-like Higgs boson for every benchmark
point. The pair production cross sections can be substantially
enhanced compared to the SM and could be observable at the
high-luminosity LHC, allowing access to the trilinear Higgs
couplings. 
\end{quote}

%%%%%%%%%%%%%%%%%%%%%%%%%%%%%%%%%%%%%%%%%%%%%%%%%%%%%%%%%%%
\newpage
\setcounter{page}{1}
\setcounter{footnote}{0}

%%%%%%%%%%%%%%%%%%%%%%%%%%%%%%%%%%%%%%%%%%%%%%%%%%%%%%%%%%%
\section{Introduction}
The discovery of the Higgs boson by the LHC experiments ATLAS
\cite{:2012gk} and CMS \cite{:2012gu} has marked a milestone for
particle physics while, at the same time, leaving many open
questions. Despite the Standard Model (SM) nature of the Higgs boson
\cite{Aad:2015mxa,Khachatryan:2014kca,Aad:2015gba,Khachatryan:2014jba}
new physics (NP) beyond the SM is called for in order to solve the
puzzles within the standard theory. The observed baryon asymmetry of the
Universe (BAU) \cite{Bennett:2012zja} is one example that requires NP
extensions. It can be generated dynamically in the early Universe
during a first order electroweak phase transition (EWPT) through the
mechanism of electroweak baryogenesis (EWBG)
\cite{Kuzmin:1985mm,Cohen:1990it,Cohen:1993nk,Quiros:1994dr,Rubakov:1996vz,Funakubo:1996dw,Trodden:1998ym,Bernreuther:2002uj,Morrissey:2012db} 
provided that all three Sakharov conditions \cite{sakharov} are
fulfilled. These are baryon number violation, C and CP violation and departure from the
thermal equilibrium. A strong first order phase transition (PT)
\cite{Trodden:1998ym,Morrissey:2012db} proceeds through bubble
nucleation and can generate a net baryon number near the bubble
wall. Diffusion through the bubble preserves this asymmetry for the
later evolution of the universe by suppressing the sphaleron
transitions in the false vacuum \cite{Manton:1983nd,Klinkhamer:1984di}. 
% 1206.2942
A first order EWPT takes places when bubbles of the broken phase
nucleate in the surrounding plasma of the symmetric phase. 
The bubble walls interact with the
various fermion species in the plasma. If there is CP
violation in the bubble wall or some CP violation in the hot {\it
  i.e.}~the symmetric phase that is disturbed by the wall, particles with opposite chirality
interact differently with the wall and CP and C asymmetries in
particle number densities can be generated in front of the bubble
wall. These asymmetries diffuse into the hot plasma ahead of the 
bubble wall biasing baryon number violating electroweak (EW) sphaleron transitions
to generate a baryon asymmetry. The latter is transferred through the
expanding wall into the broken phase. The rate of sphaleron
transitions is strongly suppressed 
in this phase such that a washout of the baryons generated before
is avoided.  The required departure from thermodynamic equilibrium
is guaranteed by the passage of the bubble walls that rapidly expand
through the cosmological plasma. Additionally, gravitational
waves produced by a strong first order EWPT \cite{Grojean:2006bp} are
potentially observable by the future space-based gravitational wave
interferometer eLISA \cite{Caprini:2015zlo}. The interplay between
gravitational waves and a strong first order PT and/or collider
phenomenology has recently been studied in \cite{Hindmarsh:2015qta,Huber:2015znp,Hashino:2016rvx,Addazi:2016fbj,Huang:2016cjm,Hashino:2016xoj,Vaskonen:2016yiu,Dorsch:2016nrg,Saeedhosini:2017dsh, Chao:2017vrq,Marzola:2017jzl,Bian:2017wfv,Hindmarsh:2017gnf,Weir:2017wfa,Kang:2017mkl,Cai:2017tmh,Chiang:2017zbz,Jinno:2017ixd,Huang:2017kzu}.
\s

In the SM in principle all three Sakharov conditions could be
realized. However, the EWPT is not of strong first order
\cite{smnot}, since this would require a SM Higgs boson mass of around 70-80~GeV
\cite{notsm2}. Additionally, the CP violation of the SM
arising from the Cabibbo-Kobayashi-Maskawa (CKM) matrix is too small
\cite{Morrissey:2012db,Gavela:1993ts,Konstandin:2013caa,amountcpviol}. Extensions
beyond the SM provide additional sources of CP violation as well as
further scalar states triggering a first order EWPT also for a SM-like
Higgs boson with a mass of 125~GeV. This is the case for the 2-Higgs Doublet
Model (2HDM) \cite{Lee:1973iz,Branco:2011iw} which belongs to the
simplest NP extensions that are compatible with present experimental
constraints. Previous studies have shown that
2HDMs provide a good framework for successful baryogenesis, both in
the CP-conserving 
\cite{baryo2HDM,Dorsch:2013wja,Dorsch:2014qja,Basler:2016obg,Dorsch:2017nza} 
and in the CP-violating case \cite{funakubo1,Cline:1995dg,funakubo2,Cline:1996mga,huberwithcp,Cline:2011mm,Dorsch:2016nrg,Haarr:2016qzq}. 
2HDMs feature five physical Higgs bosons, and the tree-level 2HDM Higgs
sector provides sources for explicit CP violation. 
Allowing for CP violation, there could in principle be a
complex phase between the vacuum expectation values (VEVs) of the two
Higgs doublets. This phase can, however, always be removed by
a change of basis \cite{Ginzburg:2002wt} so that without loss of
generality it can be set to zero. This may not be the case any more at
finite temperature. CP violation 
might be generated spontaneously only in the bubble wall around the
critical temperature \cite{Cline:1995dg} and provide a source for the
generation of the matter-antimatter asymmetry through EWBG. The
CP-violating phase which at zero 
temperature is just another parameter of the theory, during EWPT becomes a
spatially varying field, and its value depends on the position
relative to the bubble walls. In order to study the effect of CP violation on the
generation of the baryon asymmetry the detailed form of the
spatially varying field has to be determined. \s
%can be computed by solving the
%relevant equations of motion including CP violation. 

%The generated chiral asymmetry 
%is converted into a baryon asymmetry in the false vacuum. This
%asymmetry  is transferred to the true vacuum when it passes the bubble wall
%\cite{Konstandin:2013caa}, provided there is departure from the
%thermal equilibrium.

%\bibitem{Hindmarsh:2013xza}
%  M.~Hindmarsh, S.~J.~Huber, K.~Rummukainen and D.~J.~Weir,
%  %``Gravitational waves from the sound of a first order phase transition,''
%  Phys.\ Rev.\ Lett.\  {\bf 112} (2014) 041301
%  doi:10.1103/PhysRevLett.112.041301
%  [arXiv:1304.2433 [hep-ph]].
 
CP-violating Higgs sectors are strongly constrained by the electric
dipole moments (EDMs). The strongest constraint \cite{Inoue:2014nva}
is imposed by the limit on the electron EDM provided by the ACME
collaboration \cite{Baron:2013eja}. The possibility of
spontaneous CP violation generated at the EWPT which vanishes at zero
temperature may provide an attractive scenario to lift the possible tension
between the restrictions imposed by the EDMs and the requirement of a
substantial amount of CP violation by the EWBG. \s

In \cite{Basler:2016obg} we investigated the implications of a strong first
order PT in the CP-conserving or real 2HDM (R2HDM) on the LHC Higgs
phenomenology. We 
found a strong interplay between the requirement of successful baryogenesis
and LHC Higgs phenomenology. In this work, we extend our analysis to the
CP-violating 2HDM (C2HDM).\footnote{For other recent works on the link
between CP violation and electroweak baryogenesis, see
\cite{Ayala:2017gqa,Chao:2017oux,Bruggisser:2017lhc}.}
The computation of the equation of motion for
the CP-violating phase between the two Higgs doublets and the
computation of the actual baryon-antibaryon asymmetry generated
through EWBG within the framework of the C2HDM is beyond
the scope of this paper. We focus instead on the interplay between the
requirement of a strong first order phase transition and LHC
phenomenology in the presence of explicit CP violation in the
tree-level 2HDM Higgs sector. We investigate the possible
spontaneous generation of a CP-violating phase at the EWPT. We
furthermore analyse in detail the effect of
higher order corrections on the trilinear Higgs self-couplings
extracted from the one-loop corrected effective
potential.\footnote{Recent investigations on the interplay between a
  strong first order phase transition and the size of the trilinear
  Higgs self-couplings can also be found in \cite{Huang:2017jws,
    Jain:2017sqm, Reichert:2017puo}.} We 
discuss the impact of the requirement of a strong phase transition on
their size and the resulting implications for LHC phenomenology,
namely Higgs pair production. We present several benchmark scenarios,
emphasizing the specific features of C2HDM parameter points compatible
with all constraints and a strong phase transition. \s

For the purpose of this paper we compute the one-loop
corrected effective potential at finite temperature
\cite{ColemanWeinberg,Quiros:1999jp,Dolan:1973qd} including daisy
resummations for the bosonic masses \cite{Carrington:1991hz}. For the
numerical analysis the parameter space of the C2HDM is scanned and
tested for compatibility of the model with the theoretical and
experimental constraints. Subsequently, the implication of a strong
first order PT on the surviving parameter sets is determined and
interpreted with respect to collider phenomenology. The former
necessitates the minimisation of the  loop-corrected Higgs potential
at increasing temperature in order to find the vacuum expectation value
$v_c$ at the critical temperature $T_c$, which is defined as the temperature
where two degenerate global minima exist. A value of
$v_c/T_c$ larger than one is indicative of a strong first order PT
\cite{Quiros:1994dr,Moore:1998swa}.\footnote{Discussions on the
  gauge dependence of $v_c/T_c$ can be found {\it e.g.}~in
  \cite{Dolan:1973qd,Patel:2011th,Wainwright:2011qy,Garny:2012cg}.} 
In order to be able to perform an efficient scan, like in \cite{Basler:2016obg}, we
renormalise the loop-corrected potential in such a way that not only
the VEV and all physical Higgs boson masses, but also all mixing
matrix elements remain at their tree-level values. In our analysis, we
will focus on the C2HDM with type I and type II couplings of the Higgs
doublets to the fermions. We will discard parameter points inducing a 
2-stage PT \cite{Land:1992sm,Hammerschmitt:1994fn}. Our analysis
reveals a strong link between the demand for a strong first order PT and testable
implications at the collider experiments. \s

The outline of the paper is as follows: In section
\ref{sec:effpot} we set our notation and present the loop-corrected
effective potential of the C2HDM at finite temperature. Our
renormalisation procedure is described in section \ref{sec:renorm}. Section
\ref{sec:numerical} is dedicated to the description of the numerical analysis. It
includes the outline of the minimisation procedure of the effective
potential and the details of the scan in the C2HDM parameter space as
well as of the applied theoretical and experimental
constraints. Sections \ref{sec:results}-\ref{eq:trilt2}  contain our
results. In Section \ref{sec:Ihleqh125} we analyse the type I C2HDM
with the lightest Higgs boson being the SM-like Higgs state. We first
investigate the spontaneous generation of a CP-violating phase and its
relation to explicit CP violation in the tree-level potential. We then 
present the parameter  regions compatible with the applied
constraints and a strong first order PT and analyse the
implications for collider phenomenology. In Section
\ref{sec:triltype1} we discuss in detail the role of the trilinear Higgs
self-couplings in the EWPT, the impact of the next-to-leading order
(NLO) corrections derived 
from the effective potential and the
implications of the requirement of a strong phase transition on the
Higgs self-couplings and their corrections. In Section
\ref{sec:Ihheqh125} we briefly summarise the results for the type I
C2HDM with the next-to-lightest Higgs boson representing the SM-like
Higgs scalar. Sections \ref{sec:IIhleq125} and \ref{eq:trilt2} are dedicated to the type II
C2HDM with the lightest Higgs boson being SM-like, and we present our
results in analogy to the type I case. Section \ref{sec:conclusions}
contains our conclusions. 
%The paper is
%accompanied by an appendix 
%containing the formulae for the masses of the relevant
%particles and, where appropriate, for the daisy resummed mass
%corrections.

%%%%%%%%%%%%%%%%%%%%%%%%%%%%%%%%%%%%%%%%%%%%%%%%%%%%%%%%%%%
\section{The effective potential in the C2HDM}\label{sec:effpot}
In this section we provide the loop-corrected effective potential at finite
temperature for the CP-violating 2HDM. We start by setting
our notation.

%%%%%%%%%%%%%%%%%%%%%%%%%%%%%%%%%%%%%%%%%%%%%%%%%%%%%%%%%%
\subsection{The CP-violating 2-Higgs-Doublet Model\label{sec:c2hdm}}
The tree-level potential of the C2HDM for the two $SU(2)_L$ scalar doublets
\beq
\Phi_1 = \left( \begin{array}{c} \phi_1^+ \\ \phi_1^0 \end{array} \right)
\quad \mbox{and} \quad
\Phi_1 = \left( \begin{array}{c} \phi_2^+ \\ \phi_2^0 \end{array}
\right) \;,
\eeq
reads
\begin{align}
\begin{split}
V_{\text{tree}} &= m_{11}^2 \Phi_1^\dagger \Phi_1 + m_{22}^2
\Phi_2^\dagger \Phi_2 - \left[m_{12}^2 \Phi_1^\dagger \Phi_2 +
  \mathrm{h.c.} \right] + \frac{1}{2} \lambda_1 ( \Phi_1^\dagger
\Phi_1)^2 +\frac{1}{2} \lambda_2 (\Phi_2^\dagger \Phi_2)^2 \\
&\quad + \lambda_3 (\Phi_1^\dagger \Phi_1)(\Phi_2^\dagger\Phi_2) +
\lambda_4 (\Phi_1^\dagger \Phi_2)(\Phi_2^\dagger \Phi_1)
+ \left[ \frac{1}{2} \lambda_5 (\Phi_1^\dagger\Phi_2)^2  +
  \mathrm{h.c.} \right] \; .
\end{split}\label{eq:treepot}
\end{align}
It incorporates a softly broken $\mathbb{Z}_2$ symmetry, under which
the doublets transform as $\Phi_1\rightarrow \Phi_1,\ \Phi_2
\rightarrow -\Phi_2$. This ensures the absence of tree-level
Flavor Changing Neutral Currents (FCNC). The hermiticity of the
potential $V_{\text{tree}}$ forces all parameters to be real apart from the soft
$\mathbb{Z}_2$ breaking mass parameter $m_{12}^2$ and the quartic
coupling $\lambda_5$. For $\mbox{arg}(m_{12}^2)=\mbox{arg}(\lambda_5)$
the complex phases of these two parameters can 
be absorbed by a basis transformation. If furthermore the VEVs
of the two doublets are assumed to be real, we are in the real or
CP-conserving 2HDM. Otherwise, we are in the C2HDM, for which we will
adopt the conventions of \cite{Fontes:2014xva} in the following. \s 
%After EWSB, the two doublet $\Phi_i$ develop
%VEVs, which can be complex if CP violation is allowed. Since the
%relative phase between the VEVs can be rotated away by a global phase
%transition in the field $\Phi_2$ \cite{}, without loss of generality, 
%we will set this phase to zero at $T=0$. The expansion of the two
%doublets about the real VEVs $v_1$ and $v_2$ is given by

After EW symmetry breaking the
two Higgs doublets acquire VEVs $\bar{\omega}_i \in \mathbb{R}$, about
which the Higgs fields can be expanded in terms of the charged CP-even
and CP-odd fields $\rho_i$ and $\eta_i$, and the neutral CP-even and
CP-odd fields $\zeta_i$ and $\psi_i$, $i=1,2$. In the general 2HDM, there are
three different types of minima, given by the normal EW-breaking one, a
CP-breaking minimum, and a charge-breaking (CB) vacuum. In
Refs.~\cite{Ferreira:2004yd,Barroso:2007rr,Ivanov:2007de} it has been
shown that at tree level minima that break different symmetries cannot
coexist. If a normal minimum exists,
all CP or CB stationary points are proven to be saddle points. These
statements may not be true any more at higher orders, as recent
studies have shown for the Inert 2HDM at one-loop level in the effective potential
approach \cite{Ferreira:2015pfi}. Consequently, we allow
for the possibility of a CP-breaking vacuum as well as a
charge-breaking one. Through the VEVs $\bar{\omega}_{\text{CP}}$ and 
$\bar{\omega}_{\text{CB}}$ we include the possibility of generating at
one-loop and/or non-zero temperature a global minimum that is
CP-violating and/or charge breaking. As a charge-breaking VEV breaks
electrical charge conservation inducing a massive photon, this
unphysical configuration of the vacuum will not further be discussed
in the numerical analysis. Denoting the VEVs of the normal vacuum by 
$\bar{\omega}_{1,2}$ and the CP- and charge-breaking VEVs by
$\bar{\omega}_{\text{CP}}$ and $\bar{\omega}_{\text{CB}}$, 
respectively, the expansion of the two Higgs doublets about the VEVs is given by
\begin{align}
\Phi_1 & = \frac{1}{\sqrt{2}} \begin{pmatrix}
\rho_1 + \mathrm{i}\eta_1 \\ \bar{\omega}_1 + \zeta_1 +
\mathrm{i}\psi_1 \label{eq:model_3}
\end{pmatrix} \\
\Phi_2 & = \frac{1}{\sqrt{2}} \begin{pmatrix}
\bar{\omega}_{\text{CB}} + \rho_2 + \mathrm{i} \eta_2 \\ \bar{\omega}_2 +\mathrm{i} \bar{\omega}_{\text{CP}}
+ \zeta_2 + \mathrm{i} \psi_2
\end{pmatrix} \,, \label{eq:model_4}
\end{align}
where, without loss of generality, the complex part of the VEVs and
the charge-breaking VEV have been rotated to the second doublet
exclusively. The VEVs of our present vacuum at zero 
temperature\footnote{While strictly speaking $T=2.7$~K (corresponding
  to about $10^{-13}$~GeV in natural units) there is no
  discernible numerical difference to the choice T=0.} are denoted as
\beq
v_i \equiv \bar{\omega}_i|_{T=0} \;, \quad i=\mbox{1,\,2,\,CP,\,CB}\;,
\eeq
with 
\beq
v_{\text{CP}} = v_{\text{CB}} = 0 \;.
\eeq
The VEVs of the normal vacuum, $v_{1,2}$, are related to the SM VEV $v
\approx 246$~GeV by
\beq
v_1^2 + v_2^2 \equiv v^2 \;.
\eeq
The angle $\beta$ defines the ratio of $v_1$ and $v_2$,
\beq
\tan\beta = \frac{v_2}{v_1} \;,
\eeq
so that
\beq
v_1 = v \cos \beta \qquad \mbox{and} \qquad
v_2 = v \sin\beta\ . \label{eq:v1v2}
\eeq
The minimum conditions of the potential Eq.~\eqref{eq:treepot},
\begin{align}
\left. \frac{\partial V_{\text{tree}}}{\partial
  \Phi_i^\dagger}\right|_{\Phi_j =\langle\Phi_j\rangle }
  &\overset{!}{=} 0 \qquad i,j\in\{1,2\}  \label{eq:mincond} \;,
\end{align}
where the brackets denote the Higgs field values in the minimum, {\it
i.e.}~$\langle\Phi_i\rangle = (0, v_i/\sqrt{2})$ at $T=0$, result in
\begin{subequations}
\label{eq:eqsmincond}
\begin{align}
m_{11}^2 & =  \mbox{Re}(m_{12}^2) \frac{v_2}{v_1} - \frac{v_1^2}{2} \lambda_1 -
           \frac{v_2^2}{2} \lambda_{345}  \label{eq:eqsmincond1}\\ 
m_{22}^2 & = \mbox{Re}(m_{12}^2) \frac{v_1}{v_2} -\frac{v_2^2}{2} \lambda_2 - \frac{v_1^2}{2}
\lambda_{345} \label{eq:eqsmincond2} \\
2 \,\mbox{Im}(m_{12}^2) & = v_1 v_2 \, \mbox{Im}(\lambda_5) \;,
\label{eq:eqsmincond3}
\end{align}
\end{subequations}
where we have introduced the abbreviation
\beq
\lambda_{345} \equiv \lambda_3 + \lambda_4 + \mbox{Re}(\lambda_5) \;.
\eeq
Equations (\ref{eq:eqsmincond1}) and (\ref{eq:eqsmincond2}) can be
used to trade the parameters $m_{11}^2$ and $m_{22}^2$ for $v_1$ and
$v_2$, while Eq.~(\ref{eq:eqsmincond3}) leads to a relation between
the two sources of CP violation in the scalar potential so that one of
the ten parameters of the C2HDM is fixed. Introducing
\beq
\zeta_3 = -\psi_1 \sin\beta + \psi_2 \cos \beta \;,
\eeq
the neutral mass eigenstates $H_i$ ($i=1,2,3$) are obtained from the
C2HDM basis $\zeta_1$, $\zeta_2$ and $\zeta_3$ through the rotation
\beq
\left( \begin{array}{c} H_1 \\ H_2 \\ H_3 \end{array} \right) = R
\left( \begin{array}{c} \zeta_1 \\ \zeta_2 \\ \zeta_3 \end{array} \right)
\;.
\label{eq:eigenstatesrot}
\eeq
The corresponding Higgs masses are obtained from the mass matrix
\beq
({\cal M}^2)_{ij} = \left\langle \frac{\partial^2 V}{\partial \zeta_i
  \partial \zeta_j} \right\rangle \;,
\label{eq:massmat}
\eeq
through the diagonalisation with the matrix $R$, 
\beq
R {\cal M}^2 R^T = \mbox{diag} (m_{H_1}^2, m_{H_2}^2, m_{H_3}^2) \;.
\eeq
The Higgs bosons are ordered by ascending mass as $m_{H_1} \le m_{H_2}
\le m_{H_3}$. With the abbreviations $s_i \equiv \sin \alpha_i$ and
$c_i \equiv \cos \alpha_i$, where
\beq
-\frac{\pi}{2} \le \alpha_i < \frac{\pi}{2}  \;,
\eeq 
the mixing matrix $R$ can be parametrised as
\beq
R =\left( \begin{array}{ccc}
c_{1} c_{2} & s_{1} c_{2} & s_{2}\\
-(c_{1} s_{2} s_{3} + s_{1} c_{3})
& c_{1} c_{3} - s_{1} s_{2} s_{3}
& c_{2} s_{3} \\
- c_{1} s_{2} c_{3} + s_{1} s_{3} &
-(c_{1} s_{3} + s_{1} s_{2} c_{3})
& c_{2}  c_{3}
\end{array} \right) \;.
\label{eq:rmatrix}
\eeq
Exploiting the minimum conditions of the potential at zero
temperature, we use the following set of 9 independent parameters of the
C2HDM \cite{ElKaffas:2007rq},
\beq
v \;, \quad t_\beta \;, \quad \alpha_{1,2,3}
\;, \quad m_{H_i} \;, \quad m_{H_j} \;, \quad m_{H^\pm} \quad
\mbox{and} \quad \mbox{Re}(m_{12}^2) \;.
\label{eq:c2hdmparams}
\eeq
The $m_{H_i}$ and $m_{H_j}$ denote any two among the
three neutral Higgs boson masses, and the mass of the third Higgs boson is
obtained from the other parameters
\cite{ElKaffas:2007rq}. For the analytic relations between the above
parameter set and the coupling parameters $\lambda_i$ of the 2HDM
Higgs potential, see \cite{Fontes:2014xva}. \s

The limit of the CP-conserving 2HDM is obtained for $\alpha_2=
\alpha_3=0$ and $\alpha_1 = \alpha + \pi/2$ \cite{Khater:2003wq}. The
mass matrix Eq.~(\ref{eq:massmat}) becomes block diagonal in this case
leading to the pure pseudoscalar $A$, which is identified with
$H_3$, and the CP-even mass eigenstates $h$ and $H$, which are
obtained from the gauge eigenstates $\zeta_{1,2}$ through the rotation
with the angle $\alpha$,
\beq
\left( \begin{array}{c} H \\ h \end{array} \right) =
\left( \begin{array}{cc} c_\alpha & s_\alpha \\ -s_\alpha &
    c_\alpha \end{array} \right) \left( \begin{array}{c} \zeta_1 \\
    \zeta_2 \end{array} \right) \;.
\eeq

The imposed $\mathbb{Z}_2$ symmetry ensures that each of the up-type quarks,
down-type quarks and charged leptons can only couple to one of the
Higgs doublets so that FCNCs at
tree level are avoided. Table~\ref{tab:types} lists the possible
different 2HDM types given by type~I, type~II, lepton-specific and
flipped. 
\begin{table}[t!]
 \begin{center}
 \begin{tabular}{lcccc}
     \toprule
    & Type I & Type II & Lepton-Specific & Flipped \\
   \midrule
Up-type quarks & $\Phi_2$ & $\Phi_2$ & $\Phi_2$ & $\Phi_2$ \\
Down-type quarks & $\Phi_2$ & $\Phi_1$ & $\Phi_2$ & $\Phi_1$ \\
Leptons & $\Phi_2$ & $\Phi_1$ & $\Phi_1$ & $\Phi_2$ \\
   \bottomrule
  \end{tabular}
\caption{Classification of the Yukawa sector in the 2HDM according to the couplings of the fermions to the Higgs doublets.\label{tab:types}}
   \end{center}
 \end{table}
For the exact form of the C2HDM Higgs couplings to the SM particles in
terms of the input parameters, we refer to
Refs.~\cite{Fontes:2015mea,Muhlleitner:2017dkd,Grober:2017gut}. In
this work we focus on C2HDMs of type~I and type~II. 

%%%%%%%%%%%%%%%%%%%%%%%%%%%%%%%%%%%%%%%%%%%%%%%%%%%%%%%%%%%
\subsection{One-loop effective potential at finite
  temperature}\label{sec:onelooppot}
The form of the one-loop effective potential at finite temperature for the C2HDM
case does not change with respect to the one introduced for the
CP-conserving 2HDM in Ref.~\cite{Basler:2016obg}. For convenience of
the reader, we briefly repeat the main ingredients, also in order to
set our notation. \s

The one-loop contribution $V_1$ to the effective potential consists of
the Coleman-Weinberg (CW) contribution $V_{\text{CW}}$
\cite{ColemanWeinberg} already present at zero temperature, and the
contribution $V_{T}$  for the thermal corrections at finite
temperature $T$. The one-loop corrected effective potential reads
\begin{equation}
 V = V_{\text{tree}} + V_1
\equiv V_{\text{tree}} + V_{\text{CW}} + V_{T} \label{eq:effoneloop} \,,
\end{equation}
with the tree-level potential given in Eq.~(\ref{eq:treepot}) after replacing
the doublets $\Phi_{1,2}$ with their classical constant field
configuration 
\beq
\Phi_1^c = \left( \begin{array}{c} 0 \\ \frac{\omega_1}{\sqrt{2}} \end{array}
\right)
\quad \mbox{and} \quad 
\Phi_2^c = \left( \begin{array}{c}
                    \frac{\omega_{\text{CB}}}{\sqrt{2}}
                    \\ \frac{\omega_2 +
i\omega_\text{CP}}{\sqrt{2}} \end{array}\right) \;. 
\eeq 
In the $\overline{\text{MS}}$ scheme the Coleman-Weinberg
potential for a particle $i$ reads \cite{Quiros:1999jp}
\begin{equation}
 V_{\text{CW}} (\{\omega\})= \sum_i \frac{n_i}{64\pi^2} (-1)^{2s_i} \,
 m_i^4(\{\omega\}) \,
 \left[\log\left(\frac{m_i^2(\{\omega\})}{\mu^2}\right)-c_i\right]\, .
\label{eq:cwpot}
\end{equation}
The sum extends over the Higgs and Goldstone bosons, the massive
gauge bosons, the longitudinal photon and the fermions $f$, with the
exception of the neutrinos, which we assume to be 
massless, $i=h, H, A, H^\pm, G^0, G^\pm, W^\pm, Z, \gamma_L,
f$. Here, $m_i^2$ denotes the respective eigenvalue for the particle
$i$ of the mass matrix squared expressed through the tree-level relations in terms of
$\omega_i$ ($i=1,\,2,\,\mbox{CP},\,\mbox{CB}$). 
The sum also extends over the Goldstone bosons and the photon. 
Although in the Landau gauge applied here the Goldstone bosons are
massless at $T=0$, they can become massive for field configurations
different from the tree-level VEVs at $T=0$, which are required in the
minimisation procedure. This is also the case for the photon, as we
allow for non-physical vacuum configurations with a non-zero charge
breaking VEV. Furthermore, the Goldstone bosons and the
longitudinal photon can become massive due to the temperature
corrections discussed below. Because of the Landau gauge we need
not consider any ghost contributions. The spin of the particle is
denoted by $s_i$ and the number of degrees of freedom by $n_i$. 
For the scalars $\Phi = H_i, G^0, H^+,H^-,G^+,G^-$, the
charged leptons\footnote{Because of the CB-breaking VEV we have to take into 
  account different masses for the charge conjugated particles.} $l^+$
and $l^-$, the quarks and antiquarks $q$ and $\bar{q}$, the
longitudinal and transversal gauge bosons
$V_L=Z_L,W^+_L,W^-_L,\gamma_L$ and $V_T = Z_T,W^+_T,W^-_T,\gamma_T$,
they are 
\beq
\begin{array}{llllllllll}
n_{\Phi} &=& 1\;,& \quad n_{l^+} &=& 2 \;, & \quad n_{l^-} &=& 2\;, \\
n_{q} &=& 6 \;, & \quad n_{\bar{q}} &=& 6 \;, & 
\quad n_{V_T} &=& 2\;, & \quad n_{V_L} = 1 \;. 
\end{array}
\eeq
%For the neutral scalars $\Phi^0\equiv
%h,H,A, G^0$, the charged scalars $\Phi^\pm \equiv H^\pm, G^\pm$,
%the leptons $l$, the quarks $q$ and the longitudinal and transversal
%gauge bosons, $V_L \equiv Z_L, W_L, \gamma_L$ and $V_T \equiv Z_T,
%W_T, \gamma_T$, they are
%\beq
%\begin{array}{llll}
%n_{\Phi^0} = 1\;,& \quad n_{\Phi^\pm} = 2\;, & \quad n_l = 4\;, &\quad
%n_q = 12\;, \\
%n_{W_T} = 4 \;, & \quad n_{W_L} = 2\;,  & \quad n_{Z_T} = 2 \;, &
%\quad n_{Z_L} = 1 \;, \\
%n_{\gamma_T} = 2 \;, & \quad n_{\gamma_L} = 1 \;.
%\end{array}
%\eeq
In the $\overline{\text{MS}}$ scheme employed here the constants $c_i$ read 
\begin{equation}\label{eq:looppot_3}
 c_i= \begin{cases}
       \frac{5}{6}\ , & i=W^\pm,Z,\gamma\ \\
       \frac{3}{2}\ , & \text{otherwise}\ .
      \end{cases}
\end{equation}
The renormalisation scale $\mu$ is fixed to $\mu=v= 246.22\
\text{GeV}$. \s

The thermal corrections $V_T$ comprise the daisy resummation
\cite{Carrington:1991hz}
of the $n=0$ Matsubara modes of the longitudinal components of the
gauge bosons $W_L^+, W_L^-, Z_L, \gamma_L$ and the bosons $\Phi$,
so that their masses receive Debye corrections at non-zero
temperature. The potential $V_T$ can be cast into the form
\cite{Dolan:1973qd,Quiros:1999jp}
\beq
V^T = \sum_k n_k \, \frac{T^4}{2\pi^2} \, J^{(k)}_{\pm} \;,
\label{eq:looppot_4}
\eeq
with $k = W_L^+, W_L^-, Z_L, \gamma_L,W_T^+,
W_T^-,Z_T,\gamma_T,\Phi,l^+,l^-,q,\bar{q}$. Since the
Goldstone bosons and the photon acquire a mass at finite temperature,
they have to be included in the sum. 
Denoting the mass eigenvalue including the thermal corrections
for the particle $k$ by $\overline{m}_k$, we have for $J_\pm^{(k)}$ (see {\it
  e.g.}~\cite{Carena:2008vj})
\beq
J^{(k)}_{\pm} = \left\{
\begin{array}{ll}
J_- \left( \frac{m_k^2}{T^2} \right) - \frac{\pi}{6} (\overline{m}_k^3 -
m_k^3) & k = W_L^+, W_L^-, Z_L, \gamma_L, \Phi \\
J_- \left( \frac{m_k^2}{T^2} \right) & k = W_T^+, W_T^-, Z_T, \gamma_T\\
J_+ \left( \frac{m_k^2}{T^2} \right) & k=l^+,l^-,q,\bar{q}
\end{array}
\right.
\label{eq:thermalj}
\eeq
with the thermal integrals
\begin{equation}
 J_{\pm}\left(\frac{m_k^2}{T^2}\right) = \mp \int_0^\infty \text{d}x \, x^2 \log\left[1\pm e^{-\sqrt{x^2+m_k^2/T^2}}\right]\ ,\label{eq:looppot_5}
\end{equation}
where $J_+$ ($J_-$) applies for $k$ being a fermion (boson). The
masses $m_i$ depend implicitly on the temperature $T$, since for each $T$ we
determine the VEVs, respectively the field configurations,
$\bar{\omega}_i = \bar{\omega}_i (T)$, that minimise the
loop-corrected potential $V$, Eq.~(\ref{eq:effoneloop}). These field
configurations enter the tree-level mass matrices. The
$\overline{m}_k$ in addition depend explicitly on $T$ through the thermal
corrections. With the definition of $J^{(k)}_{\pm}$ Eq.~(\ref{eq:thermalj})
we follow the 'Arnold-Espinosa' approach of
Ref.~\cite{Arnold:1992rz}. A different approach has been proposed in
\cite{Parwani:1991gq}, to which we refer as 'Parwani' method. Here the
Debye corrections are included for all the bosonic thermal loop
contributions and the Debye corrected masses are also used in the CW
potential. Since the 'Parwani' method admixes higher-order
contributions, possibly leading to dangerous artifacts at one-loop
level, we apply the 'Arnold-Espinosa' method. For a discussion and
comparison of the two methods, see also
\cite{Cline:1996mga,Cline:2011mm}. \s

The minimisation procedure requires the numerical evaluation of the
integral Eq.~\eqref{eq:looppot_5} at each configuration in
$\{\omega\}$ and $T$, which is very time consuming. The
integrals $J_{\pm}$ are therefore approximated by a series in $x^2 \equiv m^2/T^2$. For
small $x^2$ we have \cite{Cline:1996mga}
 \begin{align}
 \begin{split} \label{eq:looppot_6}
J_{+,\text{s}}(x^2,n) & = -\frac{7\pi^4}{360} +
\frac{\pi^2}{24}x^2 +\frac{1}{32}x^4
\left(\log x^2-c_{+}\right) \\
& \quad - \pi^2x^2 \sum_{l=2}^{n}
\left(-\frac{1}{4\pi^2}x^2\right)^l \frac{(2l-3)!!
  \zeta(2l-1)}{(2l)!! (l+1)} \left(2^{2l-1}-1\right)
\end{split} \\
\begin{split}\label{eq:looppot_7}
J_{-,\text{s}}(x^2,n) & =  - \frac{\pi^4}{45} +
\frac{\pi^2}{12}x^2 -
\frac{\pi}{6}\left(x^2\right)^{3/2} -
\frac{1}{32}x^4
\left(\log x^2-c_{-} \right) \\
& \quad + \pi^2 x^2 \sum_{l=2}^{n}
\left(-\frac{1}{4\pi^2}x^2 \right)^l \frac{(2l-3)!!
  \zeta(2l-1)}{(2l)!! (l+1)}  \ ,
\end{split}
\end{align}
with
\beq
c_+= 3/2+2\log\pi -2\gamma_E \quad \mbox{and} \quad c_-=c_+ +2\log 4 \;,
\eeq
where $\gamma_E$ denotes the Euler-Mascheroni constant, $\zeta(x)$ the
Riemann $\zeta$-function and $(x)!!$ the double factorial. For large
$x^2$ we use for both the fermions and the bosons \cite{Cline:1996mga}
\begin{equation}\label{eq:looppot_8}
 J_{\pm,\text{l}}(x^2,n) = -\exp\left(-\left(x^2\right)^{1/2}\right)
 \left(\frac{\pi}{2}\left(x^2\right)^{3/2} \right)^{1/2}
 \sum_{l=0}^{n} \frac{1}{2^l l!} \frac{\Gamma(5/2+l)}{\Gamma(5/2-l)}
 \left(x^2\right)^{-l/2} \ ,
\end{equation}
with $\Gamma(x)$ denoting the Euler Gamma function. For the
interpolation between the two approximations the point is determined where the
derivatives of the low- and high-temperature expansions can be
connected continuously. At this point a small finite shift to the small
$x^2$ expansion is added such that also the two expansions themselves are
connected continuously. Denoting the values of $x^2$ where the connection
  is performed, by $x^2_+$ and $x^2_-$ and the corresponding shifts by
  $\delta_\pm$ for the fermionic and bosonic contributions,
  respectively, they are given by
\beq
\begin{array}{lcllcl}
x^2_+ & = & 2.2161\;, & \qquad \delta_+ & = & -0.015603 \;,
\\
x^2_- & = & 9.4692 \;, & \qquad \delta_- & = & 0.0063109
\;.
\end{array}
\label{eq:looppot_10}
\eeq
For small $x^2$ the exact result is well approximated by including terms of up
to order $n=4$ in the expansion $J_{+,\text{s}}$ for fermions, for
bosons this is the case for $n=3$ in $J_{-,\text{s}}$. For large $x^2$, the integral is well
approximated by $n=3$ in both the fermion and the boson case,
$J_{\pm,\text{l}}$. The deviation of the approximate results from
the numerical evaluation of the integrals is less than two
percent. The approximations Eqs.~\eqref{eq:looppot_6}-\eqref{eq:looppot_8} are only valid for
$m^2 \geq 0$. For bosons this is not necessarily the case as the
eigenvalues of the mass matrix of the neutral Higgs bosons can become
negative for certain configurations $\{\omega\}$
and temperatures $T$ in the minimisation procedure. In this case 
the value of the integral $J_-$, Eq.~\eqref{eq:looppot_5}, is
set to the real part of its numerical evaluation which is the relevant
contribution for the extraction of the global minimum
\cite{WeinbergWu}. In practice, the integral is evaluated numerically
at several equidistant points in $m^2/T^2 < 0$, and in the
minimisation procedure the result obtained from the linear
interpolation between these points is used. This allows for a significant
speed-up. We explicitly verified that the difference between the exact
and the interpolated result is negligible for a
sufficiently large range of $m^2/T^2$. 

%%%%%%%%%%%%%%%%%%%%%%%%%%%%%%%%%%%%%%%%%%%%%%%%%%%%%%%%%%%
\section{Renormalisation \label{sec:renorm}}
The masses and mixing angles extracted from the loop-corrected
potential differ from those extracted from the tree-level
potential. In the tests for the compatibility of the model with the
experimental constraints these corrections have to be taken into account. In order
to ensure an efficient scan over the parameter space of the model in
terms of the input parameters Eq.~\eqref{eq:c2hdmparams}, it is more
convenient to directly use the loop-corrected masses and angles
as inputs. We therefore modify the $\overline{\text{MS}}$
renormalisation applied in the Coleman-Weinberg potential Eq.~\eqref{eq:cwpot}
and choose a renormalisation prescription
by which we enforce the one-loop corrected masses and mixing matrix
elements to be equal to the tree-level ones. This follows the approach
chosen in our analysis of the CP-conserving 2HDM \cite{Basler:2016obg}.
The counterterm potential $V_{\text{CT}}$, which is added to the one-loop effective potential
Eq.~(\ref{eq:effoneloop}),
\beq
\tilde{V}= V + V_{\text{CT}} = V_{\text{tree}} + V_{\text{CW}} + V_{T}
+ V_{\text{CT}} \;, \label{eq:totalpot} 
\eeq 
reads
\beq
V_{\text{CT}} &=& \frac{\delta m_{11}^2}{2} \omega_1^2 + \frac{\delta
m_{22}^2}{2} (\omega_2^2+\omega_{\text{CP}}^2 + \omega_{\text{CB}}^2) 
- \delta \mbox{Re}(m_{12}^2) \,\omega_1\omega_2 + \delta
\mbox{Im}(m_{12}^2) \,\omega_1\omega_{\text{CP}} + \frac{\delta \lambda_1}{8}
\omega_1^4 \nonumber \\
&+& \frac{\delta \lambda_2}{8} (\omega_2^2+\omega_{\text{CP}}^2+\omega_{\text{CB}}^2)^2
+\frac{\delta\lambda_3}{4} \omega_1^2\left(\omega_{2}^2
  +\omega_{\text{CP}}^2 + \omega_{\text{CB}}^2\right) 
+ \frac{\delta\lambda_4}{4}
\omega_1^2\left(\omega_2^2+\omega_{\text{CP}}^2\right)  \nonumber\\
&+& \frac{\delta \mbox{Re}(\lambda_5)}{4} \omega_1^2
\left(\omega_2^2-\omega_{\text{CP}}^2\right) -
\frac{\delta\mbox{Im}(\lambda_5)}{2}
\omega_1^2\omega_2\omega_{\text{CP}} \nonumber \\
&+& \delta T_1 \, \omega_1 + \delta T_2 \, \omega_2
+ \delta T_{\text{CP}} \, \omega_{\text{CP}}
\;.
\eeq
In the last line we explicitly included the tadpole counterterms
$\delta T$ for the directions in field space in which we allow for the
development of a vacuum.\footnote{Since the physical vacuum is
  required to be neutral, there exist no charge breaking diagrams
  contributing to the one-loop effective potential.} 
Since we check for the compatibility with the experimental constraints
at $T=0$ we apply our renormalisation conditions at this
temperature. They are given by ($i,j=1,...,8$)
\beq
\partial_{\phi_i} \left.V_{\text{CT}} (\phi)\right|_{\phi= \langle
  \phi^c \rangle_{T=0}} &=&
- \partial_{\phi_i} \left.V_{\text{CW}} (\phi) \right|_{\phi=
  \langle\phi^c \rangle_{T=0}} \label{eq:rencond1} \\
\partial_{\phi_i} \partial_{\phi_j}\left.V_{\text{CT}}
  (\phi)\right|_{\phi= \langle\phi^c\rangle_{T=0}} &=&
- \partial_{\phi_i} \partial_{\phi_j}\left.V_{\text{CW}} (\phi)\right|_{\phi=
  \langle\phi^c\rangle_{T=0}} \label{eq:rencond2} \;,
\eeq
with
\beq
\phi_i \equiv \{ \rho_1, \eta_1, \rho_2, \eta_2, \zeta_1, \psi_1,
\zeta_2, \psi_2 \} \;,
\eeq
and $\langle \phi^c \rangle_{T=0}$ denoting the field configuration in
the minimum at $T=0$,
\beq
\langle \phi^c \rangle_{T=0} = (0,0,0,0,v_1,0,v_2,0) \;.
\eeq
The first set of conditions, Eq.~(\ref{eq:rencond1}), ensures that at $T=0$ the tree-level
position of the minimum yields a local minimum. We check
numerically if it is also the global one. The second set of
conditions, Eq.~(\ref{eq:rencond2}), guarantees that at $T=0$ both the masses and the mixing
angles remain at their tree-level values. In Ref.~\cite{Camargo-Molina:2016moz}
formulae for both the first and the second derivatives of the CW
potential have been derived in the Landau gauge basis. We employ these
formulae to calculate the required derivatives. Since the system of
equations resulting from the conditions Eqs.~(\ref{eq:rencond1}) and
(\ref{eq:rencond2}) is not sufficient to fix all renormalisation constants, one of
them is left free. In analogy to our previous paper
\cite{Basler:2016obg}, we choose to set  
\beq
\delta \lambda_4 = 0 \;.
\eeq
This finally yields the counterterms in terms of the derivatives of
the CW potential,
\begin{align}
\delta m_{11}^2 &= \frac{1}{2} \left[ H_{\zeta_1,\zeta_1}^{\text{CW}} +
  2H_{\psi_1,\psi_1}^{\text{CW}} - \frac{v_2}{v_1} \left(
    H^{\text{CW}}_{\eta_1,\eta_2}-H^{\text{CW}}_{\zeta_1,\zeta_2}\right) -
  5H^{\text{CW}}_{\rho_1,\rho1} \right] \nonumber \\ 
\delta m_{22}^2 &= \frac{1}{2} \left[ \frac{v_1}{v_2} \left(
    H^{\text{CW}}_{\zeta_1,\zeta_2} - H^{\text{CW}}_{\eta_1,\eta_2}\right) -
  \frac{v_1^2}{v_2^2}
  \left(H^{\text{CW}}_{\rho_1,\rho_1}-H^{\text{CW}}_{\psi_1,\psi_1}\right) -
  3H^{\text{CW}}_{\eta_2,\eta_2} + H^{\text{CW}}_{\zeta_2,\zeta_2}\right]  \nonumber\\
\delta\mbox{Re}(m_{12}^2) &= \frac{v_1}{v_2} \left(H^{\text{CW}}_{\psi_1,\psi_1}  -
  H^{\text{CW}}_{\rho_1,\rho_1}\right) + H^{\text{CW}}_{\eta_1,\eta_2}
\nonumber\\
\delta\mbox{Im}(m_{12}^2) &= - \left[ H^{CW}_{\zeta_1,\psi2} + 2\frac{v_1}{v_2} 
H^{CW}_{\zeta_1,\psi_1} \right] \nonumber \\
\delta \lambda_1 &= \frac{1}{v_1^2} \left( 2H^{\text{CW}}_{\rho_1,\rho_1} -
  H^{\text{CW}}_{\psi_1,\psi_1} - H^{\text{CW}}_{\zeta_1,\zeta_1}\right) \nonumber\\
\delta \lambda_2 &= \frac{1}{v_2^2} \left[ \frac{v_1^2}{v_2^2}
  \left(H^{\text{CW}}_{\rho_1,\rho_1}-H^{\text{CW}}_{\psi_1,\psi_1}\right) +
  H^{\text{CW}}_{\eta_2,\eta_2} - H^{\text{CW}}_{\zeta_2,\zeta_2}\right] \nonumber\\
\delta \lambda_3 &= \frac{1}{v_1v_2^2} \left[ \left(
    H^{\text{CW}}_{\rho_1,\rho_1} - H^{\text{CW}}_{\psi_1,\psi_1}\right) v_1 +
  \left(H^{\text{CW}}_{\eta_1,\eta_2} - H^{\text{CW}}_{\zeta_1,\zeta_2}\right)v_2
\right] \nonumber\\
\delta \lambda_4 &= 0 \nonumber \\
%\eeq
%\beq
\delta \mbox{Re}(\lambda_5) &= \frac{2}{v_2^2} \left( H^{\text{CW}}_{\psi_1,\psi_1}
  - H^{\text{CW}}_{\rho_1,\rho_1}\right) \nonumber\\
\delta\mbox{Im}(\lambda_5) &= -\frac{2}{v_2^2} H^{CW}_{\zeta_1,\psi_1}
\nonumber \\
\delta T_1 &= H^{CW}_{\eta_1,\eta_2} v_2 + H^{CW}_{\rho_1,\rho_1} v_1
- N^{CW}_{\zeta_1} \nonumber\\ 
\delta T_2 &= H^{CW}_{\eta_1,\eta_2} v_1 + H^{CW}_{\eta_2,\eta_2} v_2
- N^{CW}_{\zeta_2}  \nonumber\\
\delta T_{\text{CP}} &= \frac{v_1^2}{v_2} H^{CW}_{\zeta_1,\psi_1} +
H^{CW}_{\zeta_1,\psi_2} v_1 - N^{CW}_{\psi_2} \;, 
\end{align}
with
\begin{align}
H^{\text{CW}}_{\phi_i,\phi_j} &\equiv
\partial_{\phi_i} \partial_{\phi_j}\left.V_{\text{CW}} (\phi)\right|_{\phi=
  \langle\phi^c\rangle_{T=0}} \\
 N^{\text{CW}}_{\phi_i} &\equiv \partial_{\phi_i} \left.V_{\text{CW}}
   (\phi) \right|_{\phi=\langle\phi^c\rangle_{T=0}} \,.  
\end{align}
We note that the second derivative of the CW potential, required for
our renormalisation procedure, leads to the well-known problem of
infrared divergences for the Goldstone bosons in the Landau gauge
\cite{Cline:1996mga,Cline:2011mm,Dorsch:2013wja,Camargo-Molina:2016moz,Martin:2014bca,Elias-Miro:2014pca, Casas:1994us}. For 
the procedure on how to treat this problem, we refer to the investigation
within the CP-conserving 2HDM in
Ref.~\cite{Basler:2016obg}. We checked that by 
applying these formulae in the limit of the real 2HDM we reproduce our 
results of the R2HDM.

%%%%%%%%%%%%%%%%%%%%%%%%%%%%%%%%%%%%%%%%%%%%%%%%%%%%%%%%%%%
\section{Numerical Analysis \label{sec:numerical}}
\subsection{Minimisation of the Effective Potential \label{sec:minimization}}
The electroweak PT is of strong first order if the ratio between the VEV $v_c$
acquired at the critical temperature $T_c$, and the critical temperature $T_c$
is larger than one \cite{Quiros:1994dr,Moore:1998swa},
\beq
\xi_c \equiv \frac{v_c}{T_c} \ge 1 \;.
\eeq
The value $v$ at a given temperature $T$ is given by
\beq
v(T) = \sqrt{\bar{\omega}_1^2 + \bar{\omega}_2^2 +
  \bar{\omega}_{\text{CP}}^2 + \bar{\omega}_{\text{CB}}^2} \;,
\eeq
where $\bar{\omega}_i$ are the field configurations that minimise the
loop-corrected effective potential at non-zero temperature.
The critical temperature $T_c$ is defined as the temperature where the
potential has two degenerate minima. In order to obtain $T_c$, the
complete loop-corrected effective potential Eq.~(\ref{eq:totalpot}),
is minimised numerically for a given  
temperature $T$. If the PT is of strong first order, the VEV jumps from $v=v_c$ at
the temperature $T_c$ to $v=0$ for $T>T_c$.
For the determination of $T_c$ we employ a bisection method in the
temperature $T$, starting with the determination of the minimum at the
temperatures $T_S=0$~GeV and ending at $T_E=300$~GeV. The minimisation
procedure is terminated when the interval containing $T_c$ is smaller
than $10^{-2}\ \text{GeV}$, and the temperature $T_c$ is then set to the
lower bound of the final interval. Parameter points that
do not satisfy $\vert v(T=0) - 246.22 \mbox{ GeV} \vert \leq 2 \,
\mathrm{GeV}$ are excluded as well as parameter points where no PT is
found for $T\leq 300\ \text{GeV}$. Adding a small safety margin, we
ensure by the latter condition that possible strong first order PTs are obtained
for VEVs below 246~GeV. We furthermore only retain parameter points with $T_c > 10\
\text{GeV}$. 

%%%%%%%%%%%%%%%%%%%%%%%%%%%%%%%%%%%%%%%%%%%%%%%%%%%%%%%%%%%
\subsection{Constraints and Parameter Scan \label{sec:scan}}
The points, for which the value of $\xi_c$ is determined, have to
satisfy theoretical and experimental constraints. We use a pre-release
version of \texttt{ScannerS}~\cite{Coimbra:2013qq,Ferreira:2014dya} to
perform scans in the C2HDM parameter space in order to
obtain viable data sets. In these extensive scans we check for
compatibility with the following 
constraints. The potential is required to be bounded from below and
the tree-level discriminant of Ref.~\cite{Ivanov:2015nea} is used to
enforce that the electroweak vacuum is the global minimum of the
tree-level potential at zero temperature. 
We further require perturbative unitarity to hold at tree level. We
take into account the flavour constraints on $R_b$
\cite{Haber:1999zh,Deschamps:2009rh} and $B \to X_s \gamma$
\cite{Deschamps:2009rh,Mahmoudi:2009zx,Hermann:2012fc,Misiak:2015xwa,Misiak:2017bgg}. They
can be generalized from the CP-conserving 2HDM to the C2HDM as they
only depend on the charged Higgs boson.
These constraints are checked as $2\sigma$ exclusion bounds on
the $m_{H^\pm}-t_\beta$ plane. According to the latest calculation
of Ref.~\cite{Misiak:2017bgg} the charged Higgs boson mass is required
to be rather heavy,
\beq
m_{H^\pm} \ge 580 \mbox{ GeV} \;,
\eeq 
in the type II and flipped 2HDM. In the type I and lepton-specific
model this bound is much weaker and depends more strongly on
$\tan\beta$. Agreement with the electroweak precision measurements is
verified using the oblique parameters $S$, $T$ and $U$. The
formulae for their computation in the general
2HDM can be found in \cite{Branco:2011iw}. For the computed $S$, $T$ and $U$
values $2\sigma$ compatibility with the SM fit \cite{Baak:2014ora} is
demanded including the full correlation among the three
parameters. One of the Higgs bosons, called $h$ in the following, is
required to have a mass of \cite{Aad:2015zhl}
\beq
m_h = 125.09\mbox{ GeV} \;.
\eeq
Compatibility with the Higgs data is checked by using
{\tt HiggsBounds} \cite{Bechtle:2013wla} and the individual signal strength fits
of Ref.~\cite{Khachatryan:2016vau} for the $h$.
% Yap, das stimmt noch. Ich knnte im C2HDM das HiggsSignals interface
%aber recht fix aufsetzen, also wenn ihr wollt kann ich das
%machen. Allerdings werden die Scans dann viel laenger
%brauchen. Alternativ knnten ich nur das HS chi2 berechnen lassen,
%dann knnten wir zumindest kommentieren wie es mit den bereichen mit starkem PT aussieht.
%
The required decay widths and branching ratios are obtained from a
private implementation of the C2HDM into {\tt HDECAY v6.51}
\cite{Djouadi:1997yw,Butterworth:2010ym}, which
will be released in a future publication. Additionally, the
Higgs boson production cross sections normalized to the SM are
needed, including the most important state-of-the-art higher order
corrections. Where available, we include the QCD corrections which can be taken over
from the SM and Minimal Supersymmetric Extension of the SM
(MSSM). Electroweak corrections are consistently neglected in both the
production and decay channels, as they cannot be taken over and are
not available yet for the C2HDM. Details on how the production cross
sections are determined can be found in  \cite{Muhlleitner:2017dkd}.
This information is passed via the {\tt ScannerS}
interface to {\tt HiggsBounds} which checks for agreement with
all $2\sigma$ exclusion limits from LEP, Tevatron and LHC Higgs
searches. As mentioned above, the properties of the $h$ are checked against the fitted
values of the signal strengths given in
\cite{Khachatryan:2016vau}. For details, we again refer to
\cite{Muhlleitner:2017dkd}. We use this method for
simplicity. {Note that performing a fit to current Higgs data is
likely to give a stronger bound than this approach. 
As we include CP violation in the Higgs sector we also have to check
for compatibility with the measurements of electric dipole moments
(EDM), where the strongest constraint originates from the EDM of the
electron \cite{Inoue:2014nva}.  The experimental limit has been given by
the ACME collaboration \cite{Baron:2013eja}. For the check, we have
implemented the calculation of the dominant Barr-Zee contributions by
\cite{Abe:2013qla} and require compatibility with the bound given in
\cite{Baron:2013eja} at 90\% C.L. \s

For the scan, the SM VEV is fixed to
\beq
v= 1/\sqrt{\sqrt{2} G_F}= 246.22 \; \mbox{GeV} \;.
\eeq
The ranges chosen for the remaining input
parameters of Eq.~(\ref{eq:c2hdmparams}) are as follows. 
The mixing angle $t_\beta$ has been varied as
\beq
0.8 \le t_\beta \le 35 \;. \label{eq:tbscanc2hdm}
\eeq
The angles parametrising the mixing matrix Eq.~(\ref{eq:rmatrix}) are
chosen in the intervals
\beq
- \frac{\pi}{2} \le \alpha_{1,2,3} < \frac{\pi}{2} \;.
\eeq
For $\mbox{Re} (m_{12}^2)$ we use the range
\beq
0 \mbox{ GeV}^2 \le \mbox{Re}(m_{12}^2)  < 500\,000 \mbox{ GeV}^2 \;.
\eeq
Note that, although possible, physical parameter points with $\mbox{Re}(m_{12}^2) <0$
are extremely rare so that we neglect them in our study. This is
mainly a result of requiring absolute stability at tree level.
One of the neutral Higgs bosons $H_i$ is identified with $h$. In
type II, the charged Higgs mass is chosen in the range
\beq
580 \mbox{ GeV } \le m_{H^\pm} < 1 \mbox{ TeV } \;,
\eeq
and in type I in the range
\beq
80 \mbox{ GeV } \le m_{H^\pm} < 1 \mbox{ TeV } \;.
\eeq
The electroweak precision constraints combined with perturbative
unitarity require at least one neutral Higgs boson to be close in mass
to $m_{H^\pm}$. For an increased scan efficiency we therefore choose the
second neutral Higgs mass $m_{H_i \ne h}$ in the interval
\beq
500 \mbox{ GeV}\leq m_{H_i}<1 \mbox{ TeV}
\eeq
in type II and
\beq
30 \mbox{ GeV}\leq m_{H_i}<1 \mbox{ TeV}
\eeq
in type I. In the C2HDM the third neutral Higgs boson $m_{H_j \ne H_i,
  h}$ is not an independent input parameter and is calculated by
{\tt ScannerS}. It is, however, required to lie in the interval 
\beq
30 \mbox{ GeV } \le m_{H_j} < 1 \mbox{ TeV } .
\eeq
We further impose that the $m_{H_i\neq h}$ deviate by at least
$5\;\text{GeV}$ from $125.09\;\text{GeV}$ to avoid degenerate Higgs
signals. To improve the coverage of the CP-conserving limit we have
performed dedicated scans in the CP-conserving 2HDM and merged the
resulting CP-violating and CP-conserving samples. The scans
in the CP-conserving 2HDM were also performed using {\tt ScannerS} with 
the same constraints\footnote{With the exception of the EDM constraint
  which is trivially satisfied if CP is conserved.} and parameter
ranges. 
%In this limit we have used separate scans for the cases where
%$h$ is the heavier or the lighter of the two CP-even Higgs
%bosons. 
The final samples are composed of more than $2 \cdot 10^5$ valid
parameter points for each Yukawa type. \s
%The final samples are composed of \textcolor{red}{\bf Is this still true, Philipp?}
%\begin{equation}
%	1\times10^5 \text{ C2HDM } + 8\times10^4\text{
%          R2HDM}(h_{\text{R2HDM}}=h) +2\times10^4\text{ R2HDM}(H_{\text{R2HDM}}=h) 
%\end{equation}
%points in both type I and type II.\s

For the SM parameters we have chosen
the following values: Apart from the computation of the oblique
parameters, where we use the fine structure constant at zero momentum transfer,
\beq
\alpha^{-1}_{\text{EM}} (0) = 137.0359997 \,,
\eeq
the fine structure constant is taken at the $Z$ boson mass scale
\cite{Agashe:2014kda},
\beq
\alpha^{-1}_{\text{EM}} (M_Z) = 128.962 \;.
\eeq
The massive gauge boson masses are set to
\cite{Agashe:2014kda,Denner:2047636}
\beq
M_W = 80.385 \mbox{ GeV} \qquad \mbox{and} \qquad M_Z = 91.1876 \mbox{
GeV} \;,
\eeq
the lepton masses to \cite{Agashe:2014kda,Denner:2047636}
\beq
m_e = 0.510998928 \mbox{ MeV} \;, \quad
m_\mu = 105.6583715 \mbox{ MeV} \;, \quad
m_\tau = 1.77682 \mbox{ GeV} \;,
\eeq
and the light quark masses to
\beq
m_u = 100 \mbox{ MeV} \;, \quad m_d = 100 \mbox{ MeV} \;, \quad
m_s = 100 \mbox{ MeV} \;,
\eeq
following \cite{LHCHXSWG}.
For consistency with the ATLAS and CMS analyses the on-shell top quark mass
\beq
m_t = 172.5 \mbox{ GeV}  \label{eq:ferm1}
\eeq
has been taken, as recommended by the LHC Higgs Cross Section Working Group
(HXSWG) \cite{Denner:2047636,Dittmaier:2011ti}. The charm and
bottom quark on-shell masses are \cite{Denner:2047636}
\beq
m_c = 1.51 \mbox{ GeV} \qquad \mbox{and} \qquad
m_b = 4.92 \mbox{ GeV} \;. \label{eq:ferm2}
\eeq
We take the CKM matrix to be real, with the
CKM matrix elements given by\cite{Agashe:2014kda}\footnote{In the
computation of the loop-corrected effective potential we
choose $V_{\text{CKM}}=\mathds{1}$ for simplicity. The
impact of this choice on the counterterms and thereby on the potential
and its minimisation is negligible.}
\beq
V_{\text{CKM}} = \left( \begin{array}{ccc} V_{ud} & V_{us} & V_{ub} \\
V_{cd} & V_{cs} & V_{cb} \\ V_{td} & V_{ts} & V_{tb} \end{array}
\right) = \left( \begin{array}{ccc} 0.97427 & 0.22536 & 0.00355 \\
    -0.22522 & 0.97343 & 0.0414 \\ 0.00886 & -0.0405 &
    0.99914 \end{array} \right) \;. \label{eq:parameterscan_alex1}
\eeq

%%%%%%%%%%%%%%%%%%%%%%%%%%%%%%%%%%%%%%%%%%%%%%%%%%%%%%%%%%%
\section{Results \label{sec:results}}
In our analysis we investigate the question to which extent the
allowed parameter space of the C2HDM is constrained by the requirement
of a first order phase transition and what are the consequences for
LHC phenomenology. We compare with the case of the
CP-conserving 2HDM which has been analysed in
\cite{Basler:2016obg}. We investigate the impact on the trilinear
Higgs self-couplings and Higgs pair production.\footnote{For previous
  studies on Higgs pair production in the real 2HDM,
  see~\cite{Baglio:2014nea,Hespel:2014sla}.} We analyse the size of
the electroweak corrections to the Higgs self-couplings derived from
the effective potential. We furthermore study the possible spontaneous generation
and size of a CP-violating phase at the electroweak phase transition. 
We will show results both for the type I and the type II C2HDM and for
the cases where the lightest of the neutral Higgs bosons is
identified with the discovered Higgs boson, {\it i.e.}~$H_1 \equiv
h$, and where the next heavier one is the SM-like Higgs boson, $H_2
\equiv h$. \s 

For the interpretation of the results, we note that the strength of the
phase transition increases with the size of the couplings of the light
bosonic particles to the SM-like Higgs boson and decreases with the Higgs boson mass
\cite{Carena:2008vj}. Since in the C2HDM all non-SM-like neutral Higgs
bosons receive a VEV through mixing and hence contribute to the PT, a
strong electroweak PT requires the participating Higgs bosons either
to be light or to have a VEV close to zero. In the 
latter case we are in the alignment limit where only one of the physical
Higgs bosons has a VEV \cite{Gunion:2002zf}. In the type II C2HDM the
requirement of a light Higgs spectrum puts the model under tension,
as EW precision tests combined with perturbative unitarity enforce one
of the neutral Higgs bosons to be close to $m_{H^\pm}$. Charged Higgs
masses below 580~GeV are already excluded by $B\rightarrow X_s
\gamma$, however. \s 

Note, that in contrast to the analysis of the
CP-conserving 2HDM in \cite{Basler:2016obg}, in the C2HDM we have a
larger number of parameters to be scanned over. Also the minimisation procedure
requires more computing power due to a possible CP-violating VEV. The
result is, that the overall density of parameter points compatible
with our applied constraints is smaller than in the real
2HDM. 
%\sout{Consequently, we did not find any
%    parameter sets in the type I 2HDM where the next-to-lightest Higgs
%    boson $H_2$ nor the heaviest one $H_3$ is the SM-like Higgs boson
%    and where we have both a strong phase transition and CP violation
%    in the Higgs sector. This is not a proof of course, that this
%    scenario cannot be realized, but so far at most shows that such
%    scenarios are very sparse.} 
Consequently, we found for type I only very few
  parameter points where $H_2 = h$ and where we have both a
  strong phase transition and CP violation in the Higgs
  sector. A considerably enlarged parameter scan might lead to more
  points fulfilling these criteria. Here we content ourselves to
  demonstrate that such configurations are possible in principle.
 In the type II C2HDM, no parameter sets were found 
    where the SM-like Higgs boson is given by the heavier neutral
    Higgs bosons, due to the
constraint $m_{H^\pm} \ge 580$~GeV
\cite{Misiak:2017bgg}.\footnote{This mass configuration corresponds in
the limit of the real 2HDM to the cases where the lighter neutral
Higgs boson $h$ corresponds to the 125~GeV Higgs boson and $m_A <
m_{h}= 125$~GeV or the heavier one, $H$, represents the discovered Higgs boson 
and $m_h < m_{H} = 125$~GeV. Already in the R2HDM where
we applied the older constraint of $m_{H^\pm} \ge 480$~GeV, we
found very few scenarios in this case that are compatible with a strong
PT, {\it cf.}~Figs.~5 and 11 in \cite{Basler:2016obg}. With the
stricter lower limit of 580~GeV on the charged Higgs mass we do not
find any allowed scenarios with a strong PT any more. In the
CP-violating case where in general 
%the mass values of the Higgs bosons move
%closer in the 
we find less scenarios compatible with a strong PT
%, {\it  cf.}~Sec.~\ref{sec:t2hl}, 
the situation becomes even more severe.}
%, as
%the spectrum of the Higgs bosons participating in the PT becomes
%overall heavier, which then suppresses $\xi_c$.}
%For the same reason we did not finde type
%II parameter sets with $H_2$ being the $h$\footnote{In the type
%  II 2HDM $H_\uparrow$ is always heavier than 480~GeV, {\it
%    cf.}~\cite{Muhlleitner:2017dkd}.} and where
%$\xi_c \ge 1$ together with explicit CP
%violation. \textcolor{red}{Anything else to add here?} 

\vspace*{0.3cm}
\noindent
%%%%%%%%%%%%%%%%%%%%%%%%%%%%%%%%%%%%%%%%%%%%%%%%%%%%%%%%%%%
{\bf Validity of the global minimum and of the unitarity constraint at
NLO} \s

\noindent
Since we compute the global minima of the effective potential at NLO,
an interesting question to ask is how the inclusion of NLO effects
influences the absolute stability of the EW vacuum. In the scan of
the type I C2HDM with $H_1 \equiv h$ we found that the inclusion
of the NLO computation eliminated about 6\% of the parameter points due to
the EW vacuum no longer being the global minimum at NLO. In case $H_2 \equiv
h$, 26\% of the tree-level points did not lead to a global
minimum any more. For the type II scan we 
found that the requirement of a global minimum at NLO eliminated
9\% of the points with a valid tree-level global
minimum in case $H_1 \equiv h$. If, however, the heavier $H_2$
is the SM-like Higgs boson, there are practically no scenarios that
represent a valid minimum of the potential at NLO. It turns out that
the request of a global NLO minimum at 246~GeV implies 
reduced mass differences between the different Higgs
bosons. If, however, the $H_2$ is SM-like then the mass difference
between the lightest Higgs boson $H_1$ with $m_{H_1} < 125$~GeV and
the charged Higgs boson with a required mass above 580~GeV is too
large to allow for an NLO minimum at 246~GeV. Consequently, there are
also no scenarios with a strong first order PT in this case.
In fact, as can be read off the formula Eq.~(\ref{eq:cwpot}) for the
Coleman-Weinberg potential large masses imply large one-loop
corrections. Through our renormalisation procedure we move these large
corrections into the quartic couplings, so that they may become too
large to guarantee a stable vacuum. \s 

This discussion shows that the inclusion of the NLO effects is important in
order to correctly define the parameter regions that are 
compatible with the requirement that the EW minimum represents the
global minimum. In our analysis we only keep points compatible with a
global NLO minimum at 246~GeV. Additionally, we have to make sure that the possibly
large corrections to the Higgs self-couplings due to our
renormalisation procedure do not spoil the unitarity constraint. In
order to check this, in a first rough approximation we insert the
renormalised $\lambda_i$, derived from the loop-corrected
effective potential, into the tree-level formulae for the unitarity
constraints \cite{Branco:2011iw}. We keep only those points for which the bound of
$8\pi$ is not violated. The unitarity check further reduces the sample of
points fulfilling the experimental and global minimum constraints
by 11\% and 18\% in the type I C2HDM with $H_1 \equiv h$ and $H_2
\equiv h$, respectively, and by 9\% in the type II C2HDM with $H_1 \equiv h$. 

%%%%%%%%%%%%%%%%%%%%%%%%%%%%%%%%%%%%%%%%%%%%%%%%%%%%%%%%%%%
\section{Type I: Parameter sets with $H_1 = h$ \label{sec:Ihleqh125}}
We first present results for the type I C2HDM where the lightest of the
three neutral Higgs bosons, $H_1$, coincides with the SM-like Higgs
boson $h$. In the following we denote the lighter (heavier) of the two
non-SM-like neutral Higgs bosons by $H_\downarrow$
($H_\uparrow$) with mass $m_\downarrow$ ($m_\uparrow$) where appropriate. 

%%%%%%%%%%%%%%%%%%%%%%%%%%%%%%%%%%%%%%%%%%%%%%%%%%%%%%%%%%%
\subsection{The CP-violating phase}
We start by investigating the size of a possible CP-violating
phase that is spontaneously generated at the EWPT, and its relation to
the explicit CP violation through a complex phase of $m_{12}^2$. In
Fig.~\ref{fig:cpphase}, we depict the tangent of the CP-violating phase $\tan
\varphi^{\text{spont}} = \bar{\omega}_3 (T_c)/\bar{\omega}_2
(T_c)$ at the critical temperature $T_c$ as a function of the tangent
of the CP-violating phase at zero temperature\footnote{In the C2HDM, the only
  CP-violating source in the Higgs potential is given by a complex
  $m_{12}^2$ or alternatively a complex $\lambda_5$ which is related
  to $m_{12}^2$ through the tadpole conditions. Any other CP-violating
  phase, namely the phase of the VEV, can be absorbed by a
  redefinition of the fermion fields.} $\tan \varphi^{\text{explicit}} 
= \mbox{Im} (m_{12}^2)/\mbox{Re}(m_{12}^2)$ for the points that fulfill
all constraints\footnote{Here and in all following plots this means
  that they fulfill the experimental constraints and also the
  unitarity and global minimum constraints at NLO.} and are compatible
with a strong PT. 
\begin{figure}[t!]
\begin{center}
\includegraphics[width=0.55\textwidth]{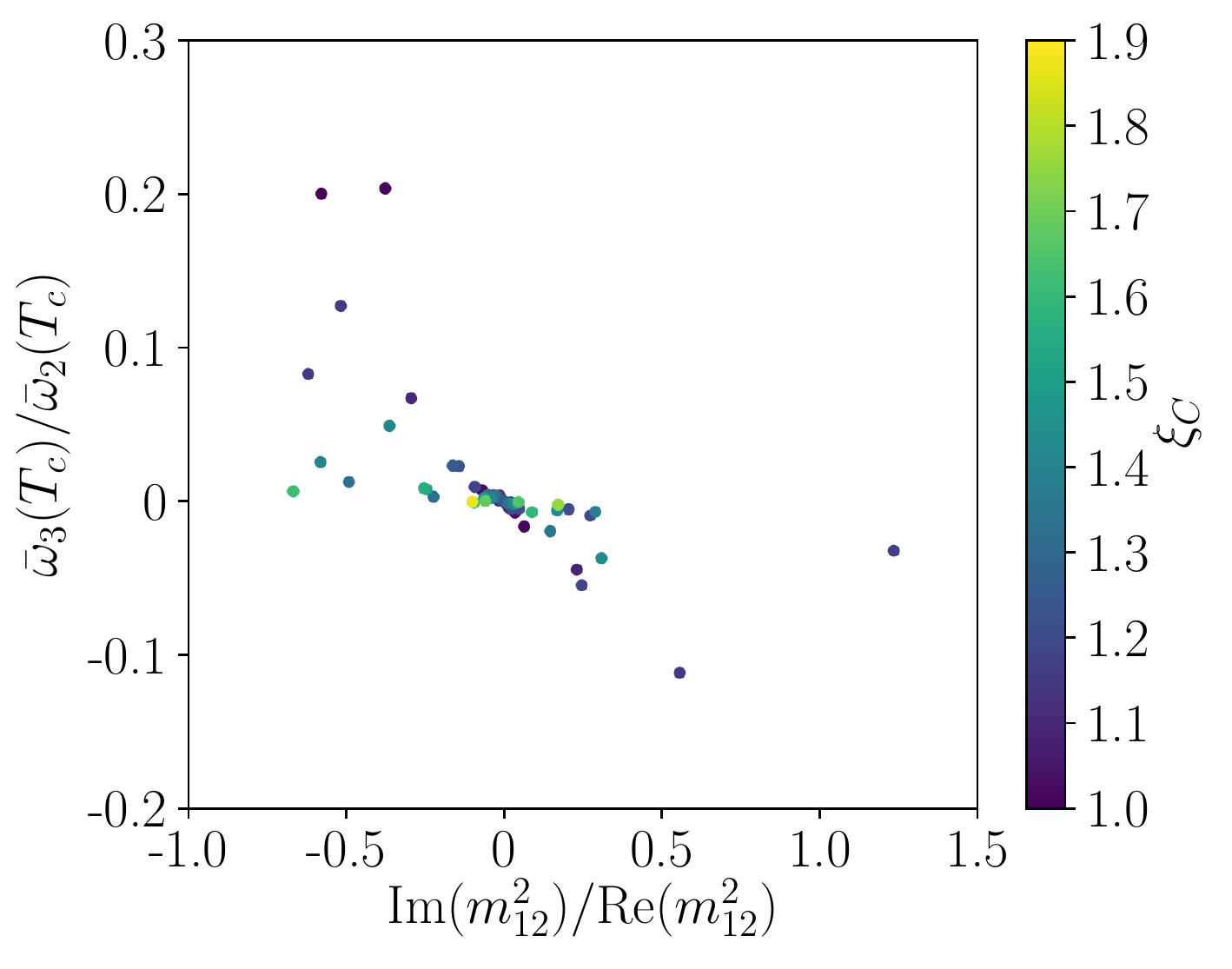}
%\includegraphics[width=0.49\textwidth]{Phase-HeatMap-AE-T1.pdf}
%\vspace*{-0.2cm}
\caption{Type I, $H_1 = h$: The value 
  $\bar{\omega}_3(T_c)/\bar{\omega}_2 (T_c)$ at the critical temperature $T_c$
  versus $\mbox{Im} (m_{12}^2)/ \mbox{Re} (m_{12}^2) \ne 0$ at
  $T=0$ for points with a strong PT. The colour code indicates the
  size of $\xi_c$. 
 \label{fig:cpphase}}
\end{center}
\vspace*{-0.4cm}
\end{figure} 
The colour code 
indicates the value of 
$\xi_c$. Note that in Fig.~\ref{fig:cpphase} we only plot C2HDM parameter points, {\it
    i.e.} $\mbox{Im} (m_{12}^2)/\mbox{Re} (m_{12}^2) \ne 0$, although
  this ratio can become very small. 
%All points comply with our applied
%  constraints and feature a strong PT.  
As can be inferred
from the figure, the phase $\varphi^{\text{spont}}$ of spontaneous CP violation
and the one of explicit CP violation, $\varphi^{\text{explicit}}$, are
correlated. A CP-violating phase at $T_c$ is generated 
spontaneously only if already in the zero-temperature potential there
is non-vanishing CP violation. 
As we set the CKM matrix to unity in the computation of the effective
potential no CP violation can be generated through loop 
effects if CP is conserved at $T=0$. The size of $\xi_c$ 
is not correlated to the size of the CP-violating phase. We observe,
however, that the maximum obtained value of $\xi_c$, which quantifies
the strength of the PT, is $\xi_c = 1.89$. This is below the value
found in the CP-conserving 2HDM, {\it  cf.}~Ref.~\cite{Basler:2016obg} and 
the following section. \s

Figure~\ref{fig:vevdevelop} shows the development of the VEVs
exemplary for one parameter point, defined by 
\beq
\begin{array}{lll}
M_{H_1} = 125.09 \mbox{ GeV}, & \qquad M_{H_2} = 163.119 \mbox{ GeV}, &
  \\
M_{H_3} = 387.461 \mbox{ GeV}, &  \qquad M_{H^\pm} = 393.035 \mbox{ GeV}, \\
\alpha_1 = 1.1376, & \qquad \alpha_2 = 0.0601, & \qquad \alpha_3 = -0.0934,\\
\tan\beta = 5.6211, & \qquad \mbox{Re} (m_{12}^2) = 4341 \mbox{ GeV}^2 \;. 
\end{array}
\eeq
We observe, in the upper left plot the generation of the
VEV $v = \sqrt{\bar\omega_1^2 + \bar\omega_2^2 +
\bar\omega_3^2}$ ($\bar{\omega}_3 \equiv \bar{\omega}_{\text{CP}}$) at
$T_c = 136.011$~GeV with a value of 
$v_c=156.245$~GeV and hence $\xi_c=1.149$. With 
decreasing temperature the VEV increases to the value 246 GeV at
$T=0$. Note, that the VEV also includes
  $\bar{\omega}_{\text{CB}}$, which, however, always turns out to be
  zero, so that we did not write it explicitly here.
The upper right and lower left plots show the development of
the absolute values of the individual VEVs, {\it i.e.} the ones of
$\bar\omega_1$ and $\bar\omega_2$, the CP-conserving 
VEVs coinciding with $v_1$ and $v_2$ at zero temperature, and of the CP-violating VEV
$\bar{\omega}_3$. We also show, in the lower right plot, the
development of $\bar\omega_3 / 
\bar\omega_2$. Because of the value of $\tan\beta$ above 1, the value
$|\bar\omega_2|$ is larger than $|\bar\omega_1|$.
The spontaneously generated CP-violating VEV $\bar\omega_3$ at the PT
amounts with $|\bar\omega_3 (T_c)| = 17.06$ to 11\% of the absolute value
of $\bar\omega_2$ and decreases monotonously to zero with decreasing
temperature, while $\bar\omega_1$ and $\bar\omega_2$ monotonously increase to
reach $\sqrt{\bar\omega_1^2 + \bar\omega_2^2} (T=0) =246$~GeV. \s
\begin{figure}[t!]
\begin{center}
\includegraphics[width=0.8\textwidth]{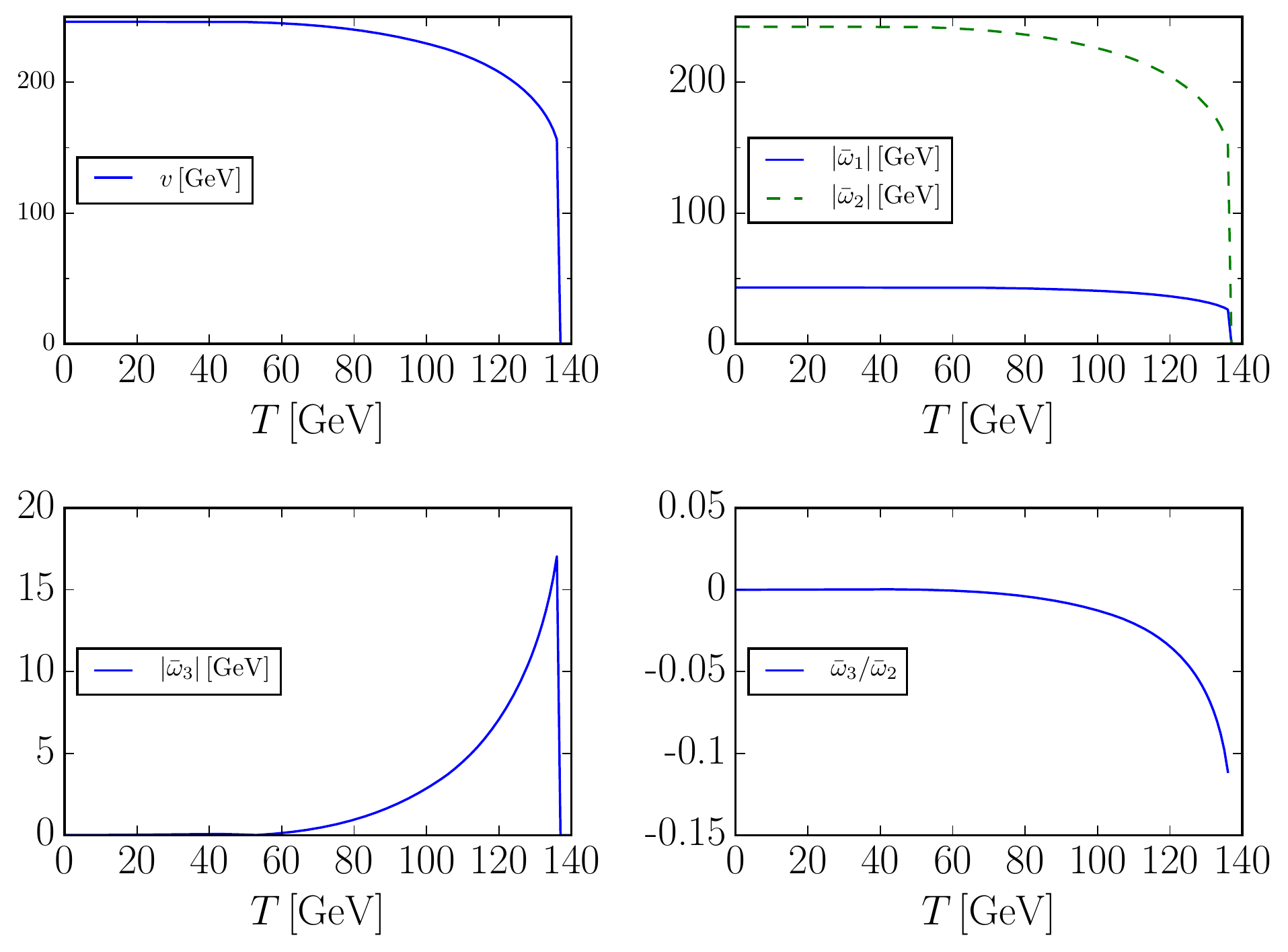}
%\vspace*{-0.2cm}
\caption{Type I, $H_1 = h$: The development of the VEV as a
  function of the temperature. In clockwise direction: the total VEV $v$; the
  absolute values of the CP-conserving VEVs, $|\bar\omega_1|$ (blue/full) and
  $|\bar\omega_2|$ (green/dashed), the absolute value of the CP-violating VEV,
  $|\bar\omega_3|$, and $\bar\omega_3/\bar\omega_2$.}
 \label{fig:vevdevelop}
\end{center}
\vspace*{-0.4cm}
\end{figure}
\begin{figure}[ht!]
\begin{center}
\includegraphics[width=0.5\textwidth]{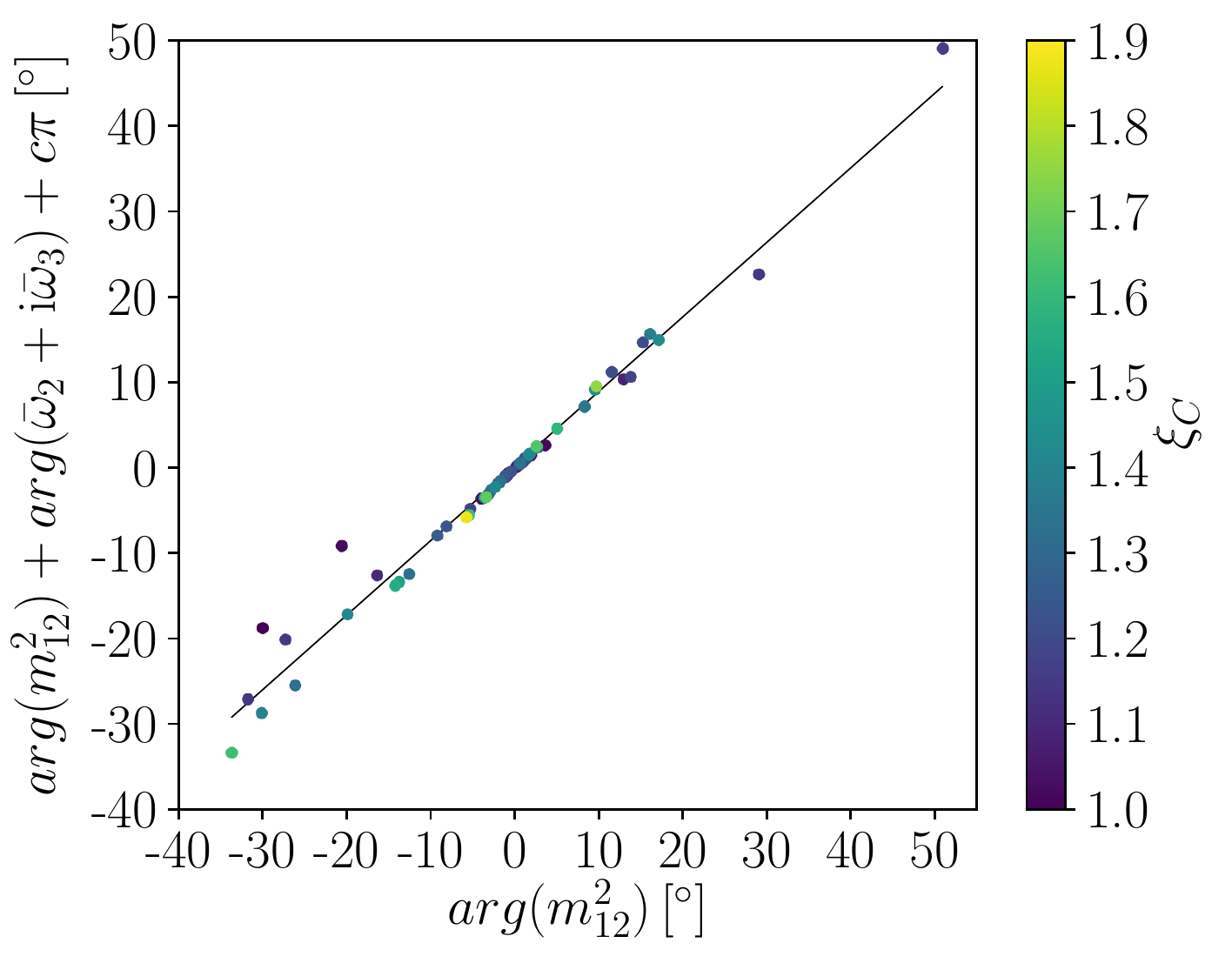}
%\vspace*{-0.2cm}
\caption{Type I, $H_1 = h$: The total CP-violating angle
  at the critical temperature $T_c$ versus the CP-violating angle at
  $T=0$. By adding $c\pi$, $c \in \{-1,0,1\}$, the angle is ensured to lie in the range
  $-90^\circ$ to $90^\circ$. The colour code indicates the size of
  $\xi_c$.  
 \label{fig:totalphase}}
\end{center}
\vspace*{-0.6cm}
\end{figure}

Figure~\ref{fig:totalphase} shows the total CP-violating angle at the PT
as a function of the CP-violating angle at
$T=0$. The former varies between about -34$^\circ$ and 49$^\circ$ for CP-violating
angles of -35$^\circ$ to 51$^\circ$ allowed at $T=0$. We observe almost
maximal CP violation which should be enough for successful
baryogenesis \cite{Cohen:1993nk,Gavela:1993ts}. 
%We
%remark, finally, that the CP-violating VEV that is spontaneously
%generated at the EWPT continously decreases to zero with decreasing
%temperature, matching the zero-temperature VEV given by $v \approx
%246$ GeV and $v_{\text{CP}} = 0$. \textcolor{red}{Should we add a plot?}

%%%%%%%%%%%%%%%%%%%%%%%%%%%%%%%%%%%%%%%%%%%%%%%%%%%%%%%%%%%
\subsection{Implications for LHC phenomenology and benchmark scenarios}
Figure~\ref{fig:mupmlowt1hl} shows the mass of the heavier non-SM-like Higgs
boson, $m_\uparrow$, versus the lighter one, $m_\downarrow$, where
the grey points pass the applied constraints and include both
CP-conserving and CP-violating points in the left plot, but only
CP-violating points in the right plot. The coloured points
additionally feature a strong first order PT. The maximum
possible value of $\xi_c$ is found to be $\xi_c = 5.7$ for all 2HDM 
points (left plot).\footnote{Barring the few
  CP-violating points, the left plot can be
  compared to the results of Ref.~\cite{Basler:2016obg}. There, we
  found $\xi_c^{\text{max}}=4.5$. The difference to
  $\xi_c^{\text{max}}$ found here (and also the
  difference in the shape of this plot and all following ones, that
  can be used for a comparison) arises from a different constraint on
  $\tan\beta$ due to different applied flavour
  constraints \cite{gfitter}. Furthermore, in the mass difference plane we now have only
    two branches instead of four in the real 2HDM, as we strictly
    order the non-SM-like Higgs masses by increasing values and not by their CP
    nature.} If we only consider CP-violating points (right plot)
the number of grey points is reduced.  The number of points
compatible with a strong PT is reduced even more, with the maximally
allowed $\xi_c$ being 1.89. The mass plots show that the
requirement of a strong PT overall prefers somewhat lighter
non-SM-like Higgs bosons. The allowed maximum masses are further reduced in
case of CP violation where $m_\uparrow$ remains 
below about 753~GeV and $m_\downarrow$ does not exceed 636~GeV. 
Through the CP mixing all Higgs bosons participate in the PT, also the heavier
ones. Additionally, a CP-violating VEV $\bar{\omega}_3$ is generated at
$T_c$ and feeds into the VEV of also the heavier Higgs bosons. With heavier
Higgs bosons participating in the PT, the PT is weakened inducing
smaller $\xi_c$ values. In order to counterbalance these effects, the
Higgs masses overall become lighter and/or they move closer together
({\it cf.}~the coloured points on the diagonal axis), thus
also distributing large portions of the VEV to the lighter among the Higgs bosons.
\s
\begin{figure}[t!]
\begin{center}
\includegraphics[width=0.49\textwidth]{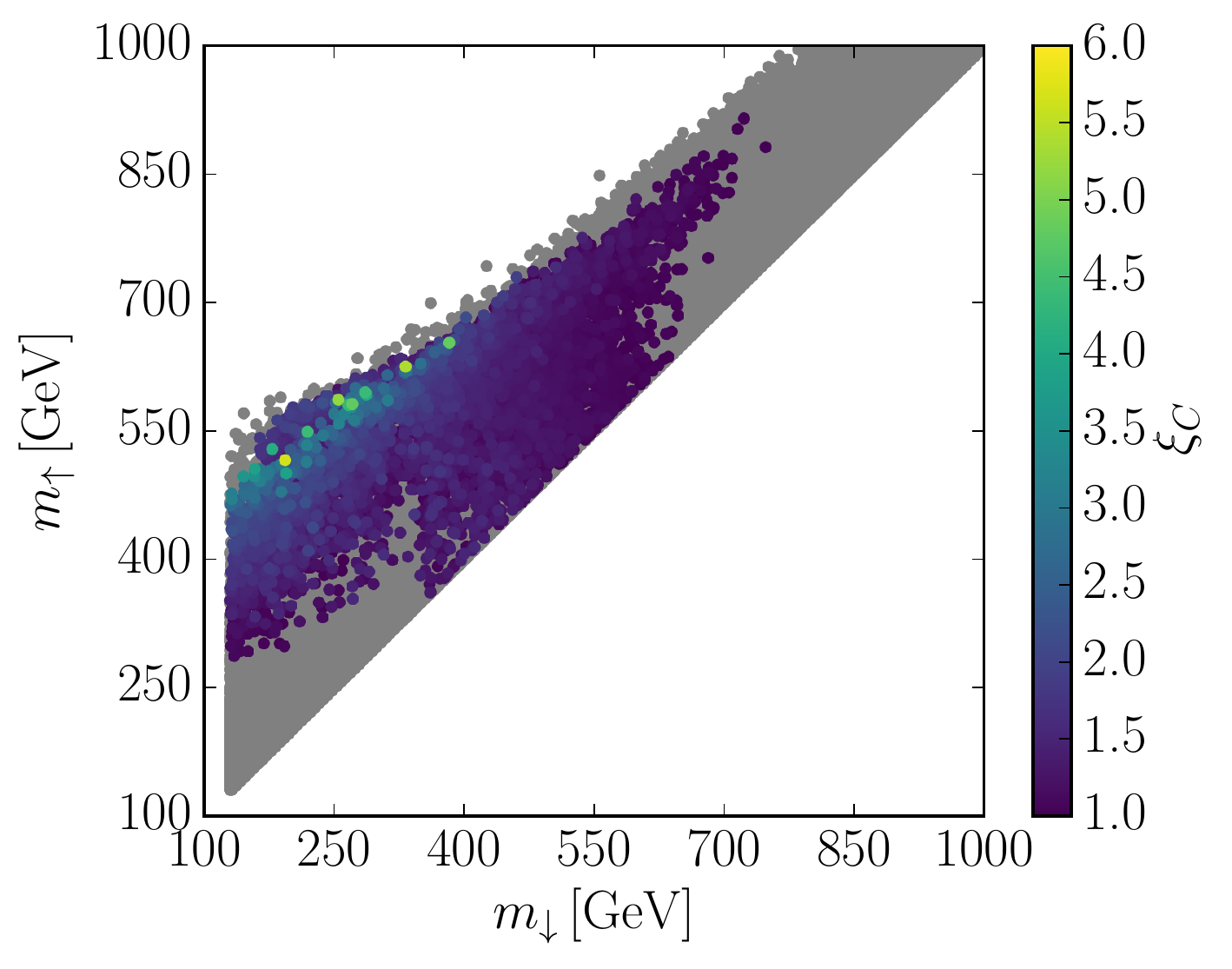}
\includegraphics[width=0.49\textwidth]{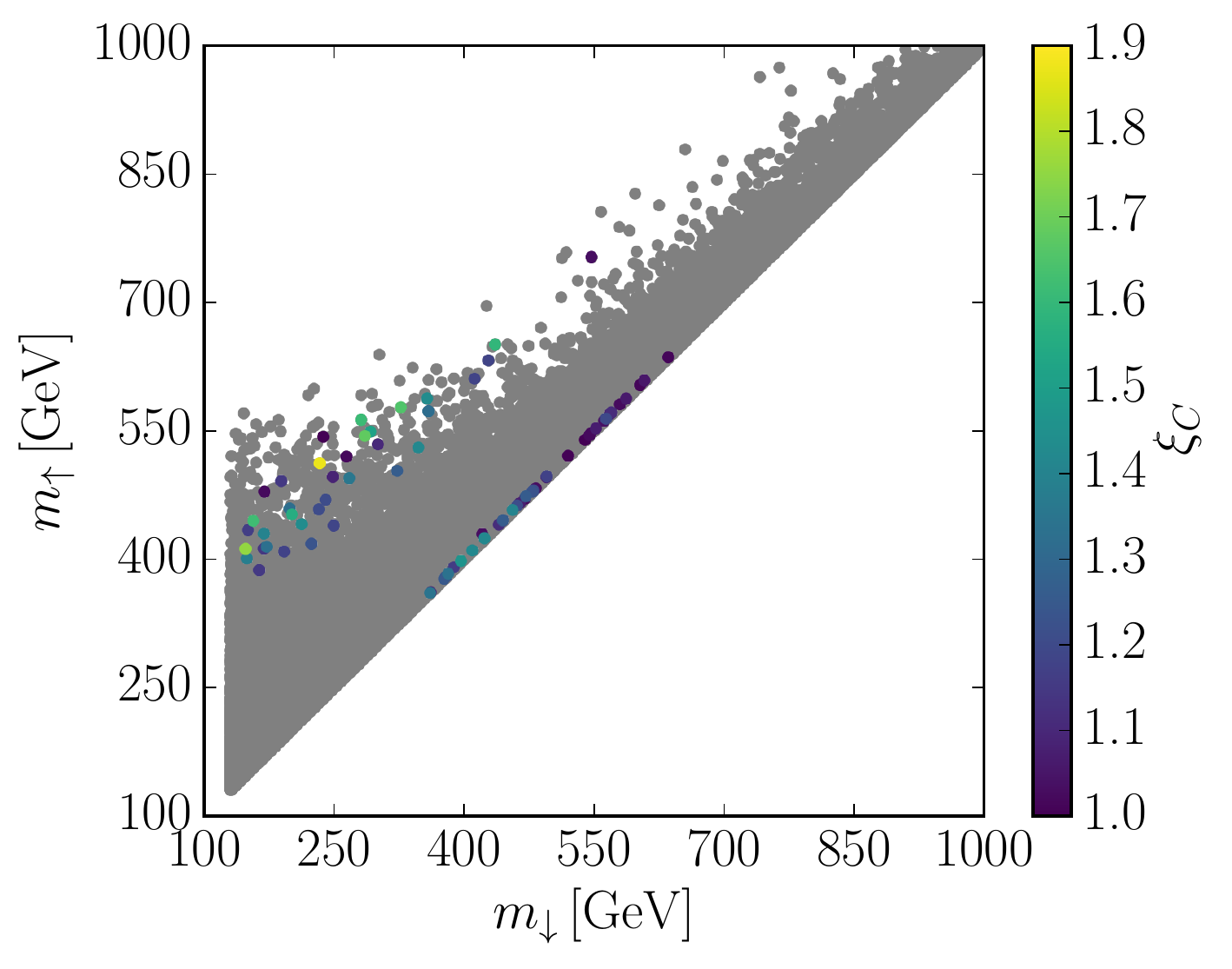}
%\vspace*{-0.2cm}
\caption{Type I, $H_1 = h$: The mass of the heavier versus the
  lighter non-SM-like Higgs boson. Left: CP-conserving and
  CP-violating points, right: only CP-violating points. Grey: points
  passing all the constraints; color: points with additionally $\xi_c \ge 1$. The
  color code indicates the value of $\xi_c$. \label{fig:mupmlowt1hl}}
\end{center}
\vspace*{-0.2cm}
\end{figure}

\begin{figure}[t!]
\begin{center}
\includegraphics[width=7.5cm]{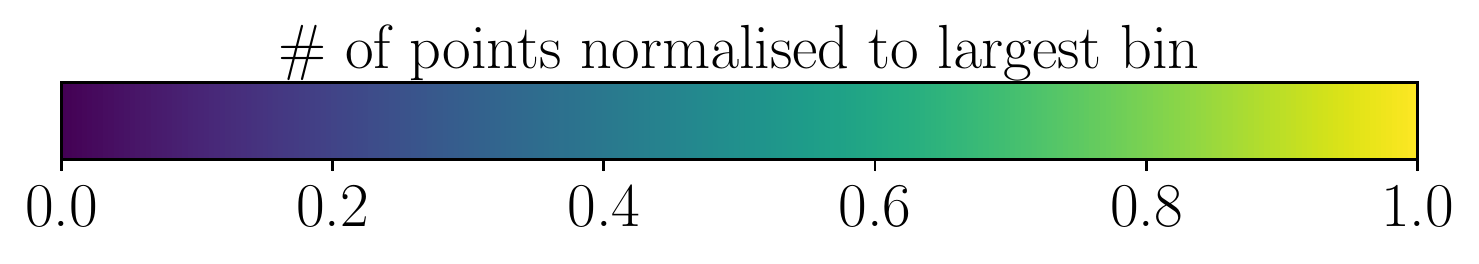}\\
\includegraphics[width=0.32\textwidth]{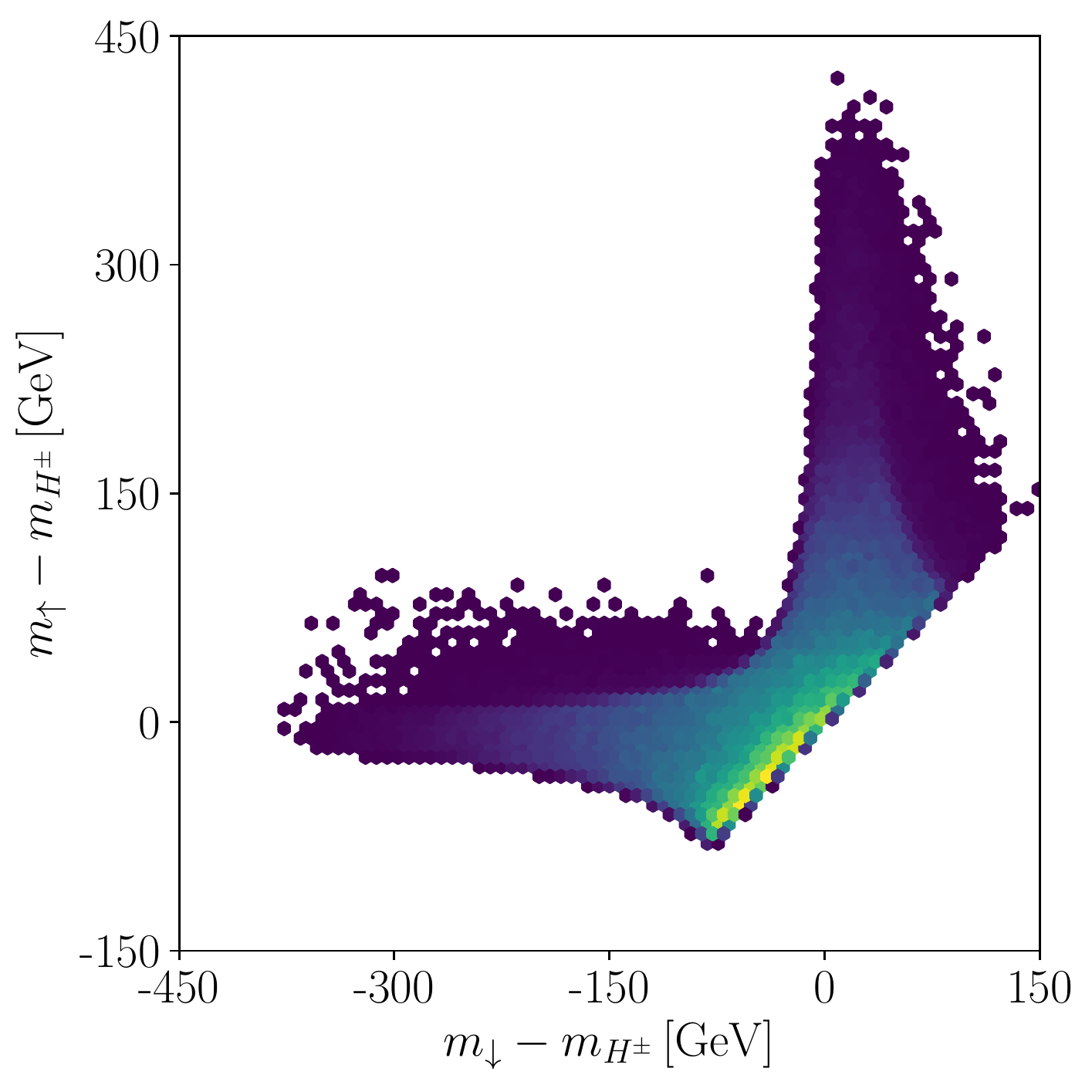}
\includegraphics[width=0.32\textwidth]{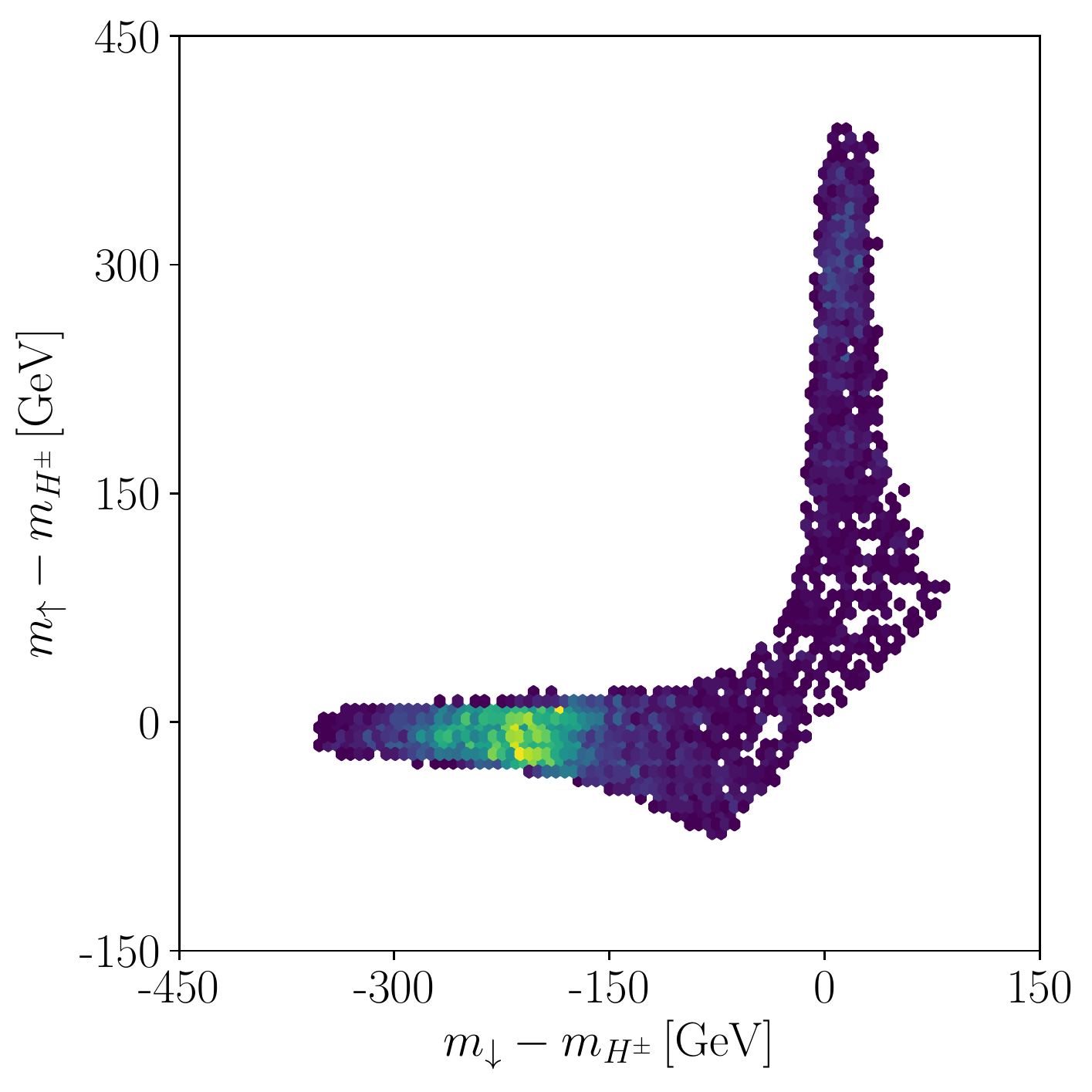}
\includegraphics[width=0.32\textwidth]{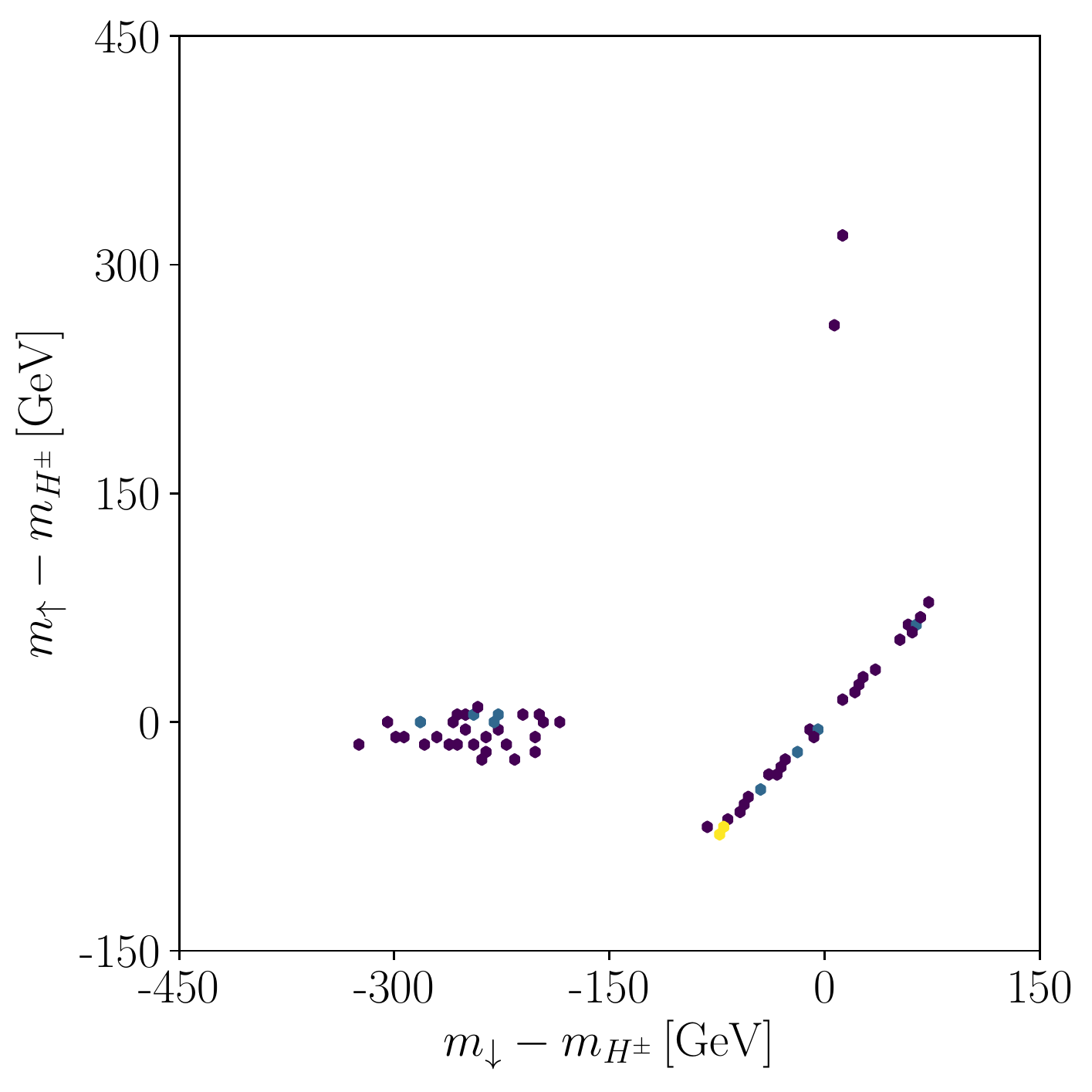}
%\vspace*{-0.2cm}
\caption{Type I, $H_1 = h$: The mass difference $m_\uparrow -
  m_{H^\pm}$ versus $m_\downarrow - m_{H^\pm}$. The colour code shows
  the relative frequency of left: all points passing the constraints;
  middle: all points with a strong PT and CP conservation; right: all points with $\xi_c
  \ge 1$ and explicit CP violation. \label{fig:tendmassdiff}}
\end{center}
\vspace*{-0.6cm}
\end{figure}
%
%\begin{figure}[b!]
%\begin{center}
%\includegraphics[width=0.49\textwidth]{DiffMass-HeatMap-Sample-T1.pdf}
%\includegraphics[width=0.49\textwidth]{DiffMass-HeatMap-AE-T1.pdf}
%%\vspace*{-0.2cm}
%\caption{Type I, $H_1 = h$: The mass difference $(m_\uparrow -
%  m_{H^\pm})$ versus $(m_\downarrow-m_{H^\pm})$. The colour code shows 
%the relative frequency of left: all points passing the constraints;
%right: all points with additionally $\xi_c \ge 1$. \label{fig:tendmassdiff}}
%\end{center}
%\vspace*{-0.6cm}
%\end{figure}
In Fig.~\ref{fig:tendmassdiff} we investigate how a strong PT affects
LHC phenomenology. The left plot shows in the plane $(m_\uparrow - m_{H^\pm})$ versus
$(m_\downarrow-m_{H^\pm})$ the frequency of the points
that pass the constraints, the middle and right plot display the frequency of the
points when additionally a strong EWPT is required and only the 
CP-conserving points (middle) or the CP-violating points (right)
are taken into account. The LHC constraints favour degenerate
non-SM-like neutral Higgs bosons that 
are lighter (by at most 70 GeV) or equal to the charged Higgs boson
mass, {\it cf.}~yellow points in Fig.~\ref{fig:tendmassdiff}
(left). In the CP-conserving case, a strong PT, however, favours a 
hierarchy between the neutral masses, see Fig.~\ref{fig:tendmassdiff}
(middle): While the heavier $H_\uparrow$ is approximately mass 
degenerate with the charged Higgs, $H_\downarrow$ is
lighter by about 200~GeV. A slight preference is also found for mass
degenerate $H_\downarrow$ and $H^\pm$ with $H_\uparrow$ being heavier
by 250-350~GeV. \s

As can be inferred from the comparison of
the middle and right plot, most of the points with a strong PT are found in
the CP-conserving limit, and we find that the strongest PT is obtained
in the alignment limit where $H_1$ is totally SM-like. In line with
Ref.~\cite{Basler:2016obg} we hence conclude, that a strong PT favours
scenarios where decays of the heavier neutral Higgs boson into the
lighter one together with a $Z$ boson are kinematically allowed and may have
a considerable branching ratio due to the involved coupling being
large in the alignment limit. We also find that the mass configuration with
a mass gap of 290-330 GeV between $H_\downarrow$ and $H_\uparrow$
induces the largest $\xi_c$ values (not shown in the plots). 
When keeping only the CP-violating points (right plot), we find three
different types of mass configurations that are compatible with a
strong PT. $(i)$ We have the parameter sets where $H_\uparrow$ and
$H^\pm$ are approximately mass degenerate 
with $H_\downarrow$ being lighter by 180 to 320~GeV. $(ii)$ We find
two points with mass degenerate $H_\downarrow$ and $H^\pm$ and $H^\uparrow$
begin heavier by 280~GeV and 330~GeV, respectively. $(iii)$ Finally,
$H_\downarrow$ and $H_\uparrow$ are approximately mass degenerate and
$H^\pm$ is either lighter, heavier or has the same mass. We
investigate the features of these points closer by identifying
benchmark points for each of these regions. We denote these points by
BPi1-3, BPii1-2 and BPiii1-3 for three mass configuration types $(i)$,
$(ii)$ and $(iii)$, respectively. In Tables~\ref{tab:cpvbenchmarksi},
\ref{tab:cpvbenchmarksii} and \ref{tab:cpvbenchmarksiii} we list the input parameters
of the 3 sets of benchmarks. We also give the derived third neutral
Higgs boson mass, the strength of the PT $\xi_c$ and the CP admixtures
of the Higgs bosons. These are quantified by the mixing matrix element
squared relating to the CP-odd neutral component of the Higgs doublets
$\zeta_3$, namely $R_{i3}^2$ ($i=1,2,3$). Finally, we give for all
benchmark points the result for the production of a SM-like Higgs pair
$h$ through gluon fusion at a c.m.~energy of $\sqrt{s}=14$~TeV including the NLO
QCD corrections in the heavy top mass limit \cite{Dawson:1998py}. We
come back to the discussion of Higgs pair production in
Section~\ref{sec:triltype1}.
In Tables~\ref{tab:cpvbpphenoi}, \ref{tab:cpvbpphenoii} and
\ref{tab:cpvbpphenoiii} we summarise 
the dominant branching ratios of the benchmark points of the three
sets, which determine their phenomenology. For $H_1$ we have SM-like
branching ratios and do not give them separately here. \s
\begin{table}[t!]
 \begin{center}
 \begin{tabular}{lccc}
     \toprule
    & BPi1 & BPi2 & BPi3  \\ 
   \midrule
$m_{H_1}$ [GeV] &  125.09 &  125.09 & 125.09 \\
$m_{H_2}$ [GeV] &  322.28 & 291.49 & 188.52 \\
$m_{H^\pm}$ [GeV] & 522.12 & 543.30 & 490.97   \\
$\mbox{Re}(m_{12}^2)$ [GeV$^2$] & 17100 & 15590 & 9053  \\
$\alpha_1$ & 1.484 & 1.366 & 1.548 \\
$\alpha_2$ & -0.018 & -0.028 & -0.085 \\
$\alpha_3$ & 0.112 & 0.086 & 0.999 \\
$\tan\beta$ & 5.97 & 5.08 & 19.97 \\ \hline 
$m_{H_3}$ [GeV] & 503.15 & 548.97 & 491.27 \\
$\xi_c$ & 1.26 & 1.52 & 1.15 \\
$R_{13}^2$ & $3.284\cdot 10^{-4}$ & $7.641 \cdot 10^{-4}$ &  $7.219
\cdot 10^{-3}$\\
$R_{23}^2$ & $0.012$ & $7.436 \cdot 10^{-3}$ & 0.702\\
$R_{33}^2$ & 0.987 & 0.992 & 0.291 \\ \hline
$\sigma_{hh}^{\text{NLO}}$ [fb] & 89.14 & 217.95 & 38.42 \\ \bottomrule
  \end{tabular}
\caption{Line 1-8: The input parameters of the benchmark points BPi1-3. Line
9 to 13: The derived 3rd neutral Higgs boson mass, the
$\xi_c$ value and the CP-odd admixtures $R_{i3}^2$. Line 14: The NLO QCD
gluon fusion $hh$ production cross section at
$\sqrt{s}=14$~TeV.} \label{tab:cpvbenchmarksi} 
   \end{center}
\vspace*{-0.2cm}
 \end{table}
\begin{table}[t!]
 \begin{center}
 \begin{tabular}{llll}
     \toprule
    & BPi1 & BPi2 & BPi3  \\ 
   \midrule
BR($H_2$) &  BR($H_2 \to H_1 H_1$) = 0.526 &  
BR($H_2 \to H_1 H_1$) = 0.400 & BR($H_2 \to WW$) = 0.787 \\ 
& BR($H_2 \to WW$) = 0.310 & BR($H_2 \to ZH_1$) = 0.294 & BR($H_2 \to
ZZ$) = 0.196\\
& BR($H_2 \to ZZ$) = 0.140 & BR($H_2 \to WW$) = 0.156 & BR($H_2 \to
b\bar{b}$) = 0.010\\\hline
BR($H_3$) &  BR($H_3 \to ZH_2$) = 0.866 & BR($H_3 \to ZH_2$)=0.940 
& BR($H_3\to ZH_2$) = 0.982 \\
& BR($H_3 \to t\bar{t}$) = 0.100 & BR($H_3 \to t\bar{t}$)=0.056 
& BR($H_3\to WW$) = 0.0075 \\
& BR($H_3 \to ZH_1$) = 0.028 & BR($H_3 \to WW$)=0.002
& BR($H_3\to ZH_1$) = 0.0044 \\ \hline
BR($H^\pm$) & BR($H^\pm \to WH_2$) = 0.906 & BR($H^\pm \to WH_2$)=0.943
& BR($H^\pm \to WH_2$) = 0.987 \\
& BR($H^+ \to t\bar{b}$) = 0.069 & BR($H^+ \to t\bar{b}$)=0.054
& BR($H^\pm \to WH_1$) = 0.011 \\
& BR($H^\pm \to WH_1$) = 0.025 & BR($H^\pm \to WH_1$)=0.002 
& BR($H^+ \to t\bar{b}$) = 0.0025 \\
   \bottomrule
  \end{tabular}
\caption{The dominant branching ratios of the BPi1-3 Higgs
  bosons.} \label{tab:cpvbpphenoi} 
   \end{center}
\vspace*{-0.6cm}
 \end{table}

\begin{table}[t!]
 \begin{center}
 \begin{tabular}{lcc}
     \toprule
    & BPii1 & BPii2 \\ 
   \midrule
$m_{H_1}$ [GeV] &  125.09 &  125.09 \\
$m_{H_2}$ [GeV] &  263.77 & 236.99 \\
$m_{H^\pm}$ [GeV] & 257.64 & 223.76 \\
$\mbox{Re}(m_{12}^2)$ [GeV$^2$] & 13823 & 8044 \\
$\alpha_1$ & 1.497 & 1.287 \\
$\alpha_2$ & $4.741\cdot 10^{-3}$ & -0.050 \\
$\alpha_3$ & -0.021 & 0.127 \\
$\tan\beta$ & 4.90 & 6.95 \\ \hline 
$m_{H_3}$ [GeV] & 519.75 & 542.95 \\
$\xi_c$ & 1.02 & 1.003 \\
$R_{13}^2$ & $2.247\cdot 10^{-5}$ & $2.459 \cdot 10^{-3}$ \\
$R_{23}^2$ & $4.545 \cdot 10^{-4}$ & $0.016$ \\
$R_{33}^2$ & 0.999 & 0.981 \\ \hline
$\sigma_{hh}^{\text{NLO}}$ [fb] & 68.28 & 30.73
\\
   \bottomrule
  \end{tabular}
\caption{Line 1-8: The input parameters of the benchmark points BPii1
  and BPii2. Line
9 to 13: The derived 3rd neutral Higgs boson mass, the
$\xi_c$ value and the CP-odd admixtures $R_{i3}^2$. Line 14: The NLO QCD
gluon fusion $hh$ production cross section at
$\sqrt{s}=14$~TeV.} \label{tab:cpvbenchmarksii}
   \end{center}
\vspace*{-0.2cm}
 \end{table} 
We start our discussion with the benchmark set BPi1-3. Denoting by
$\phi$ the heavier mass degenerate Higgs bosons $H_\uparrow=H_3$ and
$H^\pm$, the three benchmarks differ in the their mass difference 
$m_\phi - m_{H_\downarrow = H_2}$ which is about 200, 250 and 300~GeV for
BPi1, BPi2 and BPi3. Since the masses of the heavier Higgs bosons
$\phi$ are between about 490 and 550~GeV in all three scenarios, this
means that $m_{\downarrow}$ becomes successively smaller with increasing
mass difference. The interplay of the kinematically available phase
space and the CP nature of the Higgs bosons determines their branching
ratios. In particular, the decays of $H_\downarrow \equiv H_2$ turn
out to be interesting. In BPi1 and BPi2, where $H_2$
is mainly CP-even\footnote{We define in this paper the CP nature of a Higgs
  boson through its CP-odd admixture $R_{i3}^2$ and call a Higgs
  boson $H_i$ mostly CP-even (CP-odd) in case of small (large)
  $R_{i3}^2$.}, it is heavy enough to decay with substantial
branching ratio into a SM-like Higgs pair $H_1=h$. Since in BPi2 the
coupling to massive gauge bosons is smaller than in BPi1, the next
important decay in BPi1 is into $WW$, while in BPi2 it is into
$H_1 Z$. Although the coupling is $H_1 H_2 Z$ is rather small, because
both $H_1$ and $H_2$ are mostly CP-even, the decay is important as all the
other decays involve even smaller couplings or are kinematically
closed. Note, in particular, that in the CP-conserving 2HDM this decay
would be forbidden. Since $H_2$ also decays with a branching
ratio of 0.156 into $WW$ the CP-violating nature
of $H_2$ can be identified at the LHC through its decay rates: Due to
the fact that we know 
already that $H_1 \equiv h$ is mainly CP-even, as it corresponds to
the discovered Higgs boson, the properties of which have been
determined, the observation of the $H_2$ decay both into $WW$ and
$ZH_1$ clearly identifies it to be CP-violating. This idea has
  been proposed and discussed before in\cite{Branco:1999fs,
    Fontes:2015xva}.\footnote{For discussions within the NMSSM, {\it 
  cf.}~\cite{King:2015oxa}.} The reason is that the
former decay requires $H_2$ to be CP-even, whereas the latter requires
it to be CP-odd in a purely CP-conserving theory. 
In BPi3, $H_2$ has a large CP-odd admixture. Due to its small mass, however, the
off-shell decay into $ZH_1$ is less important than the on-shell
decays into  massive gauge bosons. Here we make the interesting
observation that $H_2$, despite its rather CP-odd nature, mainly
decays into massive gauge bosons as a consequence of the available phase
space. These decays are only possible because $H_2$ also
has a CP-even admixture. The heavier non-SM-like Higgs
$H_\uparrow \equiv H_3$ in all three scenarios mainly decays into
$ZH_2$. In BPi1 and BPi2, where $H_3$ is mainly CP-odd, the next
important decay is the one into top-quark pairs. In BPi3 $H_3$ is more
CP-even, so that the second important decay becomes the one into
$WW$. Note, that the decay into $ZH_1$ is less important than into
$ZH_2$ because of a much smaller involved coupling. In our scenarios, the coupling of
$H^\pm$ to $WH_1$ is smaller than the one to $WH_2$, so that the charged Higgs
decays mainly into $WH_2$ followed by the decay into $t\bar{b}$ in
BPi1 and BPi2 and by $WH_1$ in BPi3. \s
%In summary, all scenarios feature
%beyond the SM Higgs decays into Higgs pairs or gauge+Higgs boson final
%states with substantial branching ratios, that should be testable at
%the LHC. Some of the decays of our scenarios would be forbidden in a
%purely CP-conserving 2HDM. \s

%
\begin{table}[t!]
 \begin{center}
 \begin{tabular}{lll}
     \toprule
    & BPii1 & BPii2 \\ 
   \midrule
BR($H_2$) &  BR($H_2 \to WW$) = 0.464 &  
BR($H_2 \to WW$) = 0.698 \\ 
& BR($H_2 \to H_1 H_1$) = 0.336 & BR($H_2 \to ZZ$) = 0.289 \\
& BR($H_2 \to ZZ$) = 0.199 & BR($H_2 \to b\bar{b}$) = 0.005 \\\hline
BR($H_3$) &  BR($H_3 \to W^\pm H^\mp$) = 0.672 & BR($H_3 \to W^\pm H^\mp$)=0.685 
\\
& BR($H_3 \to ZH_2$) = 0.297 & BR($H_3 \to ZH_2$)=0.297
\\
& BR($H_3 \to t\bar{t}$) = 0.019 & BR($H_3 \to ZH_1$)=0.011
\\ \hline
BR($H^\pm$) & BR($H^+ \to t\bar{b}$) = 0.923 & BR($H^+ \to t\bar{b}$)=0.924
\\
& BR($H^\pm \to WH_1$) = 0.075 & BR($H^\pm \to WH_1$)=0.074
\\
   \bottomrule
  \end{tabular}
\caption{The dominant branching ratios of the BPii1 and BPii2 Higgs
  bosons.} \label{tab:cpvbpphenoii} 
   \end{center}
\vspace*{-0.2cm}
 \end{table}
The benchmark points BPii1 and BPii2 are rather similar. They differ
in the mass gap between the heavier $H_\uparrow \equiv H_3$ and the
lighter $H_\downarrow \equiv H_2$, which is now almost mass degenerate with
$H^\pm$. Denoting the latter two by $\varphi$, we have $m_{H_\uparrow} -
m_\varphi \approx 260$~GeV in BPii1 and around 310~GeV in BPii2, where
$H_\downarrow$ and $H^\pm$ are lighter than in BPii1. In both
scenarios $H_1\equiv h$ and $H_\downarrow$ are mostly CP-even and
$H_\uparrow$ is mostly CP-odd. Therefore, $H_2$ mainly decays into the
massive gauge bosons, {\it cf.}~Table~\ref{tab:cpvbpphenoii}. In BPii1
also the decay into $H_1H_1$ has a substantial branching ratio of
0.34. In BPii2, this decay is kinematically closed.\footnote{In the
  computation of the branching ratios the off-shell decays into
  massive gauge bosons and gauge+Higgs boson final states are
  included but not the ones into Higgs pair 
  final states. The computation and inclusion of the off-shell decays
  into Higgs pairs of the C2HDM is deferred to a future project.} The dominant
decays of $H_3$, which is now heavier than $H^\pm$, are into a 
gauge+Higgs boson final state, namely 
into $W^\pm H^\mp$ and $ZH_2$ with branching ratios of about 0.7 and
0.3 in both scenarios. In contrast to the parameter set $(i)$, the
charged Higgs boson is considerably lighter, so that it mainly decays
into $t\bar{b}$ and the decay into a gauge+Higgs final state is very
small. Again, we find non-SM-like decays in these scenarios, where the
possible final states reflect the mass hierarchy among the Higgs
bosons. Note, finally, that both benchmark points have a $\xi_c$ just slightly above the
strong PT value of $\xi_c =1$ and are almost ruled out.\s

\begin{table}[t!]
 \begin{center}
 \begin{tabular}{lccc}
     \toprule
    & BPiii1 & BPiii2 & BPiii3  \\ 
   \midrule
$m_{H_1}$ [GeV] &  125.09 &  125.09 & 125.09 \\
$m_{H_2}$ [GeV] & 494.834 & 420.481 & 460.698  \\
$m_{H^\pm}$ [GeV] & 503.432 & 499.906 & 385.220    \\
$\mbox{Re}(m_{12}^2)$ [GeV$^2$] & 39529 & 27614 & 20392 \\
$\alpha_1$ & 0.920 & 0.957 & 0.932  \\
$\alpha_2$ & $9.303 \cdot 10^{-3}$ & 0.012 & 0.0101  \\
$\alpha_3$ & -0.461 & -0.131 & -0.514  \\
$\tan\beta$ & 1.488 & 1.851 & 1.608  \\ \hline
$m_{H_3}$ [GeV] & 496.683 & 429.492 & 462.683 \\
$\xi_c$ & 1.18 & 1.03 & 1.21 \\
$R_{13}^2$ & $8.655 \cdot 10^{5}$ & $1.391 \cdot 10^{-4}$ & $1.011 \cdot 10^{-4}$ \\
$R_{23}^2$ & 0.198 & 0.0170 & 0.241 \\
$R_{33}^2$ & 0.802 & 0.983 & 0.748 \\ \hline
$\sigma_{hh}^{\text{NLO}}$ [fb] & 36.05 & 47.85 & 31.66 \\
   \bottomrule
  \end{tabular}
\caption{Line 1-8: The input parameters of the benchmark points BPiii1-3. Line
9 to 13: The derived 3rd neutral Higgs boson mass, the
$\xi_c$ value and the CP-odd admixtures $R_{i3}^2$. Line 14: The NLO QCD
gluon fusion $hh$ production cross section at
$\sqrt{s}=14$~TeV.} \label{tab:cpvbenchmarksiii}
   \end{center}
\vspace*{-0.4cm}
 \end{table}
\begin{table}[t!]
 \begin{center}
 \begin{tabular}{llll}
     \toprule
    & BPiii1 & BPiii2 & BPiii3  \\ 
   \midrule
BR($H_2$) &  BR($H_2 \to t\bar{t}$) = 0.972 &  
BR($H_2 \to t\bar{t}$) = 0.860 & BR($H_2 \to t\bar{t}$) = 0.948 \\
& BR($H_2 \to WW$) = 0.015 & BR($H_2 \to WW$) = 0.089 & BR($H_2 \to WW$) = 0.027 \\
& BR($H_2 \to ZZ$) = 0.007 & BR($H_2 \to ZZ$) = 0.042 & BR($H_2 \to ZZ$) = 0.013 \\ 
\hline
BR($H_3$) & BR($H_3 \to t\bar{t}$) = 0.984 & BR($H_3 \to t\bar{t}$)= 0.951
& BR($H_3\to t\bar{t}$) = 0.969 \\ 
& BR($H_3 \to Zh$) = 0.010 & BR($H_3 \to Zh$) = 0.044 & BR($H_3 \to Z h$) = 0.018 \\
\hline
BR($H^\pm$) & BR($H^+ \to t\bar{b}$) = 0.987 & BR($H^+ \to t \bar{b}$)= 0.932
& BR($H^+ \to t\bar{b}$) = 0.985 \\
& BR($H^\pm \to Wh$) = 0.011 & BR($H^\pm \to Wh$) = 0.065 & BR($H^\pm\to W h$)
= 0.013 \\
   \bottomrule
  \end{tabular}
\caption{The dominant branching ratios of the BPiii1-3 Higgs
  bosons.} \label{tab:cpvbpphenoiii} 
   \end{center}
\vspace*{-0.2cm}
 \end{table}
We find four scenarios for which all non-SM-like Higgs bosons
are approximately mass degenerate. Denoting by $\Phi$ generically
$H_\downarrow$, $H_\uparrow$ and $H^\pm$, we have for their average
mass $m_\Phi \approx 498$~GeV, $547$~GeV, $555$~GeV and 563~GeV,
respectively, for these scenarios. All four scenarios feature the same
dominant branching ratios. We exemplary give as
benchmark point BPiii1 the scenario with the lightest mass spectrum. 
The two other benchmark points feature $m_\downarrow \approx m_\uparrow$,
where the charged Higgs boson is heavier (BPiii2) or lighter
(BPiii3). In BPiii1, there is no mass gap between the non-SM-like Higgs bosons,
and in BPiii2 and BPiii3 the largest mass gap is 80 and
77~GeV, respectively. Furthermore, in all these scenarios the
couplings between $H_{2,3}$, $Z$ and 
$h$ are small. Therefore the non-SM-like Higgs bosons decay dominantly
into SM final states, which due to their mass values are $t\bar{t}$ for
the neutral Higgs bosons and $t\bar{b}$ for the charged Higgs
boson. For $H_2$ which has a significant CP-odd admixture but still is
dominantly CP-even, the next important decay channels are those into
$WW$ and $ZZ$. \s

We can therefore summarise that the requirement of a strong PT induces
Higgs spectra with mass gaps that are characterized by large Higgs
branching ratios of the non-SM-like Higgs bosons into gauge+Higgs
final states or into Higgs pairs which should be testable at the
LHC. Some of the decays of our scenarios would be forbidden in a
purely CP-conserving 2HDM. In contrast to the
CP-conserving case, due to the CP-mixing of all Higgs bosons, also
scenarios with the non-SM-like Higgs bosons being close in mass or even
mass degenerate are preferred. In these cases the dominant decays are those
into SM final states.\s

\begin{figure}[t!]
\begin{center}
\includegraphics[width=0.49\textwidth]{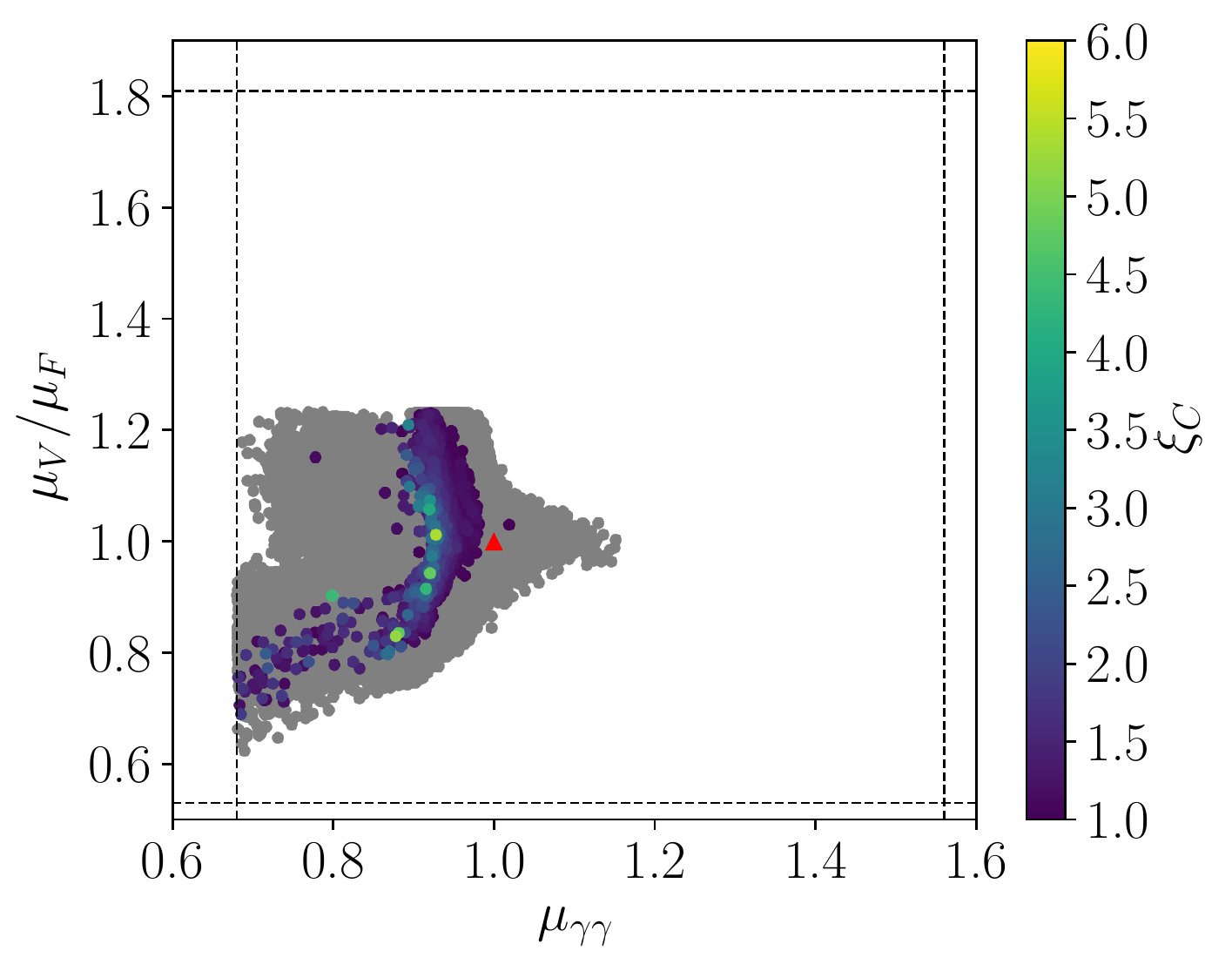}
\includegraphics[width=0.49\textwidth]{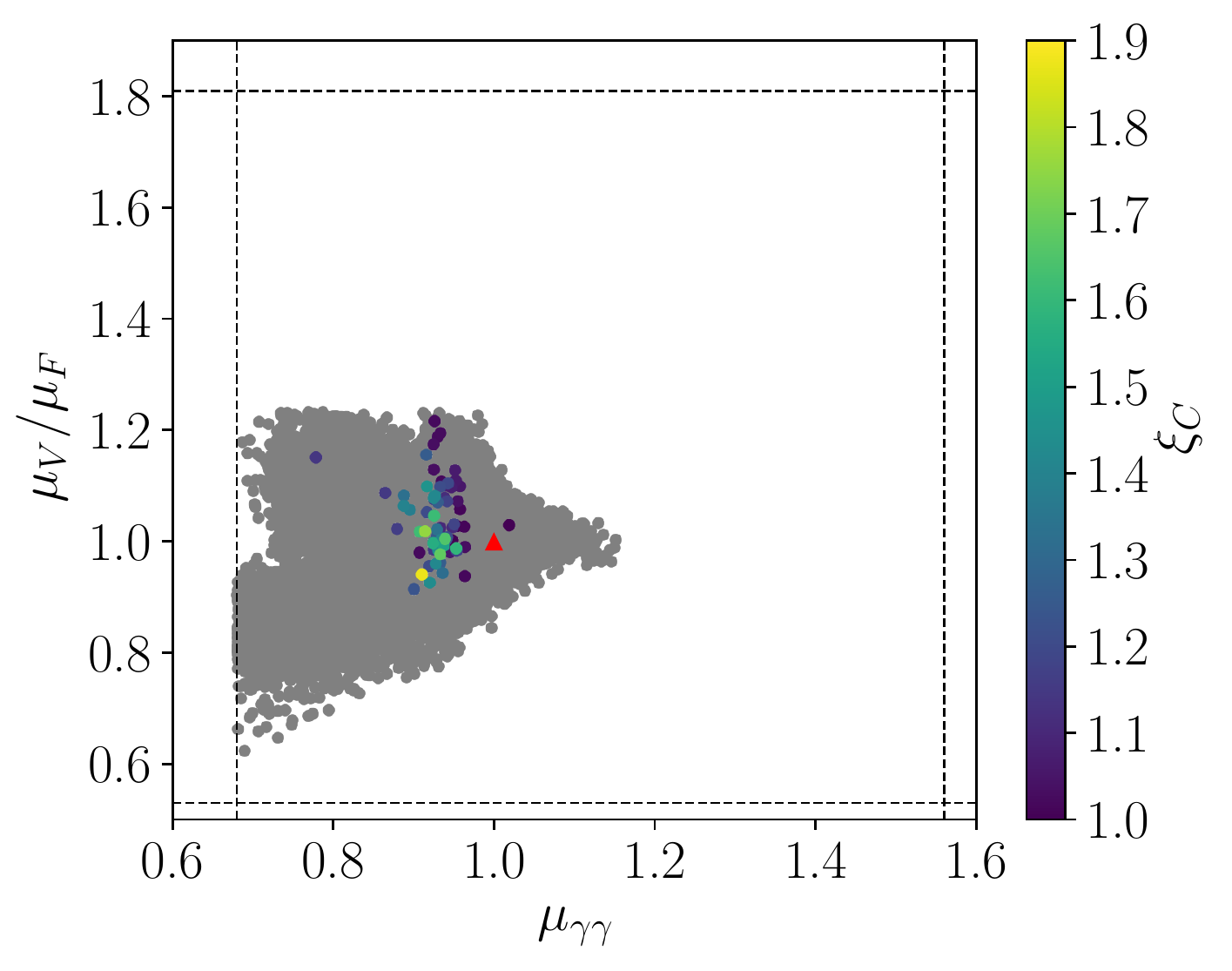}
%\vspace*{-0.2cm}
\caption{Type I, $H_1 = h$: $\mu_V/\mu_F$ versus
  $\mu_{\gamma\gamma}$. Grey: all points passing the applied
  constraints, colour: all points with additionally $\xi_c \ge 1$;
  left: all 2HDM points, right: only C2HDM points. The red triangle
  marks the SM result. 
 \label{fig:muvfmugagat1}}
\end{center}
\vspace*{-0.6cm}
\end{figure}
Figure~\ref{fig:muvfmugagat1} shows in grey the distribution of the
Higgs signal strengths for the scenarios passing the constraints and
in colour the ones that are additionally compatible with a strong
PT. The fermion initiated cross section (gluon fusion and associated
production with a heavy quark pair) of the SM-like Higgs boson
$h$ normalised to the SM is denoted by $\mu_F$, and the
normalised production cross section through massive gauge bosons
(gauge boson fusion and associated production with a vector boson) by
$\mu_V$. The value $\mu_{xx}$ is defined as
\beq
\mu_{xx} = \mu_F \frac{\mbox{BR}_{\text{C2HDM}} (h \to
  xx)}{\mbox{BR}_{\text{SM}} (H_{\text{SM}} \to xx)} \;,
\eeq
where $H_{\text{SM}}$ is the SM Higgs boson with mass
125~GeV. In the right plot we retained only the points with explicit CP
violation. Photonic rates of up to 1.15 are still allowed. When
imposing a strong PT, however, this reduces to 1.02 for 
parameter points with CP violation and below 1 for
  CP-conserving scenarios. Note that we
did not find any points with both reduced $\mu_V/\mu_F$ and
$\mu_{\gamma\gamma}$ by more than 10\% that lead to $\xi_c \ge 1$ in the CP-violating
case.  \s

In Fig.~\ref{fig:mutautaumuvvt1} we see in the
$\mu_{\tau\tau}-\mu_{VV}$ plane the distribution of points
passing the constraints and how this compares to the points when we
require a strong PT (left) and only keep those points that
are CP-violating and feature $\xi_c \ge 1$ (right). Clearly, in the
CP-violating case the points with a strong PT are much more sparse and
disfavour points with $\mu_{\tau\tau}$ above the SM value of 1 and
reduced $\mu_{VV}$. The differences in the rates that we found can be
exploited to distinguish the C2HDM from the R2HDM or to exclude the C2HDM.
%\textcolor{red}{Interestingly, in the CP-violating case we
%  find one scenario with both strongly reduced $\mu_{\tau\tau} \approx
%0.72$ and $\mu_{VV} \approx 0.82$.}
\begin{figure}[h!]
\begin{center}
\includegraphics[width=0.49\textwidth]{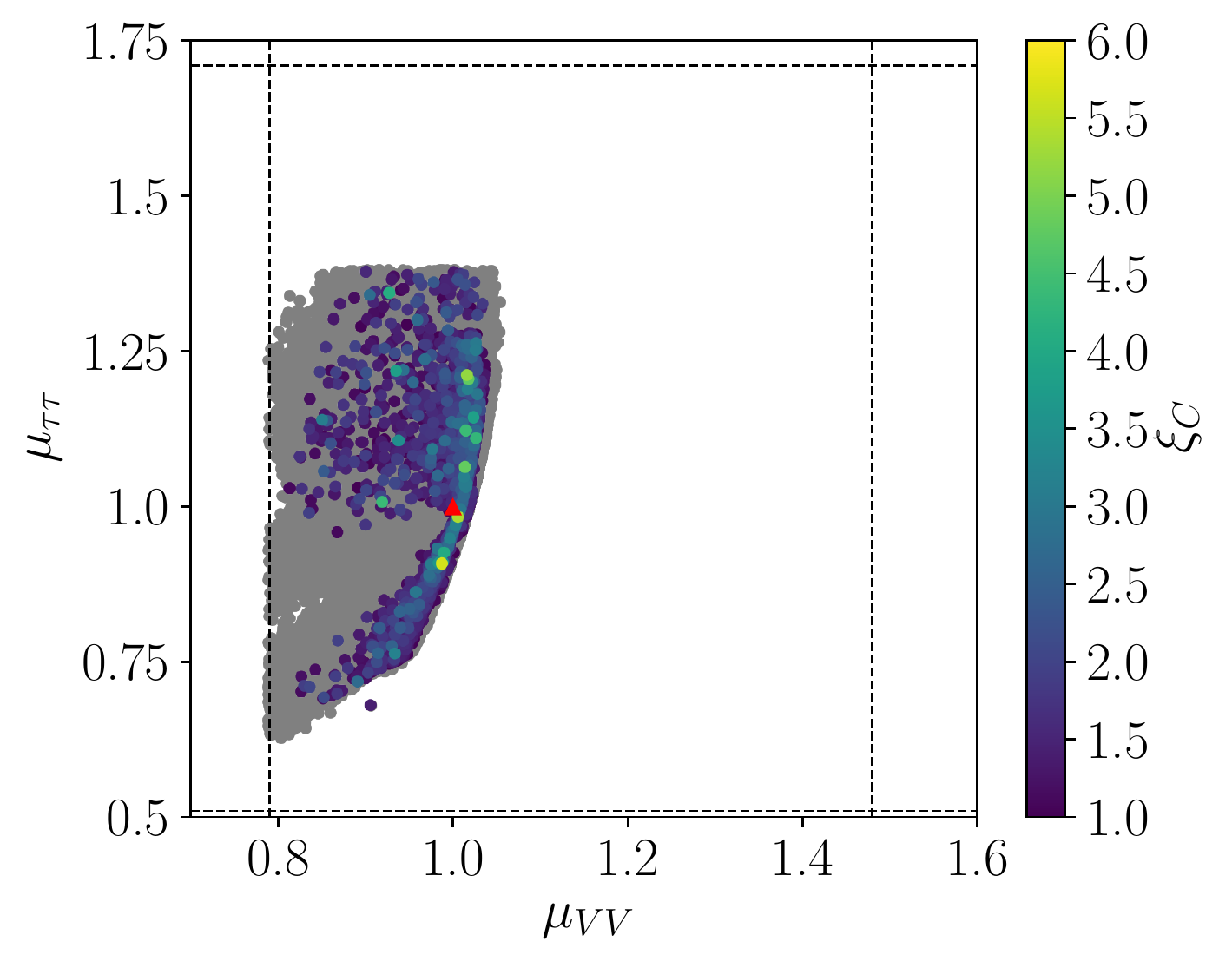}
\includegraphics[width=0.49\textwidth]{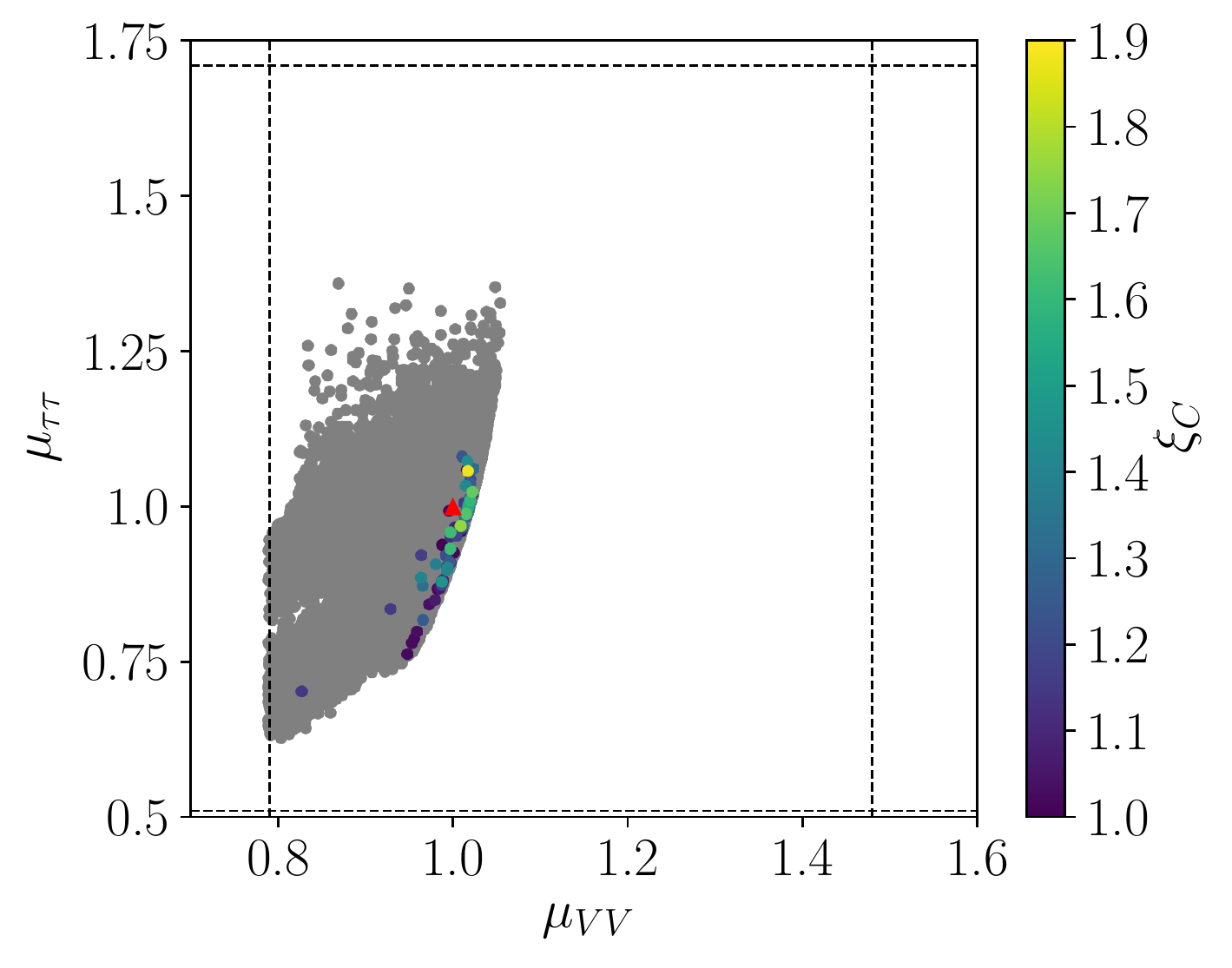}
%\vspace*{-0.2cm}
\caption{Type I, $H_1 = h$: $\mu_{\tau\tau}$ versus
  $\mu_{VV}$. Grey: all points passing the applied
  constraints, colour: all points with additionally $\xi_c \ge 1$; 
  left: all 2HDM, right: only C2HDM points. The red triangle
  marks the SM result. \label{fig:mutautaumuvvt1}}
\end{center}
\vspace*{-0.6cm}
\end{figure}

%%%%%%%%%%%%%%%%%%%%%%%%%%%%%%%%%%%%%%%%%%%%%%%%%%%%%%%%%%%
\section{Analysis of the trilinear Higgs self-couplings and Higgs pair
  production in the C2HDM Type I \label{sec:triltype1}}
Having computed the loop-corrected effective potential at finite
temperature, we now investigate the effects of
the NLO corrections on the trilinear Higgs self-coupling as well as
the interplay between a strong PT and the trilinear Higgs
self-couplings. The loop-corrected trilinear Higgs self-couplings are
obtained from the loop-corrected effective potential by performing the
third derivative of the Higgs 
potential with respect to the Higgs fields. The problem of infrared
divergences related to the Goldstone bosons in the Landau gauge is
treated analogously to the extraction of the masses from the second
derivative of the potential \cite{Camargo-Molina:2016moz}. 

%%%%%%%%%%%%%%%%%%%%%%%%%%%%%%%%%%%%%%%%%%%%%%%%%%%%%%%%%%%
\subsection{The Higgs self-coupling between three SM-like Higgs bosons}
\begin{figure}[b!]
\begin{center}
\includegraphics[width=0.49\textwidth]{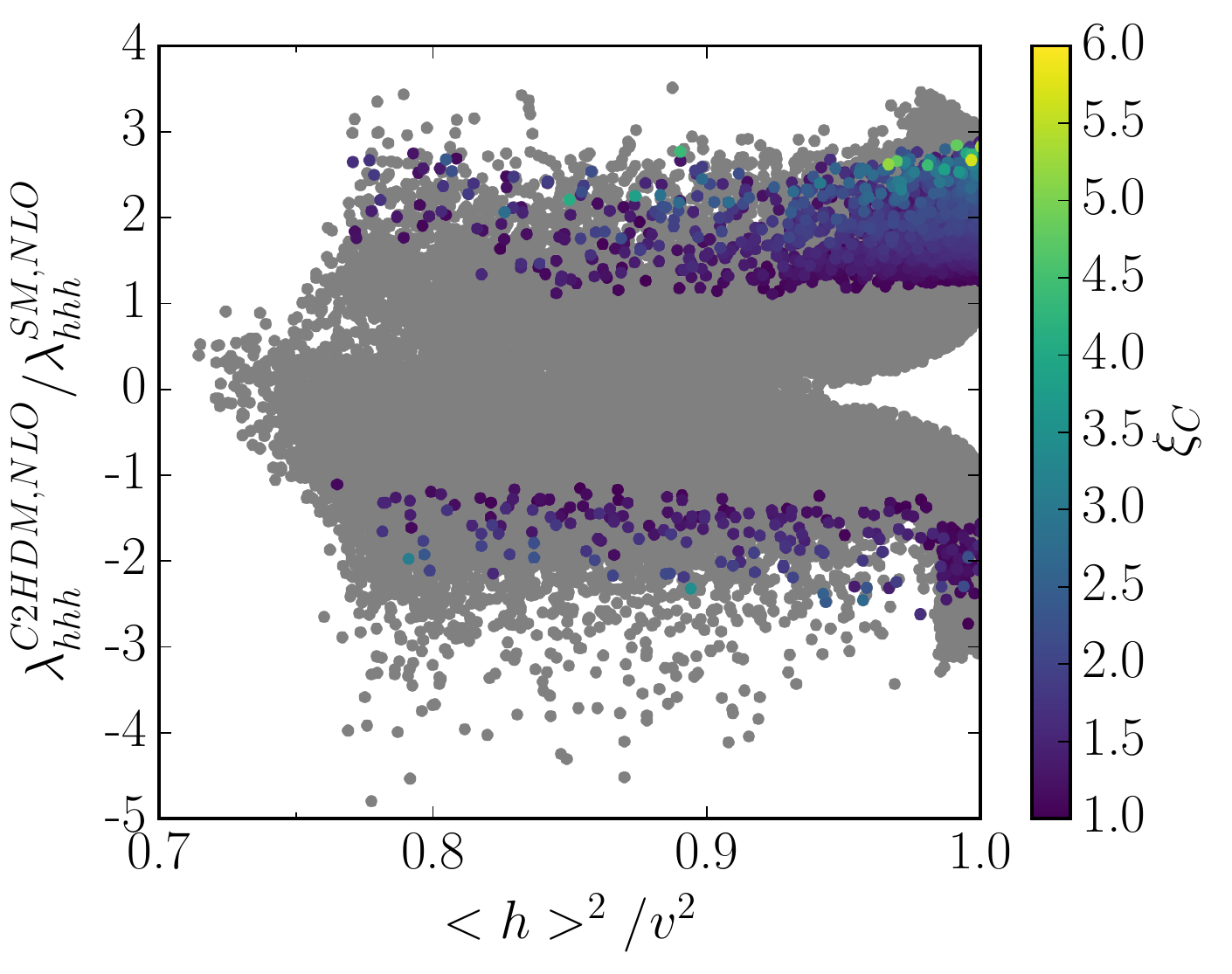}
\includegraphics[width=0.49\textwidth]{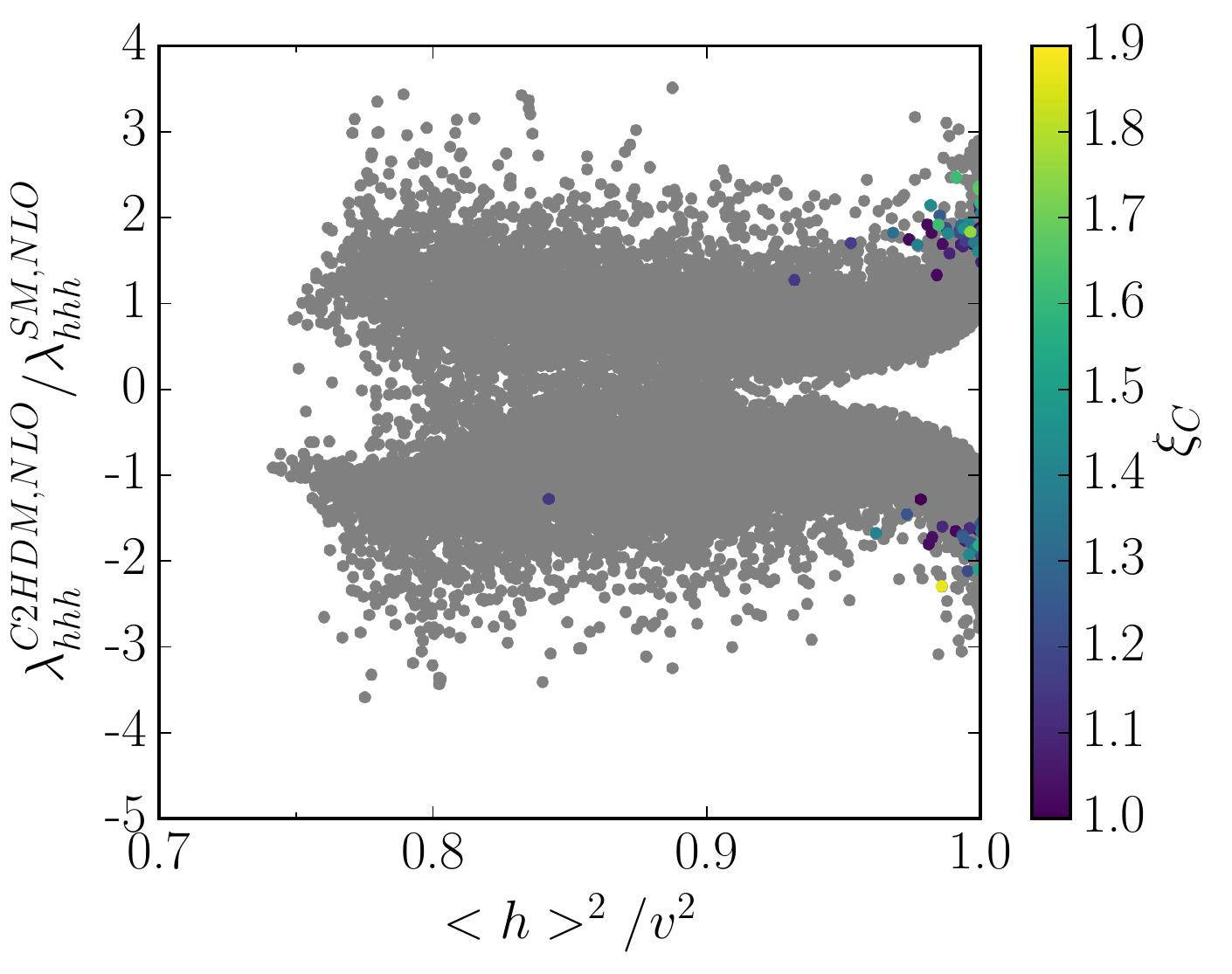}
%\vspace*{-0.2cm}
\caption{Type I, $H_1 = h$: The C2HDM trilinear Higgs
  self-coupling between three SM-like Higgs bosons $h$ normalized to the
  SM value, at NLO, as function of the fraction of the VEV squared carried by
  $h$. Grey: all points passing the constraints; colour:
  points with additionally $\xi_c \ge 1$. Left: all 2HDM points;
  right: only the C2HDM points. 
 \label{fig:trilhhh}}
\end{center}
\vspace*{-0.4cm}
\end{figure}
In Fig.~\ref{fig:trilhhh} we show the NLO trilinear Higgs coupling 
between three SM-like Higgs bosons $h$ of the C2HDM
normalized to the SM value, each at NLO, as function of the fraction
of the VEV squared (indicated by the brackets) carried by $h$.  The left plot
comprises all 2HDM points, while the right one only those with
explicit CP violation at $T=0$. For
the NLO SM value of the self-coupling we use the formula given in
\cite{Kanemura:2002vm} that takes into account the dominant top-quark
contributions at NLO. The grey points of the left plot show that in 
the 2HDM the trilinear coupling can substantially deviate from the SM
value and both be suppressed or enhanced compared to the
SM, {\it i.e.}~the present constraints do not restrict this coupling
to be close to the SM value. The maximum enhancement factor is $\pm
4.9$ (without the NLO unitarity constraints it would be 8). When
requiring a strong PT the enhancement is smaller,  
with $\lambda_{hhh}^{\text{C2HDM,NLO}}/\lambda_{hhh}^{\text{SM,NLO}}\approx 
(-2.73...-1.1)$ and $(1.1...2.9)$. 
Still, a strong PT clearly requires a large enough Higgs
self-coupling, larger than the one realized in the SM. Along these
lines, we see that the largest $\xi_c$ values are obtained for the
largest enhancement factors in this range. Including
only points with explicit CP violation the maximum enhancement
factor is found to be $\sim 3.6$ for all points passing the
constraints. The scenarios with $\xi_c >1$ reduce the maximum enhancement
factor somewhat with
$\lambda_{hhh}^{\text{C2HDM,NLO}}/\lambda_{hhh}^{\text{SM,NLO}}\approx
(-2.30...-1.28)$ and $(1.26...2.46)$. The upper bound on the trilinear
coupling is given by the  
interplay between the quartic self-couplings of the potential and the
masses of the Higgs bosons participating in the EWPT, with the latter
weakening the strength of the phase transition for too large
mass values. Therefore, for $\xi_c \ge 1$ and CP violation,  the ratio of the VEV
carried by the lighter of the non-SM-like Higgs bosons,
$H_\downarrow$, can be up to about $0.4 \, v$, in contrast to the heavier
$H_\downarrow$ with $\langle H_\uparrow \rangle = 0.12 \, v$ 
at most. The larger the fraction of the VEV carried by the Higgs
boson, the stronger is its participation in the EWPT, so that it
should not be too heavy in order not to spoil the strong
PT. The largest fraction of the VEV is carried by $h$, with
$<h> = (0.87...1)\, v$. The largest $\xi_c$ values are obtained
in the alignment limit, which is preferred by a strong PT, where the
SM-like Higgs boson carries the entire VEV. In this case the remaining two
neutral Higgs bosons do not take 
part in the PT, so that a strong first order PT requires a substantial
trilinear Higgs self-coupling for $h$.  
A conservative estimate of the prospects of the high-luminosity LHC to
measure the trilinear Higgs self-coupling of the SM, concludes that an
accuracy of about 50\% on its value might be feasible
\cite{Baglio:2012np}. This allows to distinguish some of the
C2HDM scenarios compatible with a strong PT from the SM case. \s

\begin{figure}[b!]
\begin{center}
\includegraphics[width=0.49\textwidth]{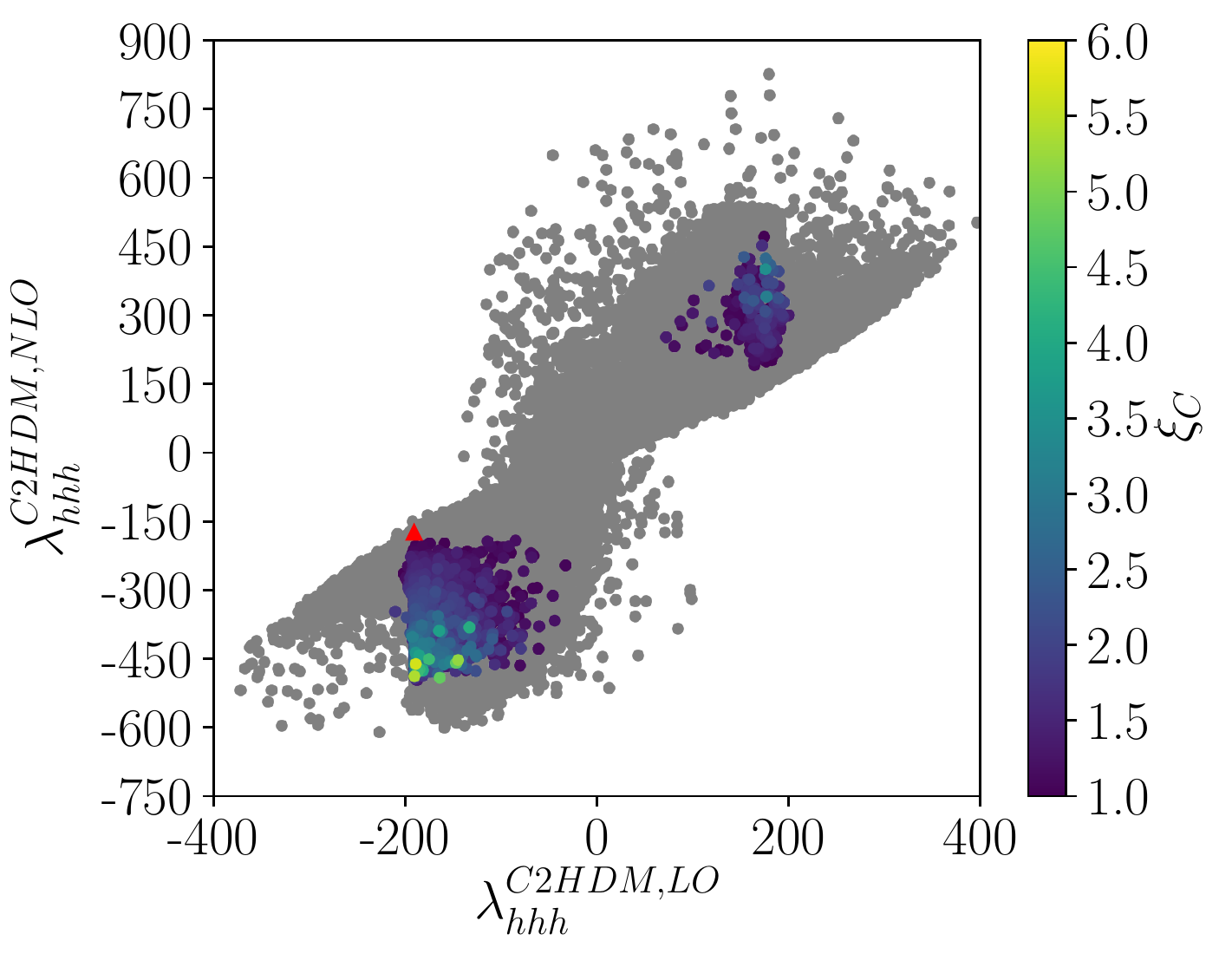}
\includegraphics[width=0.49\textwidth]{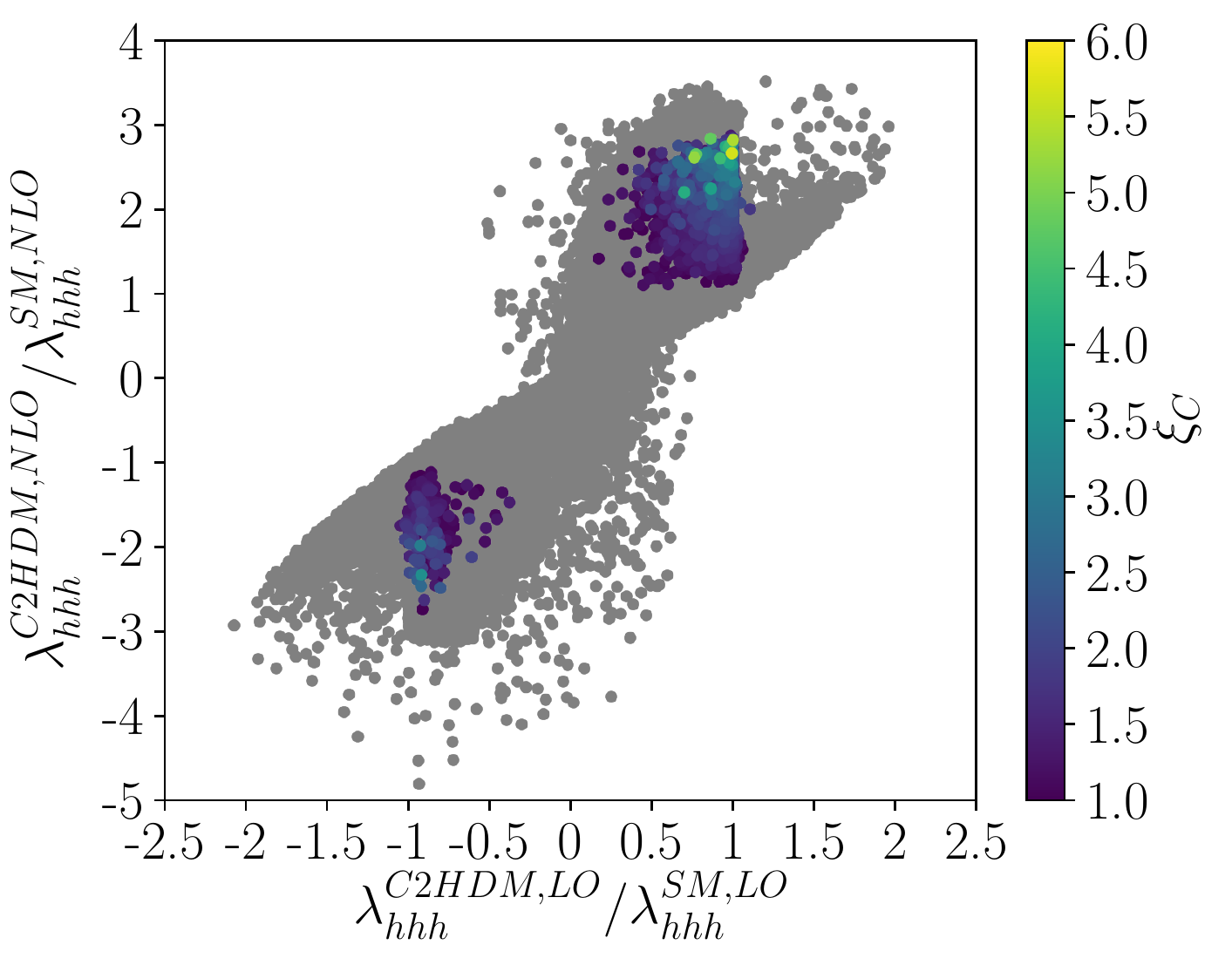}
\vspace*{-0.3cm}
\caption{Type I, $H_1 = h$: Left: The NLO C2HDM trilinear Higgs
  self-coupling between three SM-like Higgs bosons versus the LO
  value. The red triangle marks the SM point.
  Right: The ratio of the C2HDM trilinear Higgs self-coupling
  between three SM-like Higgs bosons and the SM counterpart at NLO
  versus the LO ratio. Grey: all points passing the constraints; colour: points
  with additionally $\xi_c \ge 1$. Both plots: all 2HDM points.
\label{fig:hhhnlo}}
\end{center}
\vspace*{-0.4cm}
\end{figure}
%\begin{figure}[b!]
%\begin{center}
%\includegraphics[width=0.49\textwidth]{C2HDM-TripleSM3-Tree-vT-T1-noscale.pdf}
%\includegraphics[width=0.49\textwidth]{C2HDM-TripleSM3-Tree-vT-T1.pdf}
%%\vspace*{-0.2cm}
%\end{center}
%\vspace*{-0.4cm}
%\end{figure}
The impact of the NLO corrections on the trilinear Higgs
self-coupling between three SM-like Higgs bosons in the C2HDM is shown
by Fig.~\ref{fig:hhhnlo}. The left plot shows the NLO coupling as
function of the leading order (LO) coupling. The NLO corrections can both suppress and
enhance the tree-level coupling. The corrections can be substantial.
%and enhancements by up to a factor 4.8 are possible while still being
%compatible with all constraints. 
For the points with a strong PT, the increase can be
by up to a factor 8.3, while it is 2.5 for the parameter point with the 
largest $\xi_c$. As outlined in \cite{Kanemura:2002vm}, 
large corrections, beyond the top loop contribution also present in
the SM, arise from Higgs loop contributions in the 2HDM. They increase with the fourth
power of the Higgs boson mass, $m_\Phi^4$, where $\Phi$ stands
generically for the 2HDM Higgs bosons $H,A,H^\pm$. And they are suppressed by
\beq
(1-M^2/m_\Phi^2)^3 \;, \label{eq:suppr}
\eeq
with $M^2 \equiv m_{12}^2/(\sin\beta \cos
\beta)$. The masses of the heavy Higgs bosons schematically take the
form
\beq
m_\Phi^2 = M^2 + f(\lambda_i) v^2 + {\cal O}(v^4/M^2) \;,
\eeq
where $f(\lambda_i)$ denotes a linear combination of the quartic
couplings of the Higgs potential. In case of small $M^2$ large Higgs
masses are generated through large values of $\lambda_i$. In this case
the loop contributions to the Higgs bosons do not decouple and
increase with $m_\Phi$, see also \cite{Krause:2016xku}. In case $M^2
\gg f(\lambda_i) v^2$ the loop 
contributions decouple in the limit $m_\Phi^2 \approx M^2 \to \infty$
due to the suppression factor Eq.~(\ref{eq:suppr}). For a strong PT we
need large Higgs boson self-couplings, respectively large couplings
$\lambda_i$, inducing the observed large corrections in the
non-decoupling regime. On the other hand, the masses of the Higgs
bosons participating in the EWPT should not become too heavy, thus
restricting the size of the quartic couplings and hence also of the
enhancement through the NLO corrections. This explains why in the
regime compatible with a strong PT the enhancement of 
$\lambda_{hhh}$ remains below the maximum enhancement factor
compatible with the applied constraints. 
%\textcolor{red}{
%The largest enhancements are reached for parameter
%sets in the limit of the real 2HDM. - True?}
The right plot of Fig.~\ref{fig:hhhnlo} displays the ratio between the C2HDM
coupling and the SM counterpart at NLO versus this ratio at LO. The
NLO ratio deviates substantially from the LO ratio for a large fraction
of the parameter points, showing that the
NLO effects in the two models can be quite different. This was to be
expected as in the C2HDM further Higgs bosons contribute to the loop
corrections and their impact can be quite sizeable due to large Higgs
self-couplings and/or light masses. Inspecting only pure CP-violating
scenarios, we find overall less scenarios compatible with the
constraints and a strong PT and the size of the NLO corrections is
somewhat reduced. 
%\tc{blue}{y-Achse: R2HDM Sample:-600 bis +800, C2HDM Sample : -600 bis +620 , x-Achse: R2HDM Sample: -290 bis + 370 , C2HDM Sample: -370 bis + 400. Also grob
%gleich f\"ur die grauen Punkte. PT Sample: y-Achse : R2HDM: -495 bis
%470, C2HDM : -425 bis 396 , x-Achse: R2HDM: -210 bis 200, CPV : -190
%bis 190. Also auch hier grob gleich, jedoch grenzwertig} 

%%%%%%%%%%%%%%%%%%%%%%%%%%%%%%%%%%%%%%%%%%%%%%%%%%%%%%%%%%%%
\subsubsection{The benchmark scenario BP3HSM}
To quantify the impact of the PT on the Higgs self-coupling and on
Higgs pair production, we exemplary give one benchmark point in the
CP-violating case, BP3HSM, with the input parameters ($M_{H_3}$ is
derived from the input parameters)
\beq
\begin{array}{llll}
\mbox{\underline{BP3HSM:}} & M_{H_1} = 125.09 \mbox{ GeV}, & \qquad M_{H_2}
=356.779 \mbox{ GeV},  & 
  \\
& M_{H_3} = 587.511 \mbox{ GeV}, &  \qquad M_{H^\pm} = 581.460 \mbox{ GeV}, \\
& \alpha_1 = 1.470, & \qquad \alpha_2 = 0.0223, & \qquad 
\alpha_3 = -0.097,\\ 
& \tan\beta = 4.17, & \qquad \mbox{Re} (m_{12}^2) = 29939 \mbox{ GeV}^2 \;. 
\end{array}
\eeq
The strength of the phase transition and the CP-odd admixtures are
\beq
\xi_c = 1.43\; , \quad R_{13}^2 = 4.962 \cdot 10^{-4} \;, \quad
R_{23}^2 = 9.292 \cdot 10^{-3} \;, \quad R_{33}^2 = 0.990 \;.
\eeq
This means that $H_1$ and $H_2$ are mostly CP-even and $H_3$ is mostly
CP-odd. The main branching ratios of the
non-SM-like Higgs bosons are ($H_\downarrow = H_2, H_\uparrow = H_3$)
\beq
\mbox{BR} (H_\uparrow \to Z H_\downarrow ) &=& 0.830 \;, \quad
\mbox{BR} (H_\uparrow \to t\bar{t} ) = 0.102 \;, \quad
\mbox{BR} (H_\uparrow \to Zh) = 0.061 \;, 
\\
\mbox{BR} (H_\downarrow \to hh ) &=& 0.514 \;, \quad 
\mbox{BR} (H_\downarrow \to WW) = 0.309 \;, \quad 
\mbox{BR} (H_\downarrow \to ZZ) = 0.142 \;, \quad \\
\mbox{BR} (H^\pm\to W^\pm H_\downarrow ) &=& 0.835 \;, \quad
\mbox{BR} (H^+\to t\bar{b}) = 0.100 \;, \quad
\mbox{BR} (H^\pm\to W^\pm h ) = 0.065 \;, \quad
\eeq
so that the decays into gauge+Higgs or Higgs pair final states
dominate.
For this scenario the ratios of the SM-like trilinear Higgs
self-coupling to the SM coupling at LO and NLO are
\beq
\mbox{\underline{BP3HSM}}: \quad
\frac{\lambda^{\text{C2HDM,LO}}_{hhh}}{\lambda^{\text{SM,LO}}_{hhh}} =
0.793\quad \mbox{and} \quad
\frac{\lambda^{\text{C2HDM,NLO}}_{hhh}}{\lambda^{\text{SM,NLO}}_{hhh}}
= 2.135 \;, 
\eeq
with the NLO to LO C2HDM coupling ratio being
\beq
\frac{\lambda^{\text{C2HDM,NLO}}_{hhh}}{\lambda^{\text{C2HDM,LO}}_{hhh}} =
2.438 \;,  \label{eq:ewnlohhh}
\eeq
showing the importance of the loop corrections. \s

We now discuss the effect of these corrections on continuum Higgs pair
production. At the LHC the dominant process is given by gluon fusion
\cite{Glover:1987nx,Plehn:1996wb,Dawson:1998py,Baglio:2012np}. The contributing
diagrams are the triangle diagrams with the production of a
Higgs or $Z$ boson that subsequently decays into a Higgs pair, and the box
diagrams \cite{Grober:2017gut}. For the NLO cross
section of gluon fusion into the SM-like Higgs pair $h h$,
computed with a private version of {\tt HPAIR} \cite{hpair} 
adapted for the C2HDM \cite{Grober:2017gut}, we find at a c.m.~energy of 14 TeV
\beq
\sigma^{\text{NLO}} (pp \to h h) = 125.447 \mbox{ fb} \;.
\eeq
The QCD corrections computed in the heavy top mass limit yield a
$K$-factor, {\it i.e.}~a ratio of NLO to LO cross section
(the latter calculated with LO strong coupling constant and parton distribution
functions), of 
\beq
K = \frac{\sigma^{\text{NLO}}}{\sigma^{\text{LO}}} = 1.982 \;,
\eeq
showing the importance of the QCD corrections. 
The NLO cross section for SM Higgs pair production computed with full
top quark mass dependence amounts to
\cite{Borowka:2016ehy,Borowka:2016ypz, Heinrich:2017kxx} 
\beq
\sigma^{\text{NLO}} (H_{\text{SM}} H_{\text{SM}}) = 32.91 \mbox{ fb} \;.
\eeq
The C2HDM cross section is hence by a factor of 3.8 larger than in the
SM. This cross section does not include any EW corrections, and in
particular not the ones given in Eq.~(\ref{eq:ewnlohhh}). The
enhancement of 3.8 is mostly due to the resonant production of $H_\downarrow$
that subsequently decays into an $h$ pair. Without this resonant  
enhancement the cross section amounts to
\beq
\sigma^{\text{NLO}}_{\text{w/o } H_\downarrow} (pp \to h
h) = 49.996 \mbox{ fb} \;. 
\eeq
%Taking off also the resonant $H_\uparrow$ production with subsequent
%decay into $hh$, the NLO cross section is
%\beq
%\sigma^{\text{NLO}}_{\text{w/o } H_\downarrow \& H_\uparrow} (pp \to h
%h) = 49.996 \mbox{ fb} \;. 
%\eeq
The quantification of the effect of the EW corrections requires the
complete calculation of the Higgs pair production process at NLO EW,
which is clearly beyond the scope of this paper. The computation of
the loop-corrected effective trilinear Higgs self-couplings gives a
flavour of the importance of the EW corrections and the impact of the
EWPT on this value 
%indicates that these corrections may be non-negligible. 
In particular, we note
that the increase of the trilinear Higgs self-coupling may also
decrease the total size of the cross-section due to the destructive
interference between triangle and box diagrams. Electroweak
baryogenesis which requires a certain size of the trilinear Higgs self-coupling
between the SM-like Higgs bosons in order to be successful hence has a
direct influence on the size of resonant and continuum Higgs pair
production that is significant enough to be tested at the LHC (and
future colliders). \s

We finish this section by commenting on the size of the Higgs pair
production cross sections of the benchmark points given in
Tables~\ref{tab:cpvbenchmarksi}, \ref{tab:cpvbenchmarksii} and
\ref{tab:cpvbenchmarksiii}. As can be inferred from the values given
in the tables, Higgs pair production in the C2HDM is significantly 
enhanced compared to the SM value for scenarios where resonant heavy
Higgs production with subsequent decay into $hh$ is kinematically
possible. In the scenarios we looked at, the $H_\uparrow hh$ couplings
is usually very small, so that the main resonant contribution comes
from $H_\downarrow$ production. In case this is kinematically not
allowed or the $H_\downarrow hh$ coupling is also small, the cross
section value compares to the one of the SM. 

%%%%%%%%%%%%%%%%%%%%%%%%%%%%%%%%%%%%%%%%%%%%%%%%%%%%%%%%%%%%
\subsection{Further Higgs self-couplings}
The inspection of the other trilinear Higgs couplings not
involving only the SM-like Higgs boson shows the following:
The trilinear C2HDM Higgs self-couplings can be suppressed but
also be substantially enhanced compared to the SM trilinear
coupling and still be compatible with all constraints. The enhancement
factor is less important for scenarios that additionally feature a
strong PT. However, it can still be considerable, depending on the
self-coupling and the scenario. Also the NLO corrections can be
important.  
%
%The ratio of the NLO coupling to the LO
%coupling can reach several tens of percent and for almost all
%couplings is smaller than the maximum ratio for the coupling of the three
%SM-like Higgs bosons. 
Barring the case where the LO coupling is close to zero and hence the
coupling becomes effectively loop-induced, the maximum enhancement
factor for CP-violating scenarios with $\xi_c >1$ is 5.7.
%When additionally a strong PT is required the
%ratio does not exceed \tc{red}{1.6 (true?)}  and for the majority of the couplings is
%close to 1, in constrast to the case of three SM-like Higgs bosons
%coupling to each other. 
Due to the large amount of possible trilinear
couplings, we exemplary show in Fig.~\ref{fig:hsmhlhhnlo} 
the coupling between the SM-like Higgs boson and $H_\downarrow$ and $H_\uparrow$,
$\lambda_{h H_\downarrow  H_\uparrow}$. For this coupling,
the enhancement in the C2HDM compared to the SM coupling can be up to
a factor $1.5$ at NLO in accordance with all constraints, {\it
  cf.}~Fig.~\ref{fig:hsmhlhhnlo} (left). 
%Die genauen Werte sind -1.506 bis +1.523
When additionally a strong PT is
demanded, the ratio drops to values between -0.34 and 0.47 the SM
coupling. The largest $\xi_c$ value is obtained close to the alignment limit
where $\langle h \rangle \approx v$. The right plot shows
the impact of the NLO correction. For the sample compatible with all
constraints the ratio of the NLO to the LO coupling can be quite
large. The demand of a strong PT has a
considerable impact, as in this case the ratio becomes much smaller, as
can be inferred from the plot.

%%%%%%%%%%%%%%%%%%%%%%%%%%%%%%%%%%%%%%%%%%%%%%%%%%%%%%%%%%%
\section{Type I: Parameter sets with $H_2  \equiv
  h$ \label{sec:Ihheqh125}}
\begin{figure}[t!]
\begin{center}
\includegraphics[width=0.49\textwidth]{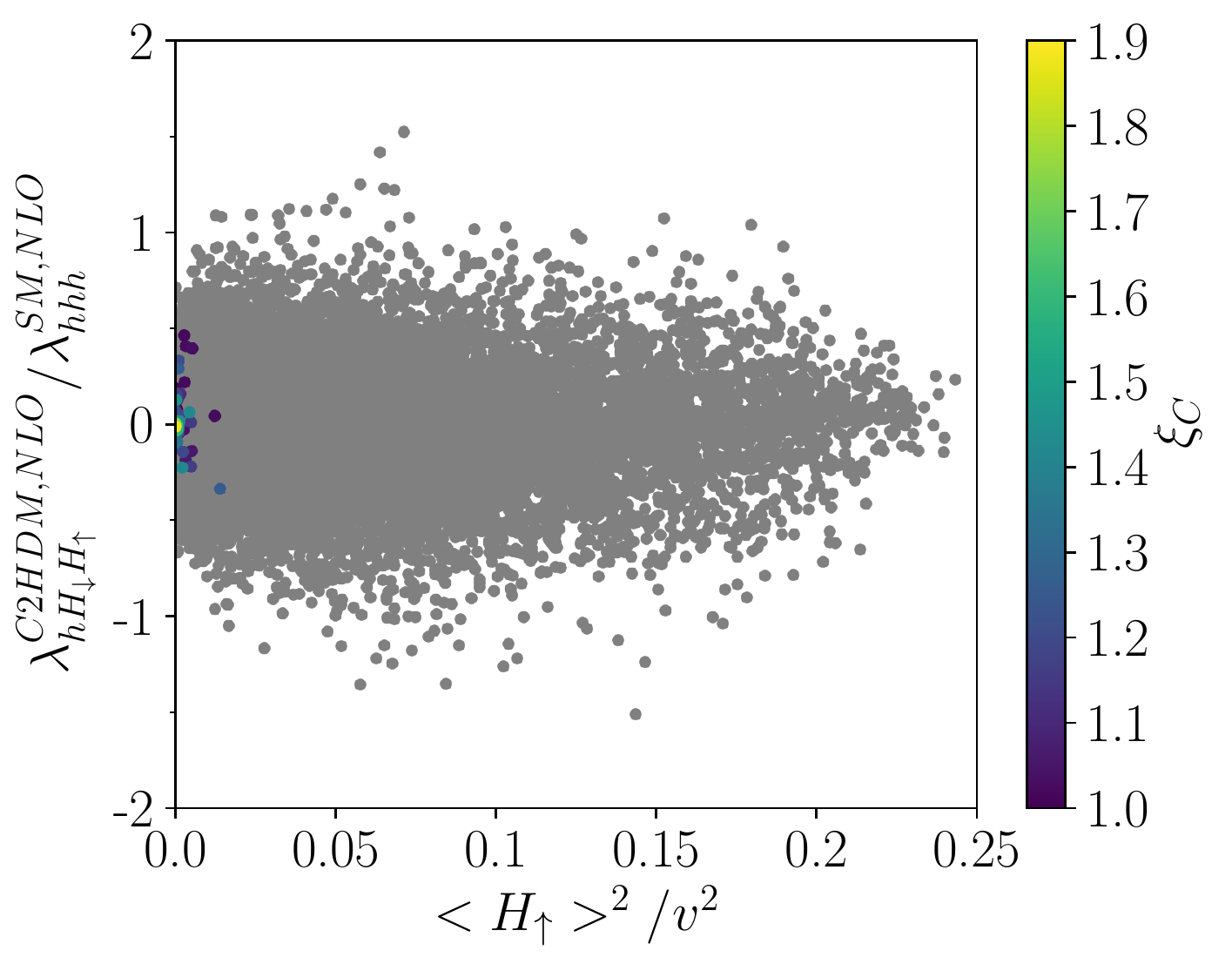}
\includegraphics[width=0.49\textwidth]{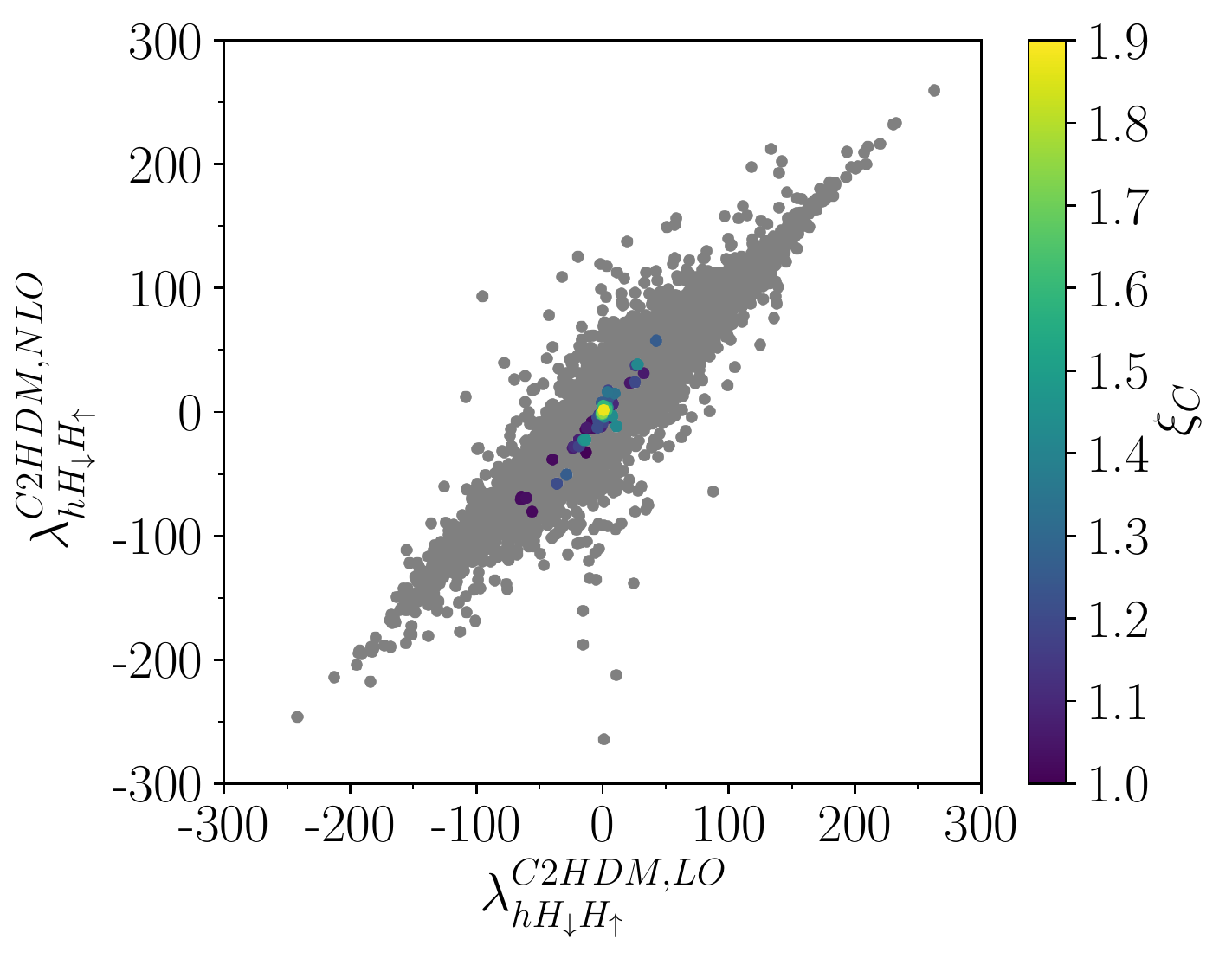}
\vspace*{-0.3cm}
\caption{Type I, $H_1 = h$: Left: 
The C2HDM trilinear Higgs self-coupling $\lambda_{h
  H_\downarrow  H_\uparrow}$ normalized to the SM value, at NLO, as a
function of the fraction of the VEV squared carried by $H_\uparrow$. Right: The NLO
C2HDM trilinear coupling $\lambda_{h  H_\downarrow
  H_\uparrow}$ versus the LO value. Grey: all points passing the
constraints; colour: points with additionally $\xi_c \ge
1$. Both plots: all 2HDM points.
\label{fig:hsmhlhhnlo}}
\end{center}
\vspace*{-0.7cm}
\end{figure} 
In this mass configuration our scan resulted in only three 
scenarios compatible with all constraints that both allow for a strong
PT and include CP violation. The results are therefore those of a real
2HDM with the heavier of the two CP-even Higgs bosons being SM-like
with a mass of 125~GeV. We reproduced the results of our previous
publication on the PT in the CP-conserving 2HDM given in
\cite{Basler:2016obg}, which we briefly summarise here: The scenarios
compatible with $\xi_c > 1$ require (neglecting a few outliers) a mass hierarchy where
$H_\uparrow$, {\it i.e.}~the pseudoscalar $A$ in the R2HDM, is mass
degenerate with $H^\pm$ and lies in the mass range $\sim$130 to $\sim$490 GeV, so
that there is a mass gap between $m_{\uparrow}$ and
$m_{\downarrow} < 125$~GeV. The reason is that due to the required
small $m_\downarrow$, coinciding with the mass of the lighter of the two
CP-even R2HDM Higgs bosons $h$, the quartic coupling $\lambda_2$ and
$m_{12}^2$ have to be small. Hence we are left with $\lambda_4$ and
$\lambda_5$ that drive the phase transition, implying large
$H_\uparrow$ and $H^\pm$ masses and the mass gap to
$H_\downarrow$. Keeping in mind that $H_2$, {\it i.e.}~the heavier of
the 2 CP-even Higgs bosons,  is SM-like induces
$\sin (\beta-\alpha) = 0$ in the limit of the R2HDM, so that this mass
hierarchy allows for $A \to hZ$ decays, involving the coupling $g_{AhZ} \sim
\cos (\beta - \alpha)$, and can be probed at the
LHC. The upper bound on the masses of the heavy Higgs bosons is given
by the fact that the Higgs bosons participating in the PT must not be
too heavy. 
\s

In the CP-violating 2HDM we barely find any points compatible with
$\xi_c >1$. We have seen that explicit CP violation comes
along with spontaneous CP violation at the PT. The thus generated
CP-violating VEV at the EWPT feeds into all Higgs bosons as they are
all mixing in the CP-violating 2HDM. As the SM-like Higgs
boson must not receive a large CP admixture it is either
$H_\downarrow$ or $H_\uparrow$ that develop a non-negligible
CP-violating VEV. Due to the above described mass hierarchy with a
heavy $H_\uparrow$ its VEV should not become too large, however, in
order not to weaken the PT. This favours the lighter $H_\downarrow$ to
receive a more important fraction of the VEV or else a hierarchy where all neutral
Higgs bosons are rather light and hence close in mass. Already in the R2HDM we see
that such hierarchies together with a strong PT are very rare, so that
the scenarios that can be found are very sparse. \s

The phenomenological implications of the R2HDM are the same as found in
\cite{Basler:2016obg} with the main feature that there are only very
few scenarios with a strong PT that yield photonic rates $\mu_{\gamma\gamma}$ beyond
0.9, although values of up to 1.45 would still be compatible with the
applied constraints. In contrast, however, to \cite{Basler:2016obg}
the rate into $\tau\tau$ can go up to the maximum allowed experimental
value of 1.4 also for $\xi_c > 1$, which is due to different, {\it i.e.}~newer, limits on
$\tan\beta$ applied in this work. The three explicitly CP-violating
scenarios lie in the same boundaries for the $\mu$-rates as the ones
of the R2HDM. As for the trilinear Higgs self-couplings, the overall picture is the
same as in the $H_1 \equiv h$ case and we content ourselves to
summarise the main features in the conclusions.

%%%%%%%%%%%%%%%%%%%%%%%%%%%%%%%%%%%%%%%%%%%%%%%%%%%%%%%%%%%
\subsection{Features of the CP-violating scenarios with strong PT}
The closer inspection of the 3 CP-violating scenarios reveals that they
all feature Higgs spectra with rather close mass values. The largest
difference between the heaviest and the lightest Higgs boson mass is
256~GeV. In Table~\ref{tab:cpvbenchmarks} we list the input parameters
of the 3 benchmark point scenarios, denoted by BPCPV1-3, featuring a strong PT in the
CP-violating case. We additionally give the derived third neutral
Higgs boson mass, $\xi_c$, the CP admixtures of the Higgs
bosons and the SM-like Higgs pair production cross section through
gluon fusion. 
%We note the rather small value of $\mbox{Re} (m_{12}^2)$
%due to the requirement of large quartic couplings and small Higgs boson
%masses for a strong PT.  
All three scenarios have $\xi_c$
values rather close to 1 underlining the difficulty in finding parameter sets inducing a
strong PT in case $H_2$ is SM-like. In BPCPV1, $H_3$ has the largest
mass of all three benchmark points with 376~GeV. A strong PT is
possible as $H_3$ receives a smaller fraction of the VEV than
$H_1$ and $H_2$. In BPCPV2 and 3, all masses are rather close with the
mass of $H_3$ being below 160~GeV, which now also carries a larger
fraction of the VEV. \s 
\begin{table}[t!]
 \begin{center}
 \begin{tabular}{lccc}
     \toprule
    & BPCPV1 & BPCPV2 & BPCPV3  \\ 
   \midrule
$m_{H_1}$ [GeV] &  119.62 &  91.31 & 118.16 \\
$m_{H_2}$ [GeV] &  125.09 & 125.09 & 125.09 \\
$m_{H^\pm}$ [GeV] & 374.95 & 191.70 & 166.56   \\
$\mbox{Re}(m_{12}^2)$ [GeV$^2$] & 1945.7 & 1124.44 & 1160.95  \\
$\alpha_1$ & -0.453769 & -0.00591145 & -0.223061 \\
$\alpha_2$ & 0.0966953 & -0.914279 & 1.41808 \\
$\alpha_3$ & 0.0658765 & -0.185891 & 0.331065 \\
$\tan\beta$ & 7.05923 & 12.3765 & 19.5889 \\ \hline 
$m_{H_3}$ [GeV] & 375.50 & 141.44 & 153.63 \\
$\xi_c$ & 1.48 & 1.02 & 1.46 \\
$R_{13}^2$ & $9.2\cdot 10^{-3}$ & 0.63 & 0.97 \\
$R_{23}^2$ & $4.2 \cdot 10^{-3}$ & 0.013 & $2.44 \cdot 10^{-3}$ \\
$R_{33}^2$ & 0.99 & 0.36 & $2.07 \cdot 10^{-2}$ \\ \hline
$\sigma_{hh}^{\text{NLO}}$ [fb] & 24.58 & 35.89 & 37.22 \\
   \bottomrule
  \end{tabular}
\caption{Line 1-8: The input parameters of the type I CP-violating
  benchmarks with $H_2 
\equiv h$ and $\xi_c > 1$, compatible with all constraints. Line
9 to 13: The derived 3rd neutral Higgs boson mass, the
$\xi_c$ value and the CP-odd admixtures $R_{i3}^2$. Line 14: The NLO QCD
gluon fusion $hh$ production cross section at
$\sqrt{s}=14$~TeV.} \label{tab:cpvbenchmarks}
   \end{center}
\vspace*{-0.4cm}
 \end{table}

The phenomenological features of the three benchmarks are summarised
in Table~\ref{tab:cpvbppheno} where we depict the dominant branching
ratios of the various Higgs bosons. For $H_2$ we have SM-like
branching ratios and do not give them separately here. 
\begin{table}[b!]
 \begin{center}
 \begin{tabular}{llll}
     \toprule
    & BPCPV1 & BPCPV2 & BPCPV3  \\ 
   \midrule
BR($H_1$) &  BR($H_1 \to b\bar{b}$) = 0.72 &  
BR($H_1 \to b\bar{b}$) = 0.72 & BR($H_1 \to b\bar{b}$) = 0.68 \\ \hline
BR($H_3$) &  BR($H_3 \to ZH_1$) = 0.88 & BR($H_3 \to WW$)=0.42 
& BR($H_3\to WW$) = 0.77 \\
& BR($H_3 \to ZH_2$) = 0.088 & BR($H_3 \to ZH_1$)=0.33 
& BR($H_3\to ZZ$) = 0.080 \\
& BR($H_3 \to t\bar{t}$) = 0.024 & BR($H_3 \to b\bar{b}$)=0.14
& BR($H_3\to ZH_1$) = 0.072 \\ \hline
BR($H^\pm$) & BR($H^\pm \to WH_1$) = 0.88 & BR($H^\pm \to WH_1$)=0.98 
& BR($H^\pm \to WH_1$) = 0.94 \\
& BR($H^\pm \to WH_2$) = 0.089 & BR($H^+ \to t\bar{b}$)=0.018 
& BR($H^+ \to t\bar{b}$) = 0.040 \\
& BR($H^+ \to t\bar{b}$) = 0.027 & BR($H^\pm \to WH_3$)=0.002 
& BR($H^\pm \to WH_2$) = 0.012 \\
   \bottomrule
  \end{tabular}
\caption{The dominant branching ratios of the BPCPV1-3 Higgs bosons.} \label{tab:cpvbppheno}
   \end{center}
\vspace*{-0.4cm}
 \end{table}
The branching ratios are in accordance with the nature of the Higgs
bosons. As can be inferred from the $R_{i3}^2$ given in
Table~\ref{tab:cpvbenchmarks} the lightest Higgs boson of BPCPV1 is mostly
CP-even whereas the heaviest one is mostly CP-odd, so that its
branching ratios into massive gauge bosons are suppressed. Instead,
the large mass gap of more than 250 GeV between $H_1$ and $H_3$
together with a maximum 
$ZH_1 H_3$ coupling, since $H_2 \equiv
h$, allows for a large branching ratio into the $ZH_1$ final
state with 88\%. Also the decay into $ZH_2$ yields close to 
10\%. With large branching ratios of $H_1$ and $H_2$ into $b\bar{b}$
and with the $Z$ boson in the final state this signature can easily be
searched for at the LHC. Instead the top pair final state plays a less
important role. The charged Higgs boson, which is almost degenerate with $H_3$,
decays mainly into $W H_1$ followed by $W H_2$. In the
scenario BPCPV2 the masses 
are overall much closer and 
both $H_1$ and $H_3$ have a considerable CP-even admixture. Consequently, $H_3$ 
dominantly decays off-shell into $WW$, with 42\%, followed by
the off-shell decay into $ZH_1$ with 33\%. The next important decay is
into $b\bar{b}$ as due to the much smaller $H_3$ mass the decay into
top quarks is kinematically closed. The charged Higgs with a larger
mass decays on-shell mostly into $WH_1$ with 98\%. The decay into $WH_2$ is
kinematically suppressed, so that the next-important decay is into
$t\bar{b}$. Due to a larger coupling to $WH_3$ than to $WH_2$ in this scenario, the
third important decay is then the one into the $WH_3$ final state. 
In BPCPV3, $H_1$ is now almost purely CP-odd, whereas
$H_3$ is mostly CP-even, so that its most important decays are now the
off-shell decays into $WW$ and $ZZ$ with 77 and 8\%, respectively. The
charged Higgs boson again mostly decays into the $WH_1$ final state
with 94\%, followed by the decay into $t\bar{b}$ with 4\%. In all
three scenarios the dominant decay of the light Higgs boson $H_1$ below the
massive gauge boson threshold is into $b\bar{b}$ with about 70\%. In
summary, all three scenarios are characterized by decays of $H_3$
and/or $H^\pm$ into gauge+Higgs boson final states, which are clearly
BSM Higgs signatures that can be searched for at the
LHC. Higgs-to-Higgs decays on the other hand play no role as the
rather small mass gaps between the Higgs bosons, induced by the
requirement of a first order phase transition, are too small. This is
also why the SM-like Higgs pair production cross sections are smaller
than or close to the SM value, {\it cf.}~the last line of
Table~\ref{tab:cpvbenchmarks}. Only for BPCPV1 the mass spectra are
such that resonant heavy Higgs production (here $H_3\equiv
H_\uparrow$) could enhance the cross section. This is not the case
here, however, as the $H_\uparrow hh$ coupling turns out to be small. 

\section{Type II: Parameter sets with $H_1  \equiv
  h$ \label{sec:IIhleq125}}
In type II, all scenarios compatible with a strong PT, where the
SM-like Higgs boson is given by the heavier neutral Higgs states are
excluded due to the lower bound on the charged Higgs mass, so that
necessarily $H_1 \equiv h$. We start our discussion of the EWPT in the
C2HDM with the investigation of the CP-violating phase. 

%%%%%%%%%%%%%%%%%%%%%%%%%%%%%%%%%%%%%%%%%%%%%%%%%%%%%%%%%%%
\subsection{The CP-violating phase}
\begin{figure}[t!]
\begin{center}
\includegraphics[width=0.55\textwidth]{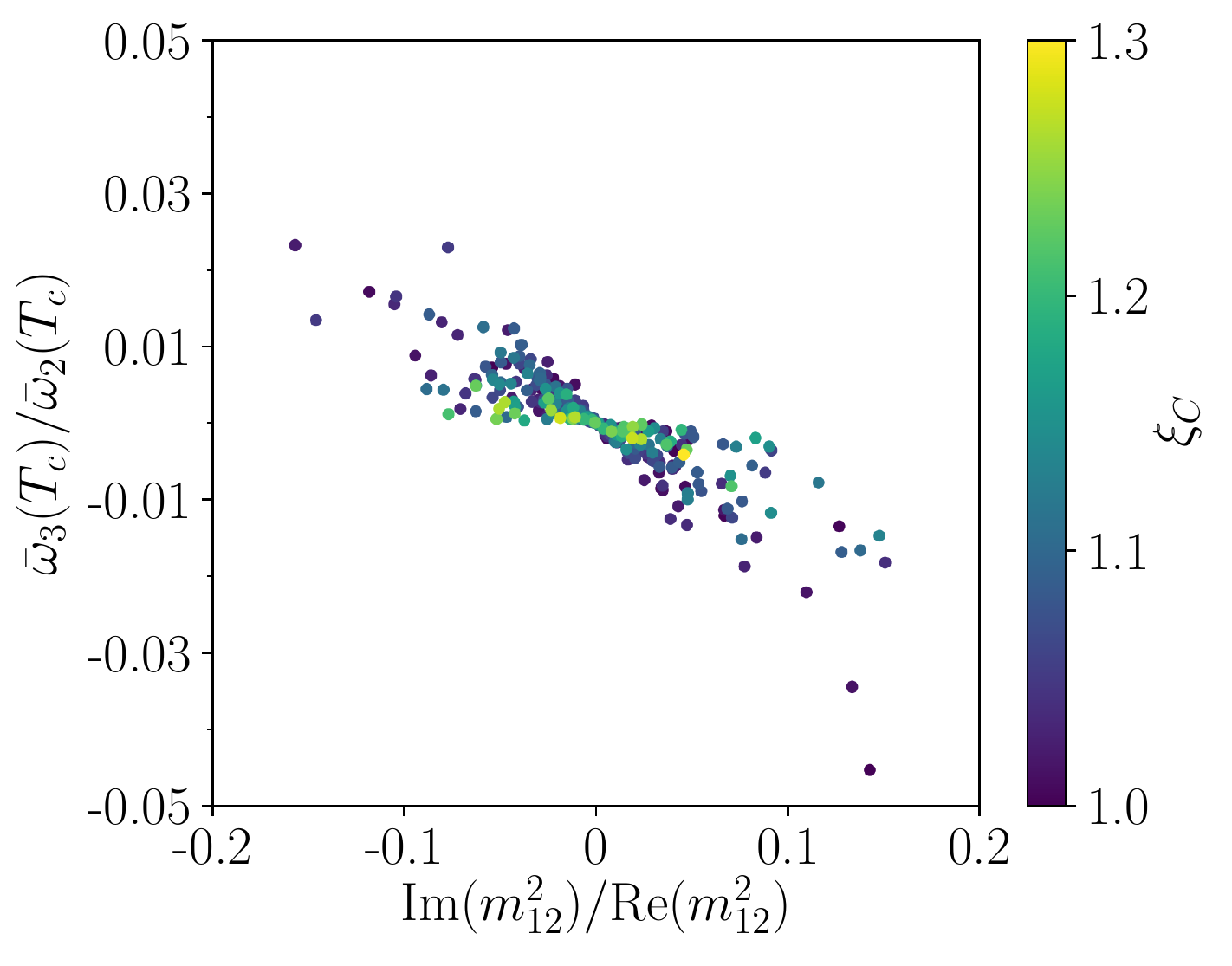}
\vspace*{-0.2cm}
\caption{Type II, $H_1 = h$: The value 
  $\bar{\omega}_3(T_c)/\bar{\omega}_2 (T_c)$ at the critical temperature $T_c$
  versus $\mbox{Im} (m_{12}^2)/ \mbox{Re} (m_{12}^2) \ne 0$ at
  $T=0$ for points with a strong PT. The colour code indicates the
  size of $\xi_c$. 
 \label{fig:t2cpphase}}
\end{center}
\vspace*{-0.6cm}
\end{figure}
\begin{figure}[t!]
\begin{center}
\includegraphics[width=0.5\textwidth]{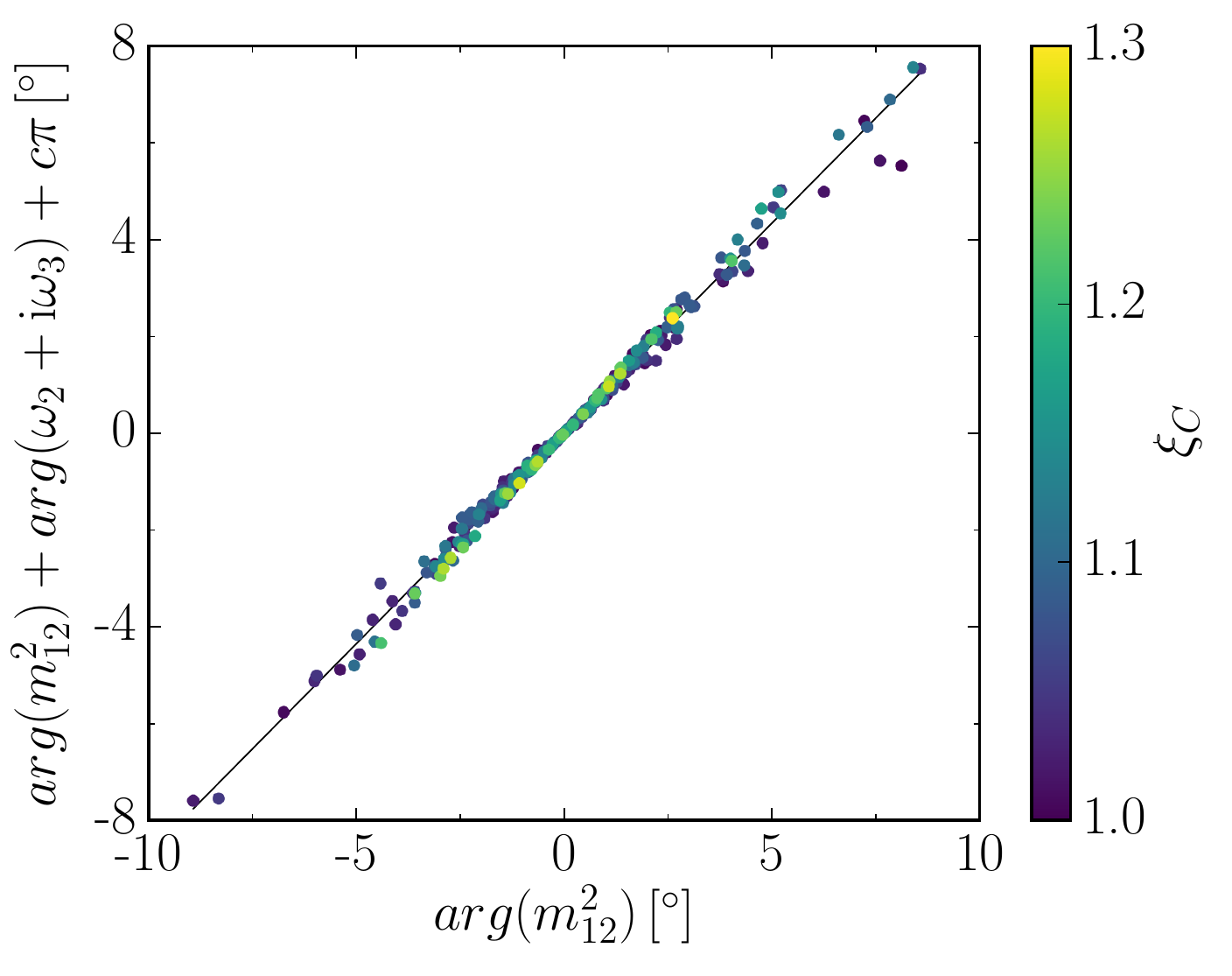}
%\vspace*{-0.2cm}
\caption{Type II, $H_1 = h$: The total CP-violating angle
  at the critical temperature $T_c$
  versus the CP-violating angle at
  $T=0$. By adding $c\pi$, $c \in \{ -1,0,1 \}$, the angle is ensured to lie in the range
  $-90^\circ$ to $90^\circ$.  
  The colour code indicates the size of $\xi_c$. 
 \label{fig:t2totalphase}}
\end{center}
\vspace*{-0.6cm}
\end{figure}
Figure~\ref{fig:t2cpphase} shows the relation between $\tan
\varphi^{\text{spont}} = \bar{\omega}_3(T_c)/\bar{\omega}_2 (T_c)$ at
the critical temperature $T_c$ 
and $\tan \varphi^{\text{explicit}} = \mbox{Im} (m_{12}^2)/ \mbox{Re} (m_{12}^2)$ at 
vanishing temperature. All points comply with our constraints and
feature a strong PT. Furthermore, only points with explicit CP violation are
included, {\it i.e.}~$\mbox{Im} (m_{12}^2)/ \mbox{Re} (m_{12}^2) \ne 0$ although
we allow it to be very small. The plot shows the correlation
between the two types of CP-violating phases, with the absolute value
of the spontaneously generated phase decreasing with increasing absolute
value of the explicitly CP-violating phase. Like in the type I, we observe that
the spontaneous generation of a CP-violating phase at the critical
temperature only appears for scenarios with explicit CP violation at
zero temperature. The color bar visualizes the size
of $\xi_c$ at $T_c$. The maximum value attained in case of CP
violation is somewhat smaller than in the type I case, with
$\xi_c^{\text{max}} = 1.3$. 
The smaller size of $\xi_c$ is to be attributed to the overall heavier
mass spectrum in the type II model due to the lower bound on the
charged Higgs mass. In the CP-violating case all
Higgs bosons mix and hence the heavier Higgs bosons receive
contributions from all VEVs, increasing their participation in the PT.
\s
%The non-SM-like Higgs bosons that participate in
%the phase transition are too heavy to trigger sizeable effects on the
%size of the strong PT. 

Figure~\ref{fig:t2totalphase} displays the total phase at $T_c$ as a
function of the CP-violating angle at $T=0$. We see that the total
phase varies between -7.6$^\circ$ and +7.6$^\circ$ and is hence much smaller
than in type I. This is closely related to the fact that at zero
temperature the applied constraints, mainly the EDM constraint,
restrict a possible CP-violating phase at $T=0$ more strongly than in type I
\cite{Muhlleitner:2017dkd}, namely to values between about -9$^\circ$ and
8.6$^\circ$. Due to the correlation between $\varphi^{\text{explicit}}$
and $\varphi^{\text{spont}}$ the total phase at $T_c$ is then smaller, so
that in the type II C2HDM smaller CP-violating effects are to be expected from this
mechanism.

%%%%%%%%%%%%%%%%%%%%%%%%%%%%%%%%%%%%%%%%%%%%%%%%%%%%%%%%%%%
\subsection{Implications for LHC phenomenology \label{sec:t2hl}}
\begin{figure}[t!]
\begin{center}
\includegraphics[width=0.49\textwidth]{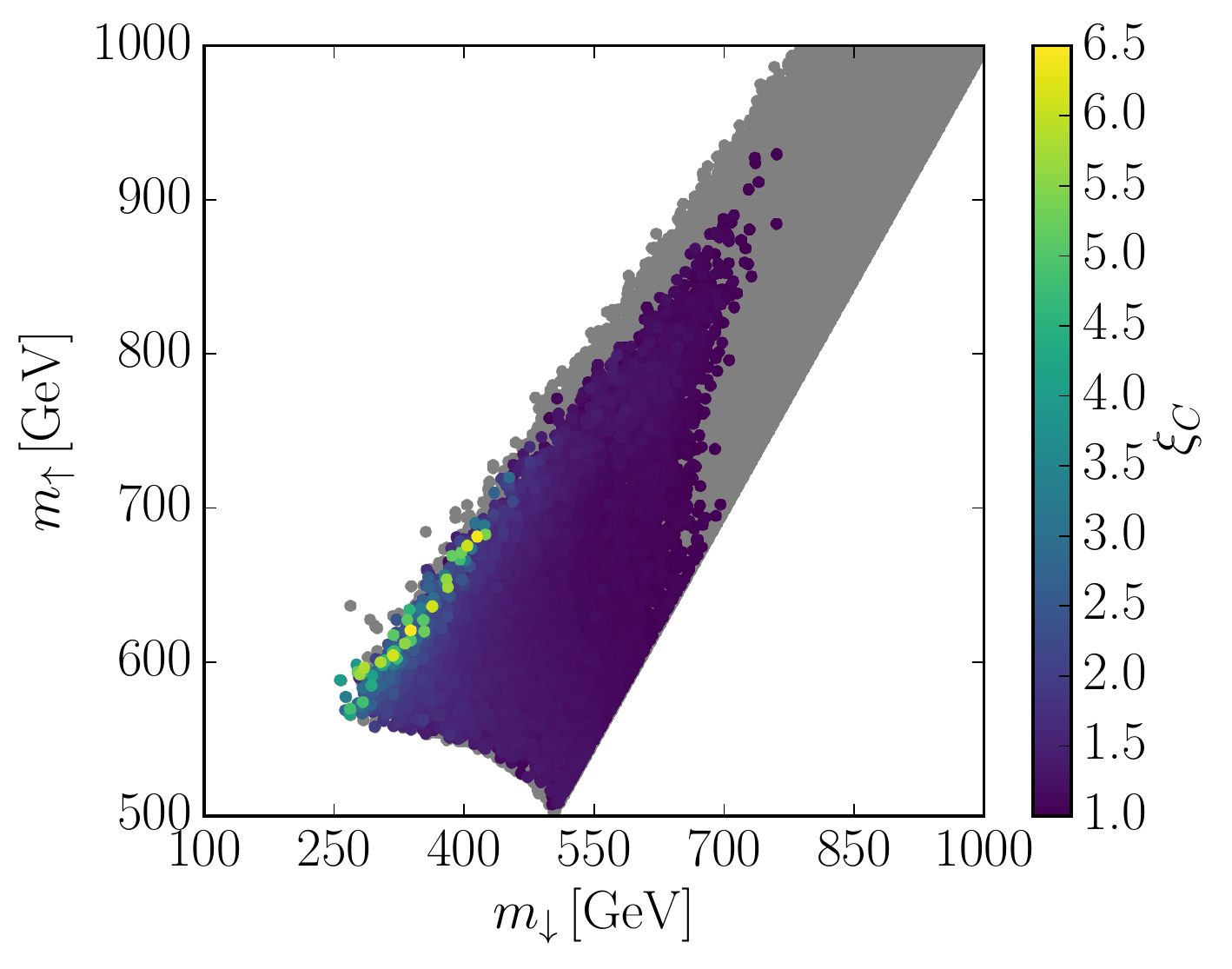}
\includegraphics[width=0.49\textwidth]{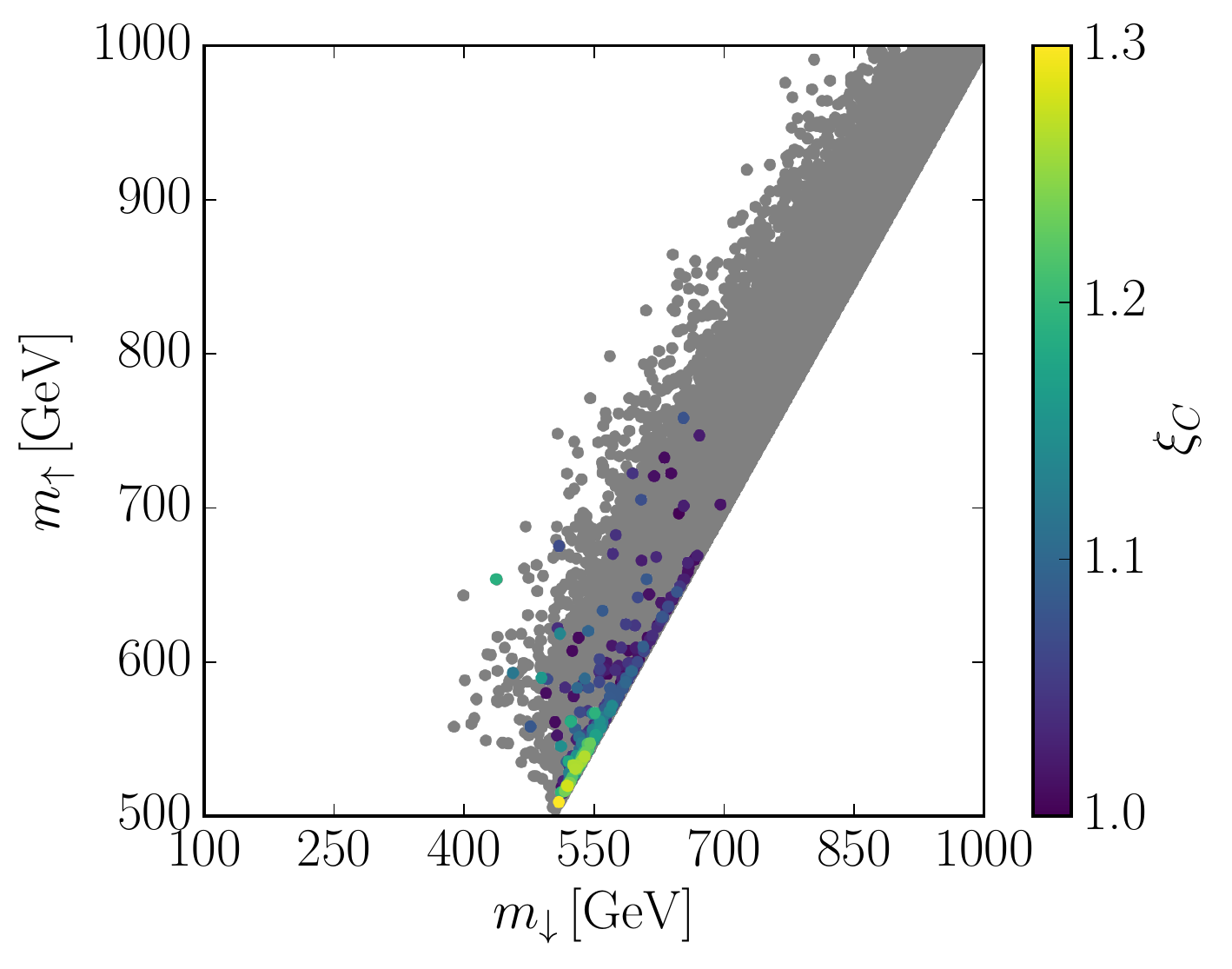}
%\vspace*{-0.2cm}
\caption{Type II, $H_1 = h$: The mass of the heavier versus the
  lighter non-SM-like Higgs boson. Left: CP-conserving and
  CP-violating points, right: only CP-violating points. Grey: points
  passing all the
  constraints; color: points with additionally $\xi_c \ge 1$. The
  color code indicates the value of $\xi_c$. \label{fig:t2masses}}
\end{center}
\vspace*{-0.6cm}
\end{figure}
Figure \ref{fig:t2masses} shows the mass values of the neutral
non-SM-like Higgs bosons compatible with all constraints (grey) that
are additionally compatible with a strong PT (color), taking all points of
the CP-conserving and CP-violating scenarios (left) and restricting to purely
CP-violating scenarios (right). The results of Fig.~\ref{fig:t2masses}
(left) basically agree with the results found in
\cite{Basler:2016obg}, taking into account 
the fact that the lower bound on $m_{H^\pm}$ has moved up to
580~GeV. Furthermore, we do not find valid points with $m_\downarrow <
250$~GeV, which would come along with large mass gap of the heavier
Higgs boson masses to $m_\downarrow$. Overall, grey points with mass
gaps between $H_\downarrow$ 
and $H_\uparrow$ above 332~GeV are not allowed any
more. This exclusion results from the unitarity check with the NLO Higgs
self-couplings. The plot confirms that for the CP-conserving parameter points
with $\xi_c^{\text{max}}= 6.5$ larger $\xi_c$ values can be obtained
than in the CP-violating case where $\xi_c^{\text{max}}= 1.3$.  
%The combination of the experimental exclusion limits on the
%Higgs masses combined with the required $\xi_c \ge 1$ ({\it cf.}~also
%the discussion in \cite{Basler:2016obg}). 
The right plot shows that
the inclusion of CP violation implies mass spectra where overall the
non-SM-like Higgs masses move closer, {\it cf.}~also
Fig.~\ref{fig:t2massdiffs}. In particular most of the points with a
strong PT feature $H_\downarrow$ and $H_\uparrow$ which are close in
mass, and the largest $\xi_c$ values are found for the  
lightest possible values that they can have within the given constraints. Again this can 
easily be understood by reminding that due to CP violation all Higgs bosons
mix and have a non-negligible VEV and that additionally in type II with $H_1 \equiv
h$ the non-SM-like neutral Higgs bosons are rather heavy. In order not
to weaken the PT too much, in case of $H_\uparrow$ having a large
portion of the VEV, it should be as light as possible and hence be
mass degenerate with $H_\downarrow$, or in case it is not mass
degenerate, the lighter of the two should acquire most of the VEV.\s

\begin{figure}[ht!]
\begin{center}
\includegraphics[width=7.5cm]{HeatColorBar.pdf}\\
\includegraphics[width=0.32\textwidth]{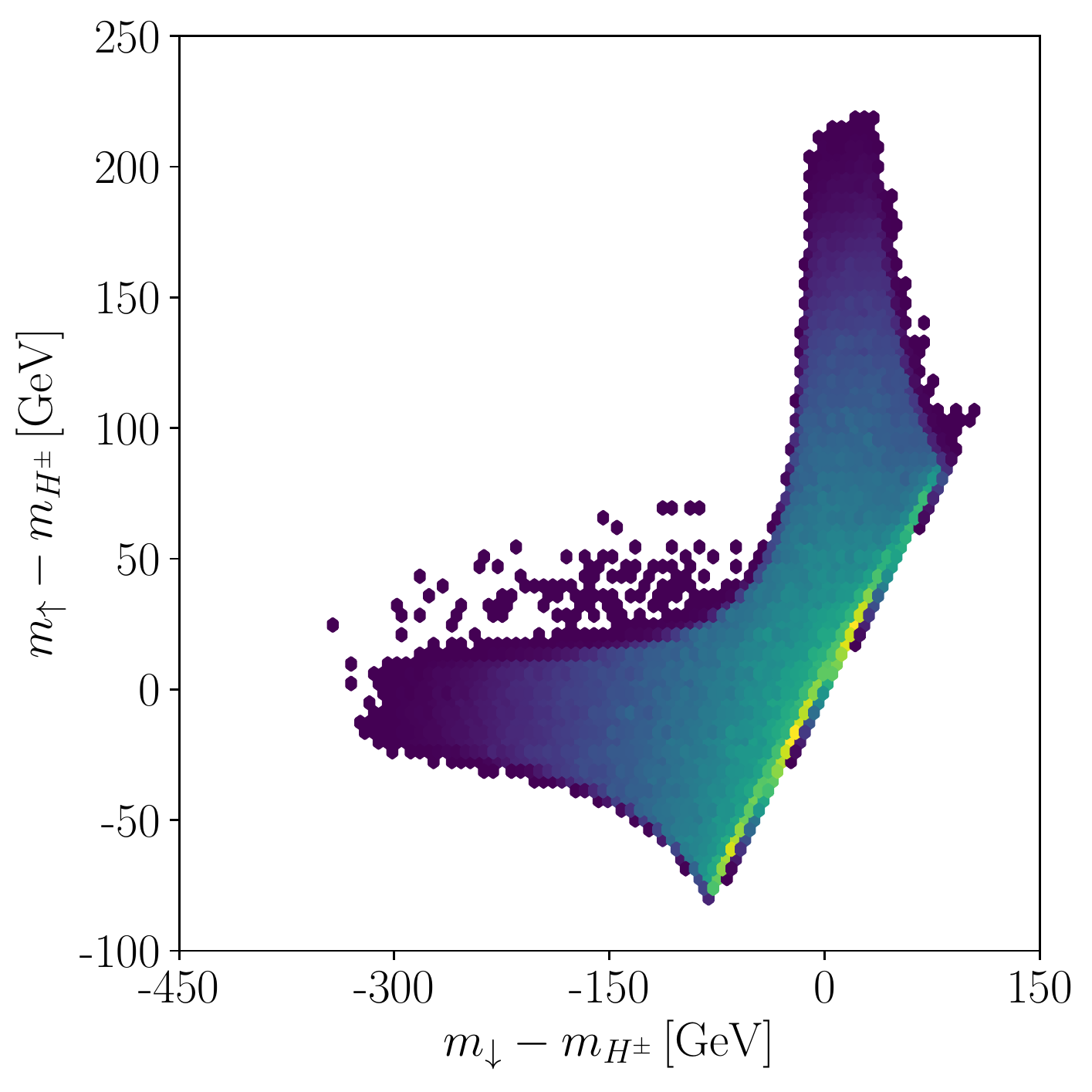}
\includegraphics[width=0.32\textwidth]{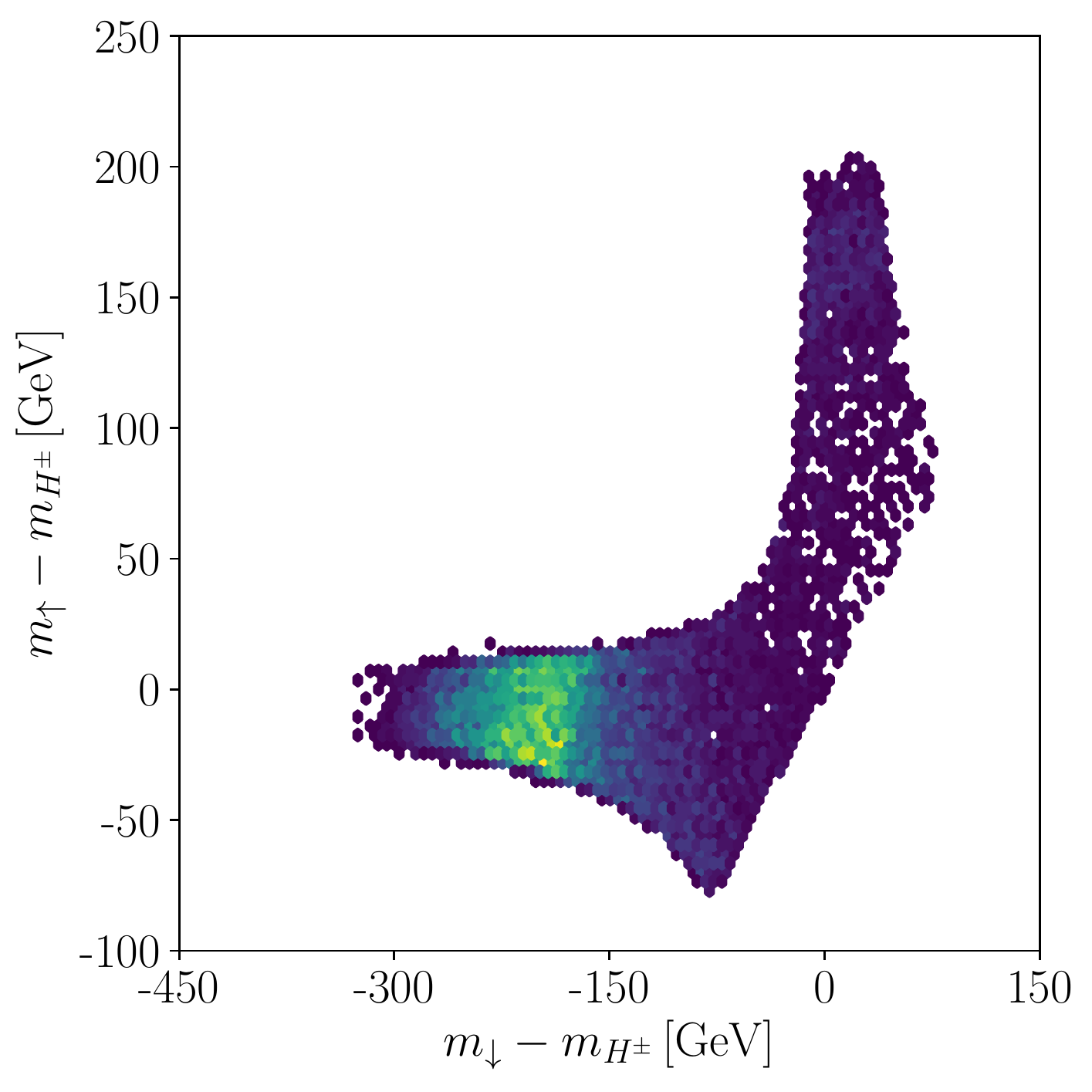}
\includegraphics[width=0.32\textwidth]{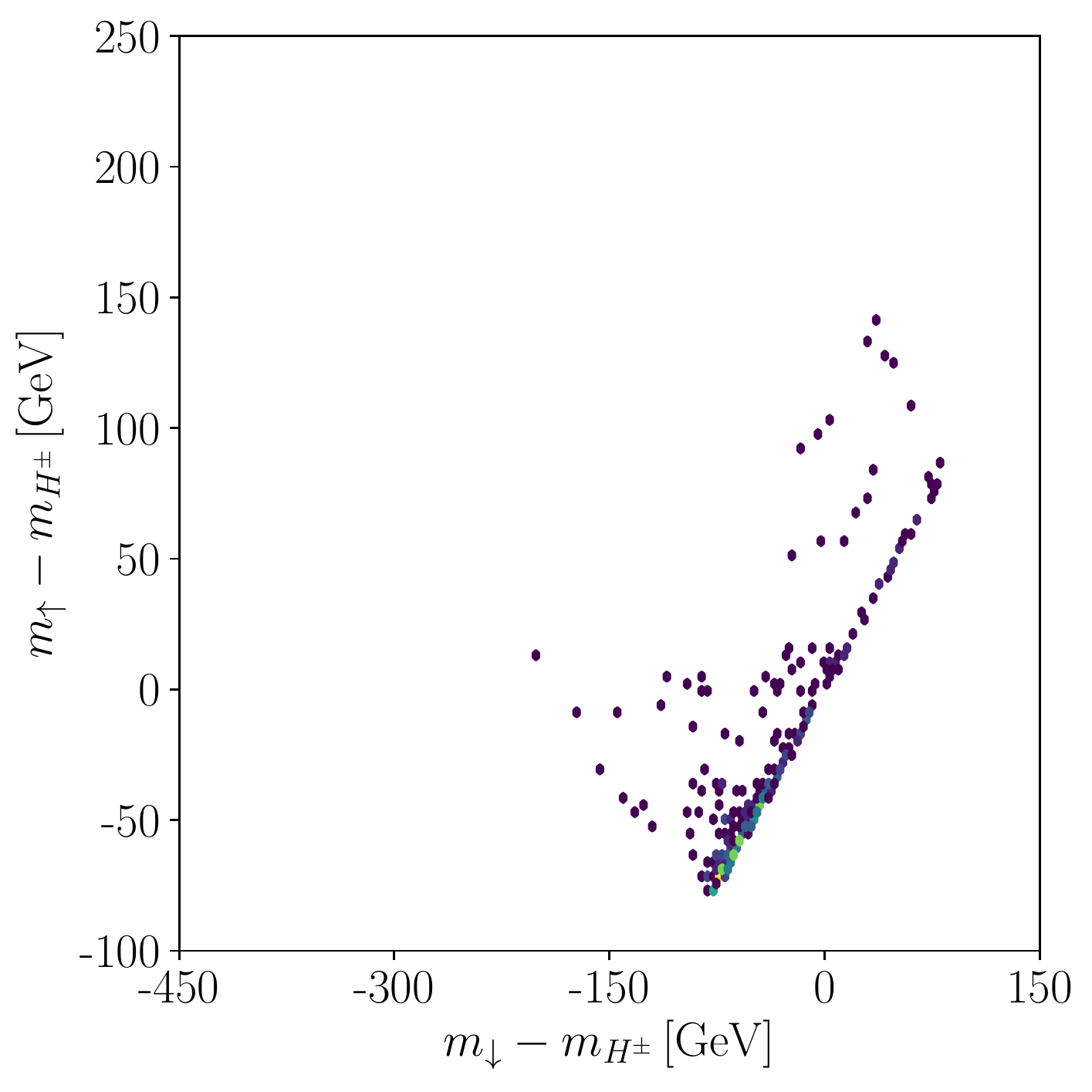}
%\vspace*{-0.2cm}
\caption{Type II, $H_1 = h$: The mass difference $m_\uparrow -
  m_{H^\pm}$ versus $m_\downarrow - m_{H^\pm}$. The colour code shows
  the relative frequency of left: all points passing the constraints;
  middle: all points with a strong PT and CP conservation;
  right: all points with $\xi_c \ge 1$ and explicit CP violation. 
\label{fig:t2massdiffs}}
\end{center}
\vspace*{-0.6cm}
\end{figure}
The implications for LHC phenomenology can be read off 
Fig.~\ref{fig:t2massdiffs}, showing the mass differences of
$H_\uparrow$ and $H^\pm$ versus the ones of $H_\downarrow$ and
$H^\pm$. The colour code shows the relative frequency for all points
passing the constraints (left), with additionally a strong PT and only
CP-conserving points (middle) and for only
CP-violating points with $\xi_c \ge 1$ (right). While the application of the
constraints favours scenarios with degenerate neutral non-SM-like
Higgs masses, the requirement of a strong PT favours a mass hierarchy
between $m_{\uparrow} \approx m_{H^\pm}$ and $m_{\downarrow}$ with a
mass gap of about 200~GeV. The comparison with the right plot shows
that these are mostly CP-conserving scenarios. This allows for decays
of $H_\uparrow \to Z H_\downarrow$ that can be searched for at the
LHC. In contrast the right plot shows that the CP-violating case
favors more degenerate masses of the non-SM-like Higgs bosons. This
has consequences for LHC phenomenology. In order to quantify this we have chosen
four exemplary benchmark scenarios featuring different mass
patterns. We have one benchmark point, BP1T2, with almost degenerate
$H_\uparrow$ and $H^\pm$ and a mass gap to the lighter
$H_\downarrow$. In BP2T2, $H_\downarrow$ and $H^\pm$ are closer in
mass and $H_\uparrow$ is the heaviest Higgs boson. The benchmarks
BP3T2 and BP4T2 feature nearly mass degenerate $H_\downarrow$ and
$H_\uparrow$ with a heavier charged Higgs boson in the former and a
lighter one in the latter case. The input parameters are given in
Table~\ref{tab:t2bps}. The overall mass spectrum is heavier than in
the type I C2HDM as expected from the lower bound on the charged Higgs
mass in type II. The dominant branching are summarised in
Table~\ref{tab:brst2bps}. These are determined by the mass pattern
together with the fact that the $H_3 VV$ ($V=W,Z$) coupling is very
small despite $R_{33}^2$ being small in BP1T2-BP3T2. Besides the
dominant decay into $t\bar{t}$, $H_3$ has a substantial
branching ratio into $ZH_2$ in BP1T2 and in into $W^\pm
H^\mp$ in BP2T2. In BP3T2 and BP4T2, the mass pattern forces $H_3$ to
mainly decay into $t\bar{t}$. This is the dominant decay channel for
$H_2$ in all four scenarios. The mass ordering of BP1T2 allows $H^\pm$
to decay with a significant branching ratio into $W^\pm H_2$ besides
the dominant decay into $t\bar{b}$. For all the other scenarios
$H^\pm$ almost exclusively decays into $t\bar{b}$. Again we find that
the fact that in the C2HDM not only mass hierarchies with
large mass gaps are preferred by the strong PT, induces also decay
patterns with SM particles in the final state. \s
\begin{table}[t!]
 \begin{center}
 \begin{tabular}{lcccc}
     \toprule
    & BP1T2 & BP2T2 & BP3T2 & BP4T2  \\ 
   \midrule
$m_{H_1}$ [GeV] &  125.09 & 125.09 & 125.09 & 125.09\\
$m_{H_2}$ [GeV] & 436.834 & 652.592 & 551.699 & 695.347 \\
$m_{H^\pm}$ [GeV] & 640.079 & 616.659 & 629.564 & 614.739 \\
$\mbox{Re}(m_{12}^2)$ [GeV$^2$] & 85376 & 121817 & 73628 & 113941 \\
$\alpha_1$ & 0.880 & 0.850 & 0.817 & 0.827 \\
$\alpha_2$ & -0.0156 & $5.945 \cdot 10^{-3}$ & $3.687\cdot 10^{-3}$ & -0.013 \\
$\alpha_3$ & 1.568 & -1.568 & -1.557 & 0.085\\
$\tan\beta$ & 1.399 & 1.224 & 1.216 & 1.182 \\ \hline 
$m_{H_3}$ [GeV] & 653.627 & 757.984 & 552.583 & 701.912 \\
$\xi_c$ & 1.190 & 1.077 & 1.169 & 1.016 \\
$R_{13}^2$ & $2.437 \cdot 10^{-4}$ & $3.534 \cdot 10^{-5}$ & $1.360\cdot 10^{-7}$
& $1.658 \cdot 10^{-4}$ \\
$R_{23}^2$ & 0.999 & 0.999 & 0.999 & $7.144 \cdot 10^{-3}$ \\
$R_{33}^2$ & $9.400 \cdot 10^{-6}$ & $7.433 \cdot 10^{-6}$ & $1.993\cdot 10^{-4}$
& 0.993 \\ \hline
$\sigma_{hh}^{\text{NLO}}$ [fb] & 62.16 & 38.97 & 57.15 & 42.78 \\
   \bottomrule
  \end{tabular}
\caption{Line 1-8: The input parameters of the type II CP-violating
  benchmarks with $H_1 
\equiv h$ and $\xi_c > 1$, compatible with all constraints. Line
9 to 13: The derived 3rd neutral Higgs boson mass, the
$\xi_c$ value and the CP-odd admixtures $R_{i3}^2$. Line 14: The NLO QCD
gluon fusion $hh$ production cross section at
$\sqrt{s}=14$~TeV.} \label{tab:t2bps}
   \end{center}
\vspace*{-0.2cm}
 \end{table}
\begin{table}[t!]
 \begin{center}
 \begin{tabular}{lll}
     \toprule
     BP1T2 & BP2T2 & BP3T2/BP4T2  \\ 
   \midrule
BR($H_2 \to t\bar{t}$) = 0.98 &  
BR($H_2 \to t\bar{t}$) = 0.99 & BR($H_2 \to t\bar{t}$) = 0.97/0.98 \\
\hline
BR($H_3 \to t\bar{t}$) = 0.54 & BR($H_3 \to t\bar{t}$)=0.76 
& BR($H_3\to t\bar{t}$) = 0.97/0.98 \\
 BR($H_3 \to ZH_2$) = 0.43 & BR($H_3 \to W^\mp H^\pm$)=0.22 
&   \\ \hline
BR($H^+ \to t\bar{b}$) = 0.58 & BR($H^+ \to t \bar{b}$)=0.99 
& BR($H^+\to t\bar{b}$) = 0.98/0.99 \\
 BR($H^\pm \to WH_2$) = 0.40 & \\
   \bottomrule
  \end{tabular}
\caption{The dominant branching ratios of the BP1T2-BP4T2 Higgs
  bosons.} 
\label{tab:brst2bps}
   \end{center}
\vspace*{-0.6cm}
 \end{table}

\begin{figure}[t!]
\begin{center}
\includegraphics[width=0.49\textwidth]{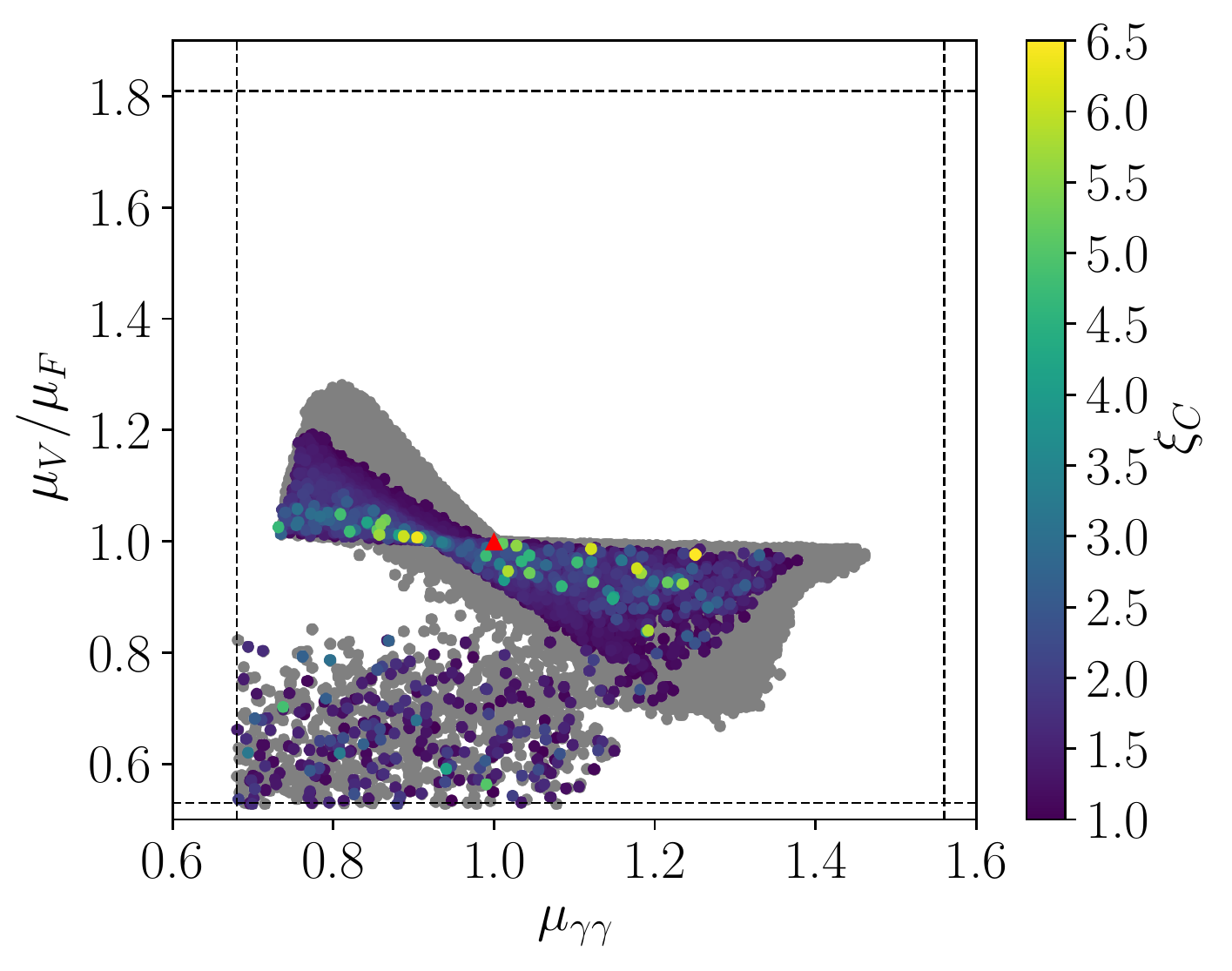}
\includegraphics[width=0.49\textwidth]{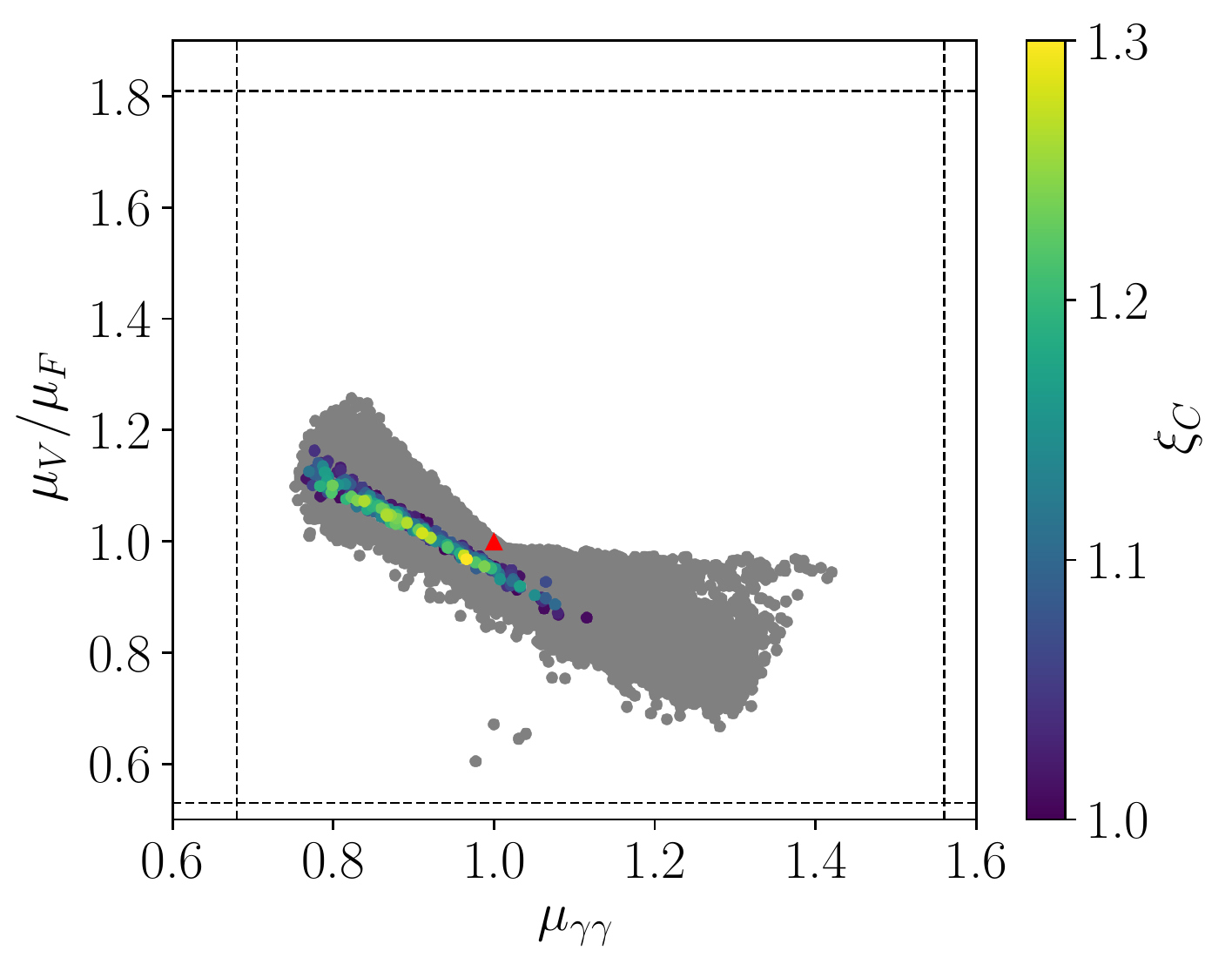}
%\vspace*{-0.2cm}
\caption{Type II, $H_1 = h$: $\mu_V/\mu_F$ versus
  $\mu_{\gamma\gamma}$. Grey: all points passing the applied
  constraints, colour: all points with additionally $\xi_c \ge 1$;
  left: all 2HDM points, right: only C2HDM points. The red triangle
  marks the SM result. 
 \label{fig:t2muvfmugaga}}
\end{center}
\vspace*{-0.2cm}
\end{figure}
\begin{figure}[t!]
\begin{center}
\includegraphics[width=0.49\textwidth]{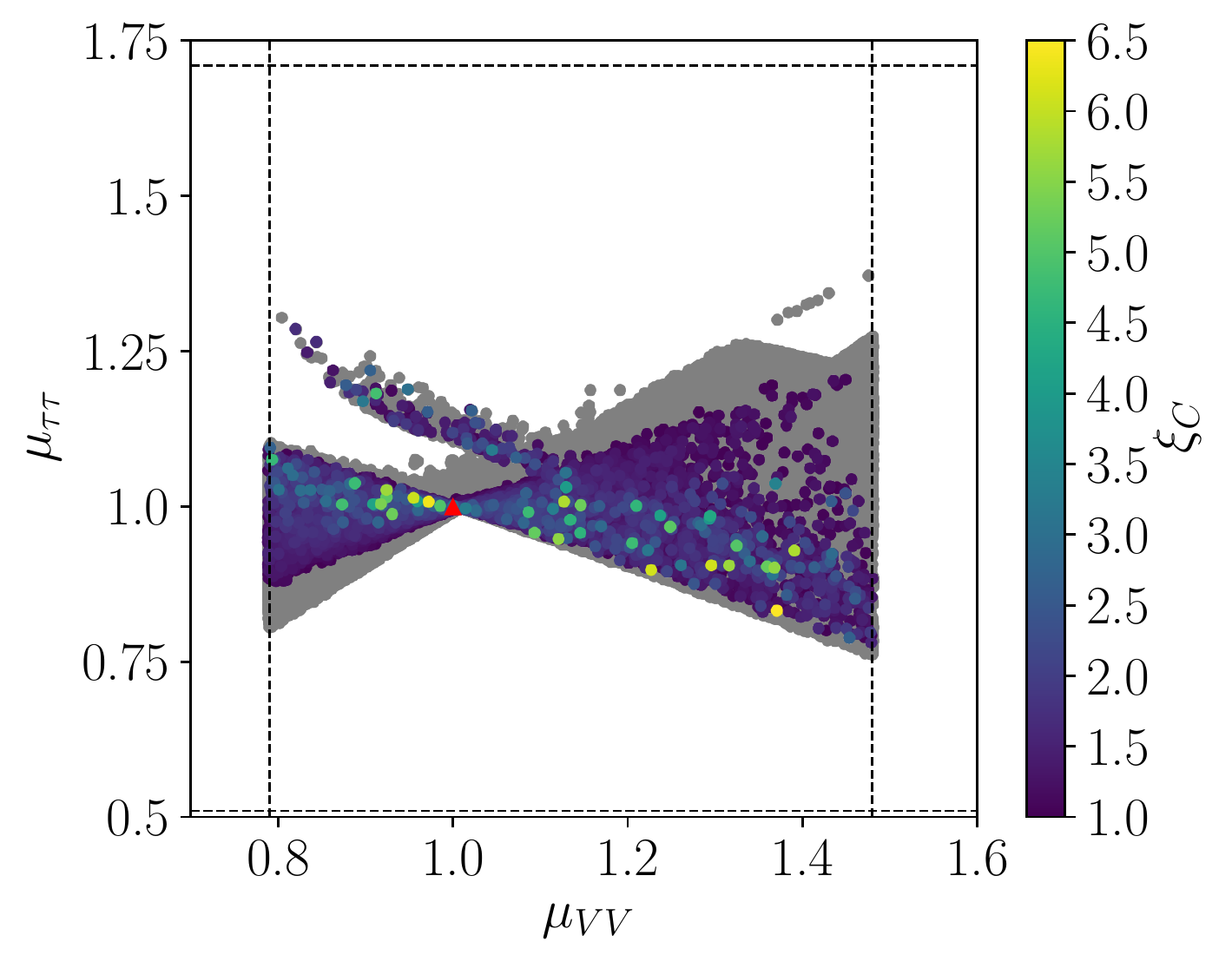}
\includegraphics[width=0.49\textwidth]{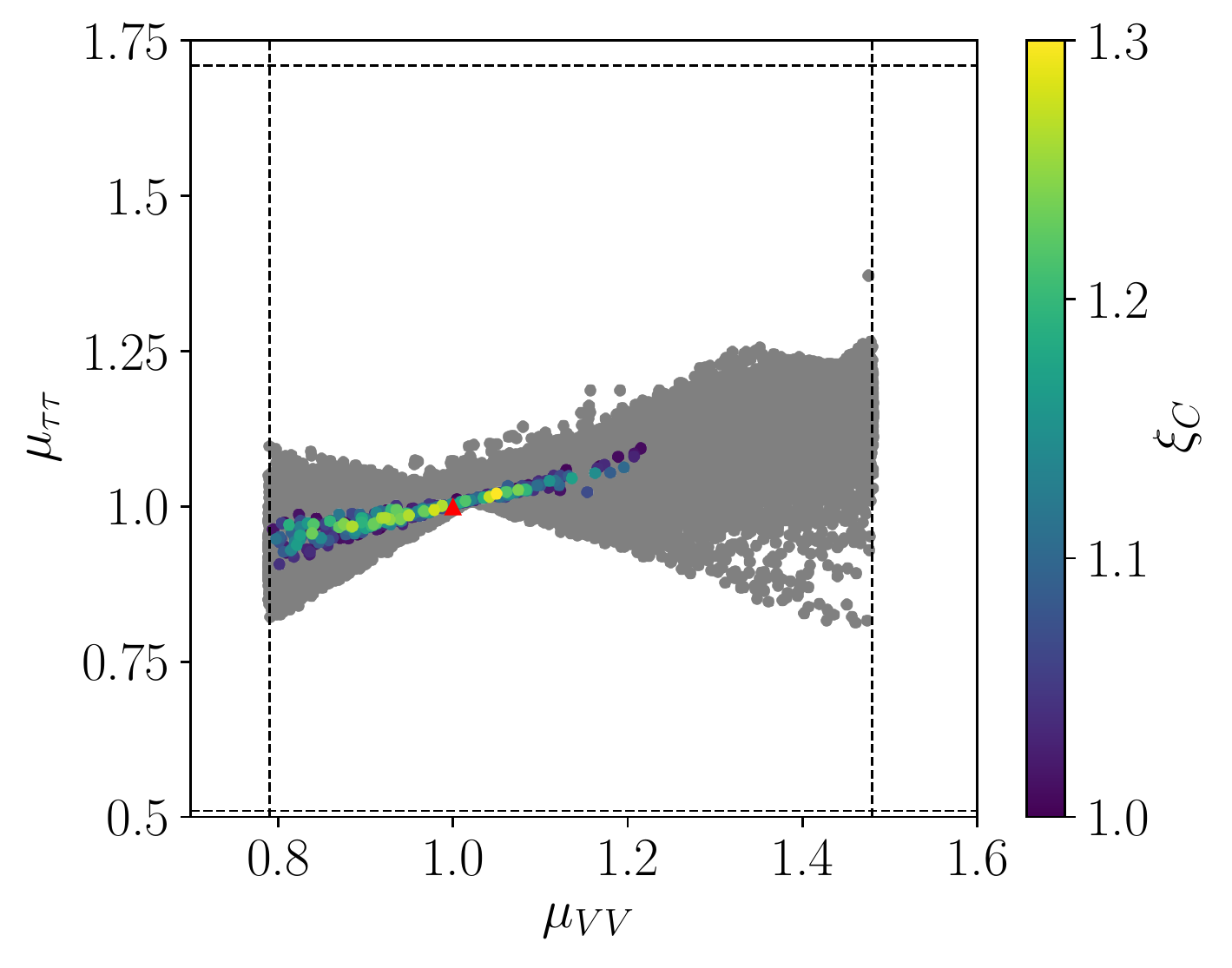}
%\vspace*{-0.2cm}
\caption{Type II, $H_1 = h$: $\mu_V/\mu_F$ versus
  $\mu_{\gamma\gamma}$. Grey: all points passing the applied
  constraints, colour: all points with additionally $\xi_c \ge 1$; 
  left: all 2HDM points, right: only C2HDM points. The red triangle
  marks the SM result.
 \label{fig:t2mutautaumuvv}}
\end{center}
\vspace*{-0.4cm}
\end{figure}
In Fig.~\ref{fig:t2muvfmugaga} we depict $\mu_V/\mu_F$ versus the
photonic rate $\mu_{\gamma\gamma}$ in grey for all points passing the
constraints and in colour for those with $\xi_c >1$. The left plot
comprises all 2HDM points, while in the right plot only the
CP-violating points are retained. The requirement of a strong PT restricts the region of
allowed $\mu$ values, and the results of the left plot reconfirm our findings of
\cite{Basler:2016obg} in the real 2HDM. Due to a more efficient scan
procedure applied here, we have now more scenarios compatible with
$\xi_c \ge 1$ in the wrong-sign limit region
\cite{Ferreira:2014naa,Ferreira:2014dya,Fontes:2014tga}  (corresponding to
the points in the lower left corner) where the
$h$ coupling to the massive gauge bosons has an opposite sign
with respect to its $b$-quark Yukawa coupling. The inclusion of CP
violation restricts the $\mu$-values further. We see a strong
correlation of $\mu_{\gamma\gamma}$ and $\mu_V/\mu_F$, with the latter
decreasing with increasing $\mu_{\gamma\gamma}$.
The wrong-sign limit is
completely excluded which, however, is already almost excluded due to the
applied constraints and not because
of a strong PT. The CP-violating 
scenarios with $\xi_c \ge 1$ preclude $\mu_V/\mu_F$ above 1.17 and
below 0.86 and $\mu_{\gamma\gamma}$ above 1.12 and below 0.76. Any
values outside these ranges point to the CP-conserving 2HDM. Note
also, that the SM point (red triangle) is not compatible with a strong
PT in the C2HDM. \s

As can be inferred from Fig.~\ref{fig:t2mutautaumuvv},
which is the same as Fig.~\ref{fig:t2muvfmugaga} but for
$\mu_{\tau\tau}$ versus $\mu_{VV}$, in the CP-violating case there is
also a strong correlation between $\mu_{\tau\tau}$ and $\mu_{VV}$. The
former increases with $\mu_{VV}$. Furthermore, no scenarios in the
wrong-sign regime (corresponding to the 
points in the upper left corner of Fig.~\ref{fig:t2mutautaumuvv}
(left)) are realized, which, again, is mostly due to the applied
constraints.
The value of $\mu_{\tau\tau}$ is restricted to values
between 0.9 and 1.10, as well as
$\mu_{VV}$ to the range between 0.8 and 1.22. \s

%%%%%%%%%%%%%%%%%%%%%%%%%%%%%%%%%%%%%%%%%%%%%%%%%%%%%%%%%%%
\section{Analysis of the trilinear Higgs self-couplings and Higgs pair
  production in the C2HDM Type II \label{eq:trilt2}}
\begin{figure}[t!]
\begin{center}
\includegraphics[width=0.49\textwidth]{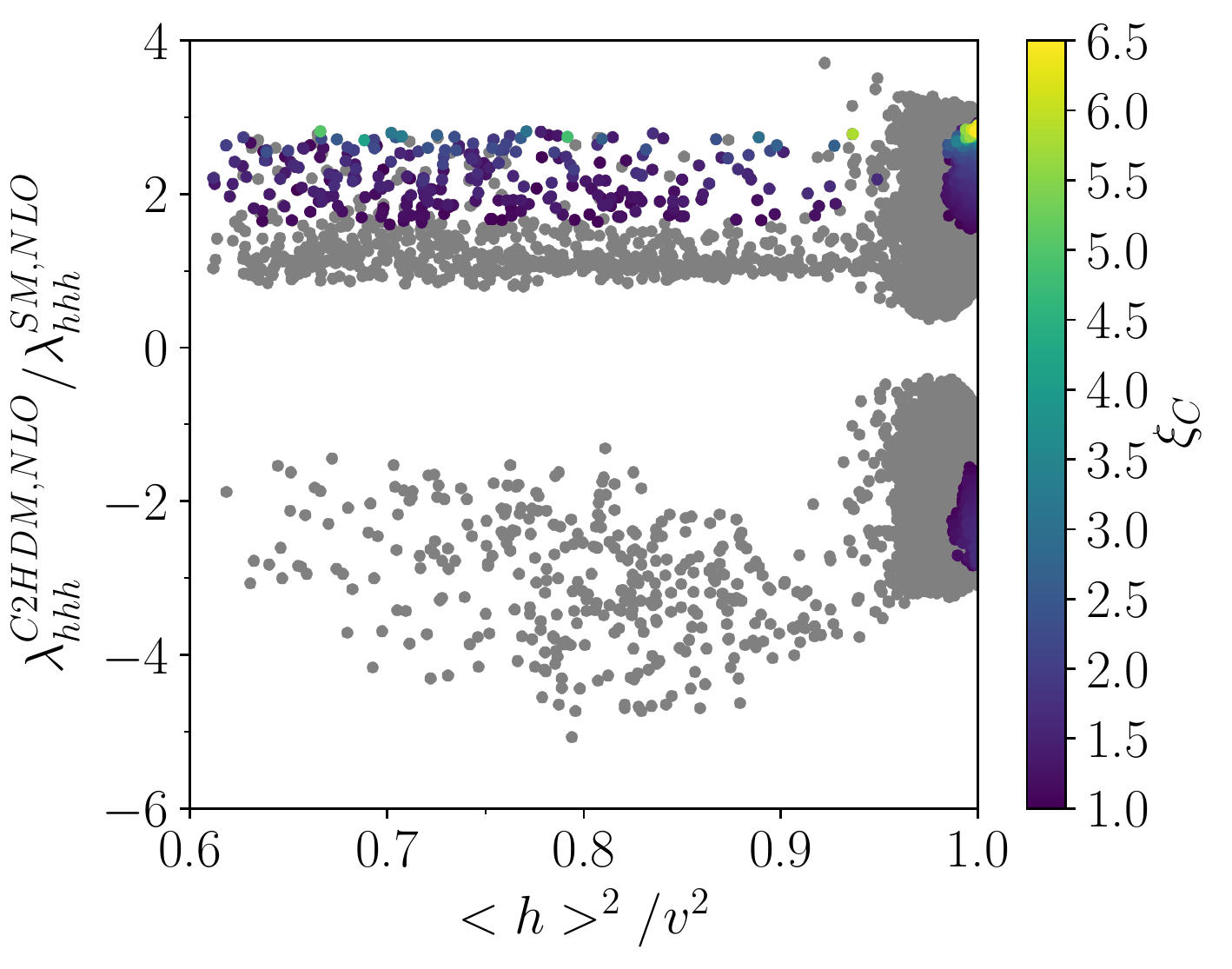}
\includegraphics[width=0.49\textwidth]{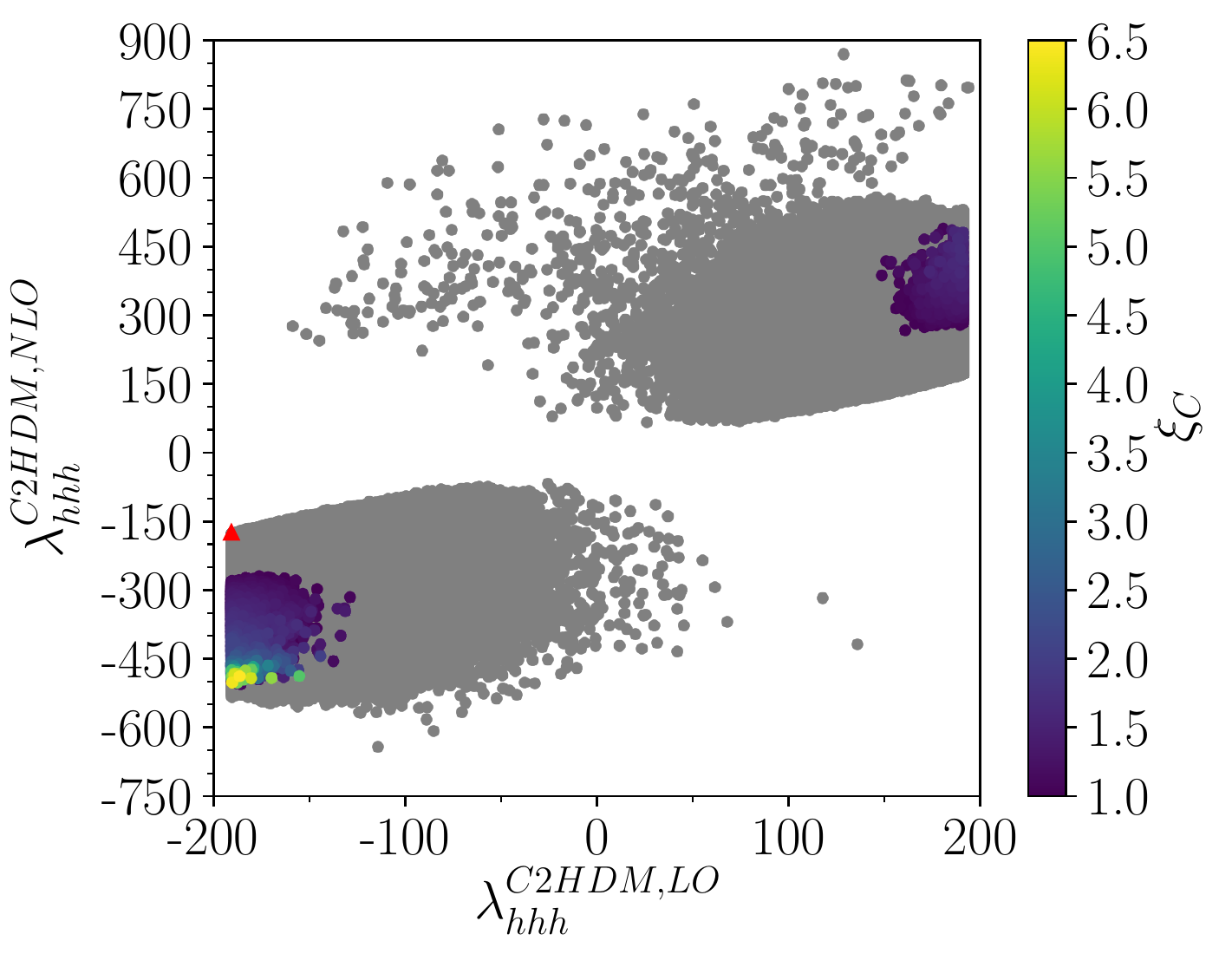}
%\vspace*{-0.2cm}
\caption{Type II, $H_1 = h$: Left: the C2HDM trilinear Higgs
  self-coupling between three SM-like Higgs bosons $h$ normalized to the
  SM value, at NLO, as function of the fraction of the VEV squared carried by
  $h$; right: the NLO coupling versus the LO
  value. Grey: all points passing the constraints; colour:
  points with additionally $\xi_c > 1$. The red triangle marks the SM point.
 \label{fig:hhhtrilt2}}
\end{center}
\vspace*{-0.4cm}
\end{figure}
We conclude our investigation with the discussion of the interplay between
the requirement of a strong PT and the trilinear Higgs self-coupling
among the SM-like Higgs bosons. 
Figure \ref{fig:hhhtrilt2} (left) depicts the values of the NLO trilinear Higgs coupling
between three SM-like Higgs bosons $h$ of the 
C2HDM normalized to the SM value, each at NLO, as function of the
fraction of the VEV squared carried by $h$, for all points passing the
constraints (grey) and for only those with a strong PT (colour). As in the
C2HDM type I, we 
observe that the C2HDM self-couplings can be enhanced with respect to
the SM, with a maximum enhancement factor of 5.1. If additionally the
PT is required to be of strong first order, the trilinear coupling has
to be somewhat larger than in the SM, but must not be above a factor of about
2.9 the SM value, in order not to weaken the PT again due to too large Higgs
masses. If only CP-violating points are taken into account this ratio
is slightly lowered to 2.4. 
%As can be inferred from the plot the
%scenarios with $\xi_c > 1$ are close to the alignment limit. 
The right plot shows the NLO coupling versus the LO one and underlines the
importance of the loop corrections. The wide spread points in the
upper half of the plot and the two isolated points in the lower right
part are due to the wrong-sign limit. In the
regions with a strong PT 
the maximum corrections amount to about a factor 3.3, and the largest
values of $\xi_c$ are found for the largest correction factor. Taking
into account only the CP-violating scenarios, the maximum enhancement is
lowered to 2.2. Note, that the sharp cut of the tree-level
couplings at large values are due to tree-level
unitarity. Interestingly, the plot shows that a
vanishing trilinear Higgs self-coupling between the SM-like Higgs
bosons is not compatible with the constraints any more, while in type
I this was not excluded. 
Concerning the trilinear Higgs self-couplings also involving
non-SM-like Higgs bosons we find the same overall features as in the
type I case. \s
%In particular, the ratio of the NLO coupling to the LO
%coupling for the majority of scenarios is close to 1, if a strong
%first order PT is required, in contrast to the coupling between three
%SM-like Higgs bosons. \s

In Table~\ref{tab:t2bps} we give the NLO QCD $hh$ production cross
sections through gluon fusion at $\sqrt{s}=14$~TeV for the four
benchmark points BP1-4T2. The size of the
non-SM-like Higgs boson masses allows for resonant enhancements 
through their on-shell production with subsequent decay into
$hh$. Depending on the size of the Higgs self-couplings between them
and $hh$, the C2HDM $hh$ cross section is more or less enhanced
compared to the SM.

%%%%%%%%%%%%%%%%%%%%%%%%%%%%%%%%%%%%%%%%%%%%%%%%%%%%%%%%%%%
\section{Conclusions \label{sec:conclusions}}
We have found that the NLO effects derived from the
effective potential have a non-negligible influence on the global
minimum and perturbativity at NLO. 
The requirement of the EW minimum to be the global minimum also at NLO
excludes ${\cal O} (5-25\%)$ of the generated scenarios compatible
with the applied constraints in the type I C2HDM with $H_1$ or $H_2
\equiv h$, denoted by C2HDM(I$_{H_1}$), and C2HDM(I$_{H_2}$) and the type
II C2HDM with $H_1 \equiv h$, denoted by C2HDM(II$_{H_1}$). In the type II C2HDM
with $H_2 \equiv h$, the mass differences enforced by the 
experimental constraints become large and, due to our renormalisation
procedure, induce large corrections to the Higgs self-couplings so
that a stable vacuum cannot be guaranteed any more so that we did not
investigate this version of the C2HDM further in this paper. Along the same
lines accordance with unitarity at NLO eliminates ${\cal
  O}(9-18\%)$ of the remaining points. \s

We showed that the presence of explicit CP violation at zero
temperature induces spontaneous CP violation at the EWPT. The size of
the induced phase correlates with the amount of explicit CP violation at
$T=0$. In type I a larger CP-violating phase at zero temperature is compatible with the
experimental constraints than in type II. Consequently, the total
CP-violating angle, including the spontaneously generated CP-violating
phase, at $T_c$ amounts up to 49$^\circ$ in the C2HDM(I$_{H_1}$) and 
up to about 8$^\circ$ in the C2HDM(II$_{H_1}$). In particular, in type I this
should be large enough to ensure successful baryogenesis. \s

We overall re-confirm our findings of our previous investigation of the PT in
the CP-conserving 2HDM for the scenarios in the CP-conserving
limit. Deviations in type I occur due to an updated limit on
$\tan\beta$ and in type II due to a more efficient scan procedure. 
A strong phase transition is found to be possible in all three
different remaining set-ups including CP violation, {\it
  i.e.}~the C2HDM(I$_{H_{1,2}}$) and the C2HDM(II$_{H_1}$). Overall, the
strength of the phase transition is smaller in the CP-violating case,
with $\xi_c^{\text{max}}=1.9$ in C2HDM(I$_{H_1}$) and
$\xi_{c}^{\text{max}}=1.3$ in C2HDM(II$_{H_1}$). In C2HDM(I$_{H_2}$)
we find only three points (BPCPV1-3) that are compatible with a strong
PT, with $\xi_c^{\text{max}}=1.48$. Although the SM-like Higgs boson has the
largest fraction of the VEV at the PT and hence mainly drives its
strength, through CP mixing all three neutral Higgs bosons, also the
non-SM-ones, receive a VEV.  Scenarios where heavy
non-SM-like Higgs bosons receive a large portion of the VEV at the
EWPT weaken its strength. Therefore either the lighter of the
non-SM-like Higgs bosons receives the larger portion of the VEV among
the two, or the overall spectrum is as light as possible. \s

\begin{table}[t!]
 \begin{center}
 \begin{tabular}{ll|l}
     \toprule
    & mass pattern & main phenomenological features  \\ %\hline \hline
   \midrule
C2HDM{I$_{H_1}$} &  \multicolumn{2}{c}{$m_\phi \equiv m_{\uparrow} \approx m_{H^\pm}$} \\ \midrule
BPi1 & $m_\phi - m_\downarrow \approx 180$~GeV &
BR$^{\text{max}}$: $H_\downarrow \to hh$, $H_\uparrow \to ZH_\downarrow$, 
$H^\pm \to WH_\downarrow$\\
BPi2 & $m_\phi - m_\downarrow \approx 260$~GeV &
BR$^{\text{max}}$: $H_\downarrow \to hh$, $H_\uparrow \to ZH_\downarrow$, 
$H^\pm \to WH_\downarrow$\\
& & combination of $H_2$ decays into $hh$, $Zh$ and $WW$ \\
& & allow for test of CP-mixed nature of $H_2$ \\
BPi3 & $m_\phi - m_\downarrow \approx 300$~GeV &
BR$^{\text{max}}$: $H_\downarrow \to WW$ despite $R_{23}^2=0.7$\\
& & $H_\uparrow \to ZH_\downarrow$, 
$H^\pm \to WH_\downarrow$ \\ \midrule
C2HDM{I$_{H_1}$} &  \multicolumn{2}{c}{$m_\varphi \equiv m_{\downarrow}
                   \approx m_{H^\pm}$} \\ \midrule
BPii1 & $m_\uparrow - m_\varphi \approx 260$~GeV & BR$^{\text{max}}$:
$H_\downarrow \to WW$, $H_\uparrow \to W^\mp H^\pm$,  $H^+ \to t\bar{b}$\\
BPii2 & $m_\uparrow - m_\varphi \approx 310$~GeV & BR$^{\text{max}}$: as
                                                BPii1, $\varphi$ lighter
   \\ \midrule
C2HDM{I$_{H_1}$} &  \multicolumn{2}{c}{$m_0 \equiv m_{\downarrow}
                   \approx m_{\uparrow}$} \\ \midrule
BPiii1 & $m_0 - m_{H^\pm} \approx 0$ GeV & BR$^{\text{max}}$: $H_{\downarrow,\uparrow} \to
t\bar{t}$, $H^+ \to t\bar{b}$ \\
BPiii2 & $m_{H^\pm} - m_0 \approx 80$~GeV & BR$^{\text{max}}$: as BPiii1 \\
BPiii3 & $m_0 - m_{H^\pm} \approx 77$~GeV & BR$^{\text{max}}$: as
                                            BPiii1 \\ \hline \hline
C2HDM{I$_{H_2}$} & \multicolumn{2}{c}{overall lighter spectrum than in
                   C2HDM{I$_{H_1}$}} \\ \midrule
BPCPV1 & $m_{H_3} \approx m_{H^\pm} \approx 375$~GeV & 
$H_\downarrow$ ($H_\uparrow$) mostly CP-even (CP-odd) \\
& $m_{\downarrow}= 120$~GeV & BR$^{\text{max}}$: 
$H_\downarrow \to b\bar{b}$, $H_\uparrow \to Z H_\downarrow$, $H^\pm
\to W^\pm H_\downarrow$ \\
BPCPV2 & max.~mass $m_{H^\pm} = 192$~GeV & 
$H_\downarrow$, $H_\uparrow$ CP-mixed \\
& $m_{\downarrow}= 91$~GeV & BR$^{\text{max}}$: 
$H_\downarrow \to b\bar{b}$, $H_\uparrow \to WW$, $H^\pm
\to W^\pm H_\downarrow$ \\
BPCPV3 & max.~mass $m_{H^\pm} = 167$~GeV & 
$H_\downarrow$ ($H_\uparrow$) mostly CP-odd (CP-even) \\
& $m_{\downarrow}= 118$~GeV & 
BR$^{\text{max}}$: 
$H_\downarrow \to b\bar{b}$, $H_\uparrow \to WW$, $H^\pm
\to W^\pm H_\downarrow$ \\ \hline \hline
C2HDM{II$_{H_1}$} & \multicolumn{2}{c}{overall heavier spectrum than in
                   C2HDM(I)} \\ \midrule
BP1T2 & $m_{\downarrow} = 437$~GeV & BR$^{\text{max}}:$ BR$(H_\downarrow \to
                               t\bar{t})$ = 0.98\\
& $m_{H^\pm} = 640$, $m_\uparrow = 654$~GeV & BR$(H_3 \to t\bar{t}/ZH_\downarrow)$
 = 0.54/0.43, \\
& & BR$(H^\pm \to t\bar{b}/WH_\downarrow)$ = 0.58/0.40
  \\
BP2T2 & $m_{H^\pm}=617$, $m_{\downarrow} = 653$~GeV & 
BR$^{\text{max}}:$ BR$(H_\downarrow \to t\bar{t})$ = 0.99\\
& $m_{\uparrow} = 758$~GeV & BR$(H_3 \to t\bar{t}/W^\pm H^\mp)$
 = 0.76/0.22, \\
& & BR$(H^+ \to t\bar{b})$ = 0.99 \\
BP3T2 & $m_\downarrow \approx m_\uparrow \approx 552$ GeV &
   BR$^{\text{max}}$ all almost exclusively: \\
& $m_{H^\pm} = 630$~GeV & $H_\downarrow \to t\bar{t}$, $H_\uparrow \to t\bar{t}$, $H^+ \to t\bar{b}$ \\
BP4T2 & $m_{H^\pm} = 615$~GeV &
   BR$^{\text{max}}$ all almost exclusively: \\
& $m_\downarrow \approx m_\uparrow \approx 698$~GeV & $H_\downarrow \to t\bar{t}$,
$H_\uparrow \to t\bar{t}$, $H^+\to t\bar{b}$ 
\\
   \bottomrule
  \end{tabular}
\caption{Main features of the CP-violating benchmark points with a
  strong phase transition.} \label{tab:summary}
   \end{center}
\vspace*{-0.6cm}
 \end{table}
 Concerning the mass hierarchies and hence the implications for
phenomenology, we confirm for scenarios
compatible with a strong PT in the CP-conserving limit 
the preference for a mass hierarchy with a
gap between the heavier and the lighter neutral non-SM-like Higgs bosons. The
heavier Higgs boson therefore mainly decays into
gauge+Higgs final states. Depending on the value of the charged Higgs
mass, which due to electroweak precision physics constraints has to be
mass degenerate with one of the neutral Higgs bosons, these final
state particles are neutral 
$(ZH_i)$ or charged $(W^\pm H^\mp)$. Such decays are a clear
indication of non-SM physics arising 
from an enlarged Higgs sector, which can be searched for at the LHC.
For CP-violating scenarios, however, no such preference is found: The
requirement of a strong phase transition combined 
with a CP-violating set-up where all Higgs bosons receive a VEV,
besides this mass pattern, also favours scenarios where all three neutral
Higgs bosons are rather close in mass. There are scenarios with one or
two of the non-SM-like neutral Higgs bosons degenerate with the charged
Higgs boson. Without mass hierarchies between the non-SM-like  Higgs
bosons (and small gauge-Higgs pair couplings involving the $h$)
they all decay into SM final states, with 
the specific nature of the final state particles determined by the
mass of the decaying Higgs boson. Among the
scenarios with a mass hierarchy we also have Higgs-to-Higgs decays and
decay patterns that allow to identify the CP-violating nature of the
Higgs boson through the combination of its decay channels. 
We have provided several benchmark scenarios for the C2HDM with a
strong PT. Their key features are summarised in
Table~\ref{tab:summary}. \s

The $\mu$-values of the SM-like Higgs boson in the C2HDM with a strong
PT turn out to be more restricted than in the CP-conserving case. This
can be exploited to exclude the C2HDM and additionally distinguish it
from the 2HDM. We summarise the distinctive  
features of the $\mu$-values in Table~\ref{tab:restric}. The regions
are excluded by the C2HDM but not by the 2HDM. Any measured value 
outside these regions rules out the CP-violating set-up for successful
baryogenesis. Note, however, that the wrong-sign-limit is mainly
disfavoured by the applied constraints and not by the requirement of
$\xi_c >1$ in contrast to the other regions given in the Table. \s
\begin{table}[t!]
 \begin{center}
 \begin{tabular}{ll|l}
     \toprule
excluded  & $\mu_V/\mu_F-\mu_{\gamma\gamma}$ &  $\mu_{\tau\tau}-\mu_{VV}$ \\ 
   \midrule
C2HDM(I$_{H_1}$) & $\mu_V/\mu_F < 0.9$ {\it and} $\mu_{\gamma\gamma} < 0.9$
&  $\mu_{\tau\tau} > 1$ {\it and} $\mu_{VV} < 1$\\
\midrule
C2HDM(II$_{H_1}$) & $\mu_{\gamma\gamma} > 1.12$ & $\mu_{VV} > 1.22$\\
& $\mu_{\gamma\gamma} < 1$ {\it and} $\mu_{V}/\mu_F < 0.82$
  (wsl) & $\mu_{VV} < 1.13$ {\it and} $\mu_{\tau\tau} >
   1.07$ (wsl) \\
& $\mu_V/\mu_F = \mu_{\gamma\gamma} =1$ (SM point) \\
   \bottomrule
  \end{tabular}
\caption{Distinguishing features between the C2HDM and 2HDM rates for
  scenarios with $\xi_c > 1$: The C2HDM excluded regions given here
  are still allowed by the 2HDM. Note that the wrong-sign limit (wsl)
  regions are excluded due to the applied constraints and not due to
  $\xi_c >1$.} \label{tab:restric} 
   \end{center}
\vspace*{-0.4cm}
 \end{table}

From the loop-corrected effective potential we derive the
NLO-corrected effective trilinear Higgs couplings. The C2HDM couplings
can be enhanced or suppressed relative to the SM case for all scenarios
compatible with the experimental and theoretical constraints. (Note,
that the maximum allowed value is constrained by the requirement of
unitarity.) The scenarios with a strong PT, however, require enhanced trilinear Higgs
self-couplings between the SM-like Higgs bosons, which is the Higgs
boson with the largest VEV and hence mainly drives
the phase transition. Including also the points in the CP-conserving
limit, the two enhancement regions are 
$\lambda_{hhh}^{\text{C2HDM,NLO}}/\lambda_{hhh}^{\text{NLO,SM}} =$
(-2.73...-1.1) and (1.1...2.9) for C2HDM(I$_{H_1}$), (-2.32...-0.94)
  and (1.04...2.58) for C2HDM(I$_{H_2}$) and (-2.83...-1.54)
  and (1.55...2.93) for C2HDM(II$_{H_1}$). For the purely CP-violating points these values
  reduce to (-2.30...-1.28) and (1.26...2.46) for
  C2HDM(I$_{H_1}$), -1.03
  and (1.03...1.54) for C2HDM(I$_{H_2}$) and (-2.35...-1.65)
  and (1.64...2.36) for C2HDM(II$_{H_1}$). Some of these
  deviations are large enough to 
  be measurable at the LHC. Note, that the upper bound on the
  coupling comes from the requirement not to weaken the PT again
  through too heavy Higgs masses. The NLO corrections are important,
  changing the LO coupling by up to a factor (1.07...8.25) in 
  C2HDM(I$_{H_1}$), a factor (0.93...2.32) in C2HDM(I$_{H_2}$) and (1.46...3.31)
  in C2HDM(II$_{H_1}$), for the scenarios compatible with $\xi_c >
  1$. Concentrating on purely CP-violating points 
  these factors change to (1.31...2.44), (0.93...1.35) and
  (1.51...2.17), respectively. The investigation
  of the trilinear self-couplings involving the non-SM-like Higgs
  bosons shows that the C2HDM values can be enhanced or
  suppressed compared to the SM, that the enhancement is less
  important for scenarios with a strong PT and that the NLO corrections are
  significant. We summarise that the requirement
  of a strong PT induces enhanced trilinear Higgs self-couplings among
  the SM-like Higgs bosons that deviate from the SM value
  significantly enough to be measurable at the LHC. The experimental
  confirmation of large self-couplings establishes a clear connection
  between a strong PT, {\it i.e.}~cosmology, and collider physics. The self-coupling
  is extracted from Higgs pair production at the LHC, with the main
  channel given by gluon fusion. We calculated for our benchmark
  scenarios the cross sections for $hh$ production including the NLO
  QCD corrections in the large top quark mass limit. Due to the
  possible resonant production of heavy 
  Higgs bosons with subsequent decay into $hh$ they are all enhanced
  compared to the SM case, so that the prospects for measuring them
  are promising. These cross sections do not include EW corrections,
  as they are not available. The derived sizes of the EW-corrected effective
  Higgs self-couplings indicate that these corrections can be
  sizeable. They could lower the Higgs pair production cross section. \s

In conclusion, the requirement of a strong PT with successful
baryogenesis demands an enlarged Higgs sector and has measurable
consequences for the C2HDM. In contrast to the real 2HDM not only mass
gaps between the Higgs bosons, but also degenerate
scenarios are favoured. The C2HDM with a strong PT predicts strong
correlations among the signal strengths of the SM-like Higgs boson. 
%The allowed $\mu$-rates of the SM-like Higgs
%boson are more constrained. 
Finally, the trilinear $hhh$ coupling must
be enhanced compared to the SM, and the additional Higgs bosons induce
$hh$ production cross sections that are larger than in the SM case and
can be measured at the LHC. The combination of successful baryogenesis
with collider phenomenology is a powerful tool to further restrict the
underlying model and to identify its true nature.

%features: BPCPV1 mass gap, $H_1$ larger portion of the VEV. BPCPV2,3 are
%closer in mass and $H_2$ features a larger fraction of the VEV.

%%%%%%%%%%%%%%%%%%%%%%%%%%%%%%%%%%%%%%%%%%%%%%%%%%%%%%%%%%%
\vspace*{0.3cm}

\section*{Acknowledgements}
The authors want to thank Pedro Ferreira, Ramona Gr\"ober, Marcel Krause, Jonas
M\"uller and Marco Sampaio for helpful discussions. PB acknowledges
financial support by the ``Karlsruhe School of Elementary Particle and
Astroparticle Physics: Science and Technology (KSETA)''.  
JW acknowledges financial support by the ``PIER Helmholtz graduate school''.

\vspace*{0.5cm}
%%%%%%%%%%%%%%%%%%%%%%%%%%%%%%%%%%%%%%%%%%%%%%%%%%%%%%%%%%%
%\section*{Appendix}
%\appendix
%
%Scenario for which we plot the temperature dependence of the VEVs.
%The main branching ratios of the
%non-SM-like Higgs bosons are
%\beq
%\mbox{BR} (H_\uparrow \to Z H_\downarrow ) &=& 0.860 \;, \quad
%\mbox{BR} (H_\uparrow \to Zh) = 0.086 \;, 
%\mbox{BR} (H_\uparrow \to t\bar{t} ) = 0.050 \;, \quad
%\\
%\mbox{BR} (H_\downarrow \to WW ) &=& 0.917 \;, \quad 
%\mbox{BR} (H_\downarrow \to b\bar{b}) = 0.042 \;, \quad 
%\mbox{BR} (H_\downarrow \to ZZ) = 0.023 \;, \quad \\
%\mbox{BR} (H^\pm\to W^\pm H_\downarrow ) &=& 0.865 \;, \quad
%\mbox{BR} (H^\pm\to W^\pm h) = 0.085 \;, \quad
%\mbox{BR} (H^+\to t\bar{b}) = 0.050 \;, \quad
%\eeq
%
%\section{Masses with thermal
%  corrections}\label{sec:app_debye_corrections}
%\textcolor{red}{Do we want to give them?}
%
%
%
%\vspace*{0.5cm}
%%%%%%%%%%%%%%%%%%%%%%%%%%%%%%%%%%%%%%%%%%%%%%%%%%%%%%%

\end{document}